\documentclass[aps,prx,notitlepage,onecolumn,longbibliography,superscriptaddress]{revtex4-1}

\usepackage[colorlinks=true,citecolor=blue]{hyperref}
\usepackage{amsmath,latexsym,amssymb,bm,color,graphicx,multirow,array,wrapfig,caption,scalerel,subcaption}
\usepackage[section]{placeins}

\newcommand{\dd}{\mathrm{d}}

\newcommand{\m}{\mathbb}

\newcommand{\cH}{\mathcal H}
\newcommand{\cT}{\mathcal T}
\newcommand{\cP}{\mathcal P}
\newcommand{\tcT}{\tilde{\mathcal T}}
\newcommand{\tcP}{\tilde{\mathcal P}}
\newcommand{\eq}[1]{Eq.~(\ref{#1})}
\newcommand{\fig}[1]{Fig.~\ref{#1}}
\newcommand{\Ref}[1]{Ref.~\onlinecite{#1}}
\newcommand{\tab}[1]{Table~\ref{#1}}
\renewcommand{\v}[1]{\boldsymbol{#1}}

\graphicspath{{figures/}}

\begin{document}

\title{Towards a complete classification of symmetry-protected topological phases for interacting fermions in three dimensions and a general group super-cohomology theory}

\author{Qing-Rui Wang}
\affiliation{Department of Physics, The Chinese University of Hong Kong, Shatin, New Territories, Hong Kong, China}
\author{Zheng-Cheng Gu}
\email{zcgu@phy.cuhk.edu.hk}
\affiliation{Department of Physics, The Chinese University of Hong Kong, Shatin, New Territories, Hong Kong, China}

\date{\today}

\begin{abstract}
The classification and construction of symmetry-protected topological (SPT) phases in interacting boson and fermion systems have become a fascinating theoretical direction in recent years. It has been shown that (generalized) group cohomology theory or cobordism theory gives rise to a complete classification of SPT phases in interacting boson/spin systems. The construction and classification of SPT phases in interacting fermion systems are much more complicated, especially in three dimensions. In this work, we revisit this problem based on an equivalence class of fermionic symmetric local unitary (FSLU) transformations. We construct very general fixed-point SPT wavefunctions for interacting fermion systems. We naturally reproduce the partial classifications given by special group super-cohomology theory, and we show that with an additional $\tilde{B}H^2(G_b, \mathbb Z_2)$ structure (the so-called obstruction-free subgroup of $H^2(G_b, \mathbb Z_2)$), a complete classification of SPT phases for three-dimensional interacting fermion systems with a total symmetry group $G_f=G_b\times \mathbb Z_2^f$ can be obtained for unitary symmetry group $G_b$.
We also discuss the procedure for deriving a general group super-cohomology theory in arbitrary dimensions. 
\end{abstract}

\maketitle

\tableofcontents

\section{Introduction}
In recent years, a new type of topological order---symmetry-protected topological (SPT) order \cite{gu09,chenScience2012,chen13}---has been proposed and intensively studied in interacting boson and fermion systems. Two-dimensional (2D) and three-dimensional (3D) topological insulators (TIs) \cite{hasan10,qi11} are the simplest examples of SPT phases, which are protected by time-reversal and charge-conservation symmetries. Although TIs were initially proposed and experimentally realized in essentially non-interacting electron systems, very recent studies have established their existence and stability even in the presence of strong interactions, by identifying non-perturbative quantum anomalies on various manifolds \cite{witten15}. 
The first attempt to systematically understand SPT phases in interacting systems was proposed in \Ref{gu09}, in which the author pointed out that the well-known spin-1 Haldane chain \cite{haldane83} was actually an SPT phase.  Later, a systematic classification of SPT phases for interacting bosonic systems in arbitrary dimensions with arbitrary global symmetry was achieved using generalized group cohomology theory \cite{chenScience2012,chen13,wen15} or cobordism theory \cite{kapustin14a}. This systematic classification essentially classifies the quantum anomalies associated with the corresponding global symmetries in interacting bosonic systems. In terms of the physical picture, it has been further pointed out that by gauging the global symmetry $G$, different SPT phases can be characterized by different types of braiding statistics of $G$-flux/flux lines in 2D/3D \cite{levin12,cheng2014,wanggu16,threeloop,ran14,wangcj15,wangj15,lin15}. Anomalous surface topological order has also been proposed as another very powerful way to identify and characterize different 3D SPT phases in interacting boson and fermion systems \cite{vishwanath13,wangc13,chen14,wangPRX2016, bonderson13,wangc13b,fidkowski13,chen14a,chong14,metlitski15}.

From the quantum information perspective, intrinsic topological phases are gapped quantum states that can be defined and classified by an equivalence class of finite depth local unitary transformations \cite{Chen2010}, which leads to the novel concept of long-range entanglement. However, in contrast to those intrinsic topological phases, all SPT phases can be adiabatically connected to a trivial disordered phase or to an atomic insulator without symmetry protection. Therefore, SPT phases contain only short-range entanglement, and can be constructed by applying local unitary transformations on a trivial product state. 
In particular, Refs.~\onlinecite{chenScience2012,chen13} introduced a systematic way of constructing fixed-point ground-state wavefunctions for bosonic SPT phases on arbitrary triangulations in arbitrary dimensions. Such a construction is fairly complete for bosonic SPT phases protected by unitary symmetry groups (up to 3D). So far, the only known example beyond this construction is the so-called $efmf$ SPT state \cite{vishwanath13}, which is protected by time-reversal symmetry in 3D. Later, it was shown that such an exotic bosonic SPT state could be realized by the Walker--Wang model \cite{walker12,burnell14}. 

The classification and systematic understanding of SPT phases in interacting fermion systems are much more complicated. One obvious way to achieve fruitful results is to study the reduction of the free-fermion classifications \cite{ff1,ff2} under the effect of interactions \cite{fidkowski10,qi13,yao13,ryu12,gu14b,you2014,morimoto15}. However, this approach misses those fermionic SPT (FSPT) phases that cannot be realized in free-fermion systems \cite{wanggu16,neupert14}. A slightly general way to understand FSPT in interacting fermion systems is to stack some additional bosonic SPT states onto a free-fermion SPT state \cite{chong14}. An arguably fairly complete classification of TIs in interacting electron systems \cite{wangc-science} can be constructed in such a way. However, there is no natural reason to believe that all FSPT phases in interacting systems can be realized by the abovementioned stacking constructions, and counterexamples can be constructed explicitly. Moreover, it has been shown that certain bosonic SPT phases become ``trivial'' (adiabatically connected to a product state) \cite{Gu2014,chong14} when embedded into interacting fermion systems. Apparently, the stacking construction cannot explain all of these subtle issues. Therefore, a systematic understanding and the construction of interacting FSPT phases are very desirable.

The first attempt to classify interacting FSPT phases in general dimensions was proposed in Ref.~\onlinecite{Gu2014}, in which a class of FSPT phases was constructed systematically by generalizing the usual group cohomology theory into the so-called special group super-cohomology theory. However, it turns out that such a construction cannot give rise to all FSPT phases except in one dimension, where the obtained classification of FSPT phases perfectly agrees with previous results \cite{chen11b,fidkowski11}). On the other hand, quantum anomalies characterized by spin cobordism \cite{kapustin14} or invertible spin topological quantum field theory (TQFT) \cite{freed14,Gaiotto2016,gaiotto16} suggest a rich diversity of FSPT phases, although it is not clear how to construct these FSPT states in an explicit and systematic way.

Alternatively, the idea of gauging fermion parity \cite{gu14b,barkeshli14,cheng15,wangcj16,lan16} provides another way to understand FSPT. 
In 2D, a fairly complete classification of FSPT can be obtained in this way, which also agrees with the anomaly classification given by spin cobordism and invertible spin TQFT \cite{kapustin14,freed14,gaiotto16,freed16,morgan16}. It has been shown that the mathematical objects that classify 2D FSPT phases with a total symmetry $G_f=G_b\times \mathbb Z_2^f$(where $G_b$ is the bosonic global symmetry and $\mathbb Z_2^f$ is the fermion parity conservation symmetry), can be summarized as three group cohomologies of the symmetry group $G_b$ \cite{cheng15,gaiotto16}: $H^1 (G_b,\mathbb Z_2 )$, $BH^2(G_b,\mathbb Z_2)$, and $H^3(G_b, U_T(1))$. $H^1 (G_b,\mathbb Z_2 )$, which corresponds one-to-one to the $\mathbb Z_2$ subgroups of $G_b$, classifies FSPT phases with Majorana edge modes. $BH^2(G_b,\mathbb Z_2)$, the obstruction-free subgroup of $H^2(G_b,\mathbb Z_2)$, is formed by elements $n_{2} \in H^2(G_b,\mathbb Z_2)$ that satisfy $Sq^2(n_{2})=0$ in
$H^{4}(G_b,U_T(1))$, where $Sq^2$ is the Steenrod square, $Sq^2:
H^{d}(G_b,\mathbb Z_2)\rightarrow H^{d+2}(G_b,\mathbb Z_2) \subset H^{d+2}[G_b,U_T(1)]$. $H^3(G_b,U_T(1))$ is the well-known classification of bosonic SPT phases. Physically, the $H^1 (G_b,\mathbb Z_2 )$ layer can be constructed by decorating a Majorana chain \cite{Kitaev2001}, which is a one-dimensional (1D) invertible fermionic TQFT, onto the domain walls of symmetry group $G_b$. The $BH^2(G_b,\mathbb Z_2)$ layer can be constructed by decorating complex fermions, which are zero-dimensional (0D) invertible TQFT, onto the intersection points of $G_b$-symmetry domain walls.
Nevertheless, the decoration scheme can suffer from obstructions, and only subgroup $BH^2(G_b,\mathbb Z_2)$ classifies valid and inequivalent 2D FSPT phases. 
Some interesting examples of SPT phases have been studied in 3D based on a Walker--Wang model construction \footnote{The fermion parity symmetry is gauged in such constructions.}, e.g., DIII-class topological superconductors \cite{fidkowski13,metlitski14,kitaevsre,queiroz16}. Unfortunately, it is impossible to construct all FSPT phases using the Walker--Wang model. It is even unclear how to reproduce all of the special group super-cohomology constructions in this way. Very recently, some new interacting FSPT phases beyond special group super-cohomology were formally proposed by using spin TQFT \cite{juven16}. However, a general principle and lattice model realization are still lacking.

\subsection{Classify FSPT phases via equivalence classes of fermionic symmetric local unitary transformations}
In this paper, we propose a general physical principle to construct all FSPT phases in 3D with the total symmetry group $G_f=G_b\times \mathbb Z_2^f$. A previous work showed that in the presence of global symmetry, symmetry-enriched topological (SET) phases can be defined and classified by equivalence classes of symmetric local unitary transformations \cite{cheng16,heinrich16}. In particular, SPT phases can be realized as a special class of SET phases whose bulk excitations are trivial and can be adiabatically connected to a product state in the absence of global symmetry.  

In Ref.~\onlinecite{Gu2015}, it was shown that fermionic local unitary (FLU) transformations can be used to define and classify intrinsic topological phases for interacting fermion systems. The Fock space structure and fermion parity conservation symmetry of fermion systems can be naturally encoded into FLU transformations. Let us first briefly review the definition of FLU transformation. Similar to the local bosonic systems, the finite-time evolution
generated by a local fermion Hamiltonian defines an
equivalence relation between gapped states in interacting fermion systems:
\begin{align}
\label{fLUdef}
 |\psi(1)\rangle \sim |\psi(0)\rangle \text{\ iff\ }
 |\psi(1)\rangle =  \cT\left[e^{-i\int_0^1 d\lambda \tilde{H}_f(\lambda)}\right] |\psi(0)\rangle
\end{align}
where $\cT$ is the path-ordering operator and $\tilde{H}_f(\lambda)$ is a local fermionic Hamiltonian defined in Fock space.  
We will
call $ \cT\left[e^{-i \int_0^1 d\lambda \tilde{H}_f(\lambda)}\right] $ a FLU evolution. It is well-known that the finite-time FLU evolution is closely
related to fermionic quantum circuits with finite
depth, which is defined through piecewise FLU operators.  A
piecewise FLU operator has the form
$ U_{pwl}= \prod_{\v i} e^{-i  H_f(\v i)}\equiv \prod_{\v i}U(\v i)$,
where $ H_f(\v i) $ is a fermionic Hermitian operator and $U(\v i)$ is the corresponding fermionic unitary operator defined in Fock space that preserve fermion parity (e.g., contains even number of fermion creation and annihilation operators) and act on a region labeled by $\v
i$. Note that regions labeled by different $\v i$'s are not
overlapping. We further require that the size of each region is less
than some finite number $l$. The unitary operator $U_{pwl}$
defined in this way is called a piecewise fermionic local
unitary operator with range $l$.  A fermion quantum
circuit with depth $M$ is given by the product of $M$
piecewise fermionic local unitary operators:
$U^M_{circ}= U_{pwl}^{(1)} U_{pwl}^{(2)} \cdots
U_{pwl}^{(M)}$.
It is believed that any FLU evolution can be
simulated with a constant depth fermionic quantum circuit and
vice versa. Therefore, the equivalence relation between gapped states in interacting fermion systems can be rewritten in terms of constant
depth fermionic quantum circuits:
\begin{equation}
|\psi(1)\rangle  \sim |\psi(0)\rangle  \text{ iff } |\psi(1)\rangle  = U^M_{circ} |\psi(0)\rangle 
\end{equation}
Thus, we can use the term FLU transformation to refer to both FLU evolution and constant depth fermionic quantum circuit. From the definition of FSPT state, it is easy to see that (in the absence of global symmetry):
\begin{equation}
  |\text{FSPT}\rangle  = U^M_{circ} |\text{Trivial}\rangle \label{FLU}
\end{equation} 
Namely, a FSPT state can be connected to a trivial state (e.g., a product state) vial FLU transformation (in the absence of global symmetry). Now let us consider the
entanglement density matrix $\rho_A$ of for a FSPT state in region $A$.
$\rho_A$ may act on a subspace of the Hilbert space in
region A, and the subspace is called the support space $\tilde{V}_A$
of region $A$. Clearly, Eq.(\ref{FLU}) implies that the support space of any FSPT in region $A$ must be one dimensional. This is simply because a trivial state (e.g., a product state) has a one dimensional support space, and any FSPT state will become a product state via a proper local basis change (induced by a FLU transformation). 

In the presence of global symmetry, we can further introduce the notion of fermionic symmetric local unitary (FSLU) transformations to define and classify fermionic SET (FSET) phases in interacting fermion systems. By FSLU transformation, we mean the corresponding piecewise FLU operator is invariant under symmetry $G_b$. More precisely, we have
$ U_{pwl}= \prod_{\v i} e^{-i  H_f(g_{\v i0},g_{\v i1},g_{\v i2},\cdots)}\equiv \prod_{\v i}U(g_{\v i0},g_{\v i1},g_{\v i2},\cdots)$ and $U(gg_{\v i0},gg_{\v i2},gg_{\v i3},\cdots)=U(g_{\v i0},g_{\v i1},g_{\v i2},\cdots)$ for any $g \in G_b$. (We note that here we choose the group element basis $g_{\v i0},g_{\v i1},g_{\v i2},\cdots$ to represent fermionic symmetric unitary operator acting on a region labeled by $\v i$.)
Again, FSPT phases are a special class of FSET phases that have trivial bulk excitation and can be adiabatically connected to a product state in the absence of global symmetry.    
Thus, we need only to enforce the FSLU transformations to be one dimensional (when acting on the support space $\rho_A$ for any region $A$) to classify all FSPT states. 

\subsection{Summary of main results}

It turns out that the novel concept of FSLU transformation allows us to construct very general fixed-point FSPT states of 2D and 3D FSPT phases. 
All of these fixed-point wavefunctions admit exactly solvable parent Hamiltonians consisting of commuting projectors on an arbitrary triangulation with an arbitrary branching structure. We begin with the 2D case, in which the discrete spin structure can be implemented by Kasteleyn orientations \cite{Kasteleyn1963,Cimasoni2007,Cimasoni2008}, allowing us to decorate Majorana chains onto $G_b$-symmetry domain walls \cite{Tarantino2016,Ware2016}. We then show how to implement the discrete spin structure on a triangulation of a 3D orientable spin manifold, which is a nontrivial generalization of 2D Kasteleyn orientation. The discrete spin structure allows us to decorate the Majorana chains onto the intersection lines of $G_b$-symmetry domain walls in a self-consistent and topologically invariant way. The fundamental mathematical data describing such a decoration scheme belong to $H^2(G_b, \mathbb Z_2)$,
subjected to an obstruction on $H^4(G_b, \mathbb Z_2)$. The obstruction can be understood through the following physical picture. As Kasteleyn orientation is not always possible for a large loop (the 3D discrete spin structure can be used to construct \emph{local} Kasteleyn orientations of small loops), complex fermion decoration on the intersection points of $G_b$-symmetry domain walls is typically required, and this is only possible when the $H^4(G_b, \mathbb Z_2)$ obstruction vanishes. Furthermore, another obstruction on $H^5(G_b, U_T(1))$ is generated by wavefunction renormalization to finally determine whether the entire decoration scheme of Majorana chains is valid for a fixed-point wavefunction in 3D. 

The precise mathematical objects that classify 3D FSPT phases with a total symmetry $G_f=G_b\times \mathbb Z_2^f$ can also be summarized as three group cohomologies of the symmetry group $G_b$: $\tilde{B}H^2 (G_b,\mathbb Z_2 )$, $BH^3(G_b,\mathbb Z_2)$, and $H^4_{\rm rigid}(G_b, U_T(1))$. $\tilde{B}H^2 (G_b,\mathbb Z_2 )$, the obstruction-free subgroup of $H^2 (G_b,\mathbb Z_2 )$, is formed by elements $\tilde n_{2} \in H^2(G_b, \mathbb Z_2)$ that simultaneously satisfy $Sq^2(\tilde n_{2})=0$ in
$H^{4}(G_b,\mathbb Z_2)$ and $\mathcal O(\tilde n_{2})=0$ in $H^{5}(G_b,U_T(1))$, where $\mathcal O$ is some unknown cohomology operation (to the best of our knowledge) that maps $\tilde n_{2}$ satisfying $Sq^2(\tilde n_{2})=0$ in $H^{2}(G_b,\mathbb Z_2)$ into an element in $H^{5}(G_b,\mathbb Z_8) \subset H^{5}[G_b,U_T(1)]$. The explicit expression of $\mathcal O$ is very complicated, and it is computed in a physical way in section~\ref{sec:3D:fSPT}. $BH^3(G_b,\mathbb Z_2)$, the obstruction-free subgroup of $H^3(G_b,\mathbb Z_2)$, is formed by elements $n_{3} \in H^3(G_b,\mathbb Z_2)$ that satisfy $Sq^2(n_{3})=0$ in
$H^{4}(G_b,U_T(1))$. We note that $BH^3(G_b,\mathbb Z_2)$ and $H^4_{\rm rigid}(G_b, U_T(1))\equiv H^4(G_b,\mathbb Z_2)/\Gamma$ were derived in the special group super-cohomology classification. Recall that $H^4(G_b,U_T(1))$ is the well-known classification of bosonic SPT phases and $\Gamma$ is a normal subgroup of $H^4(G_b,U_T(1))$ generated by $Sq^2(n_2)$, where $n_2 \in H^2 (G_b , \mathbb Z_2 )$ and $Sq^2(n_2)$ are viewed as elements of $H^4[G_b, U_T(1)]$. Physically, $\Gamma$ describes those trivialized bosonic SPT phases when embedded into interacting fermion systems.

Together with several previous works \cite{Gu2014,cheng15}, we conjecture that up to spacial dimension $d_{sp}=3$, FSPT with symmetry $G_f=G_b \times \mathbb Z_2^f$ can be classified by the general group super-cohomology class $H^{d_{sp}+1}_{f}[G_f,U_T(1)]$ defined by the exact sequences summarized in Table~\ref{supercohomology}. We note that for spacial dimension $d_{sp}>1$, general group super-cohomology theory is defined by two short exact sequences. The first short exact sequence can be understood as decoration of complex fermions onto the intersection points of the $G_b$-symmetry domain walls, which was first derived by special group super-cohomology theory. The second exact sequence can be understood as decoration of Kitaev's Majorana chains onto the intersection lines of $G_b$-symmetry domain walls, and our construction gives rise to a general scheme to compute $\tilde{B}H^{d_{sp}-1} (G_b,\mathbb Z_2 )$ (the obstruction-free subgroup of $H^{d_{sp}-1} (G_b,\mathbb Z_2 )$) in arbitrary dimensions. As an application, we also illustrate the classification results of FSPT phases for some simple symmetry group $G_b$ in all physical dimensions in Table~\ref{tbF}.

\begin{table*}[tb]
\centering
\begin{tabular}{|c|c|}
\hline
$d_{sp}$  & short exact sequence   \\
\hline
$0$ &  $0\rightarrow H^1[G_b,U_T(1)]\rightarrow H^{1}_f[G_f,U_T(1)]\rightarrow \mathbb Z_2  \rightarrow 0$ \\
\hline
$1$ &  $0\rightarrow H^2[G_b,U_T(1)]\rightarrow H^{2}_{f}[G_f,U_T(1)]\rightarrow H^{1}(G_b,\mathbb Z_2)  \rightarrow 0$ \\
\hline
\multirow{2}{*}{$2$} & $0\rightarrow H^3[G_b,U_T(1)] \rightarrow \cH^{3}[G_f,U_T(1)]\rightarrow BH^{2}(G_b,\mathbb Z_2) \rightarrow 0$  \\  
& $0\rightarrow \cH^3[G_f,U_T(1)] \rightarrow H^{3}_{f}[G_f,U_T(1)]\rightarrow H^{1}(G_b,\mathbb Z_2) \rightarrow 0$ \\
\hline
\multirow{2}{*}{$3$} & $0\rightarrow H^4_{\rm{rigid}}[G_b,U_T(1)] \rightarrow \cH^{4}[G_f,U_T(1)]\rightarrow BH^{3}(G_b,\mathbb Z_2) \rightarrow 0$  \\
& $0\rightarrow \cH^{4}[G_f,U_T(1)] \rightarrow H^{4}_{f}[G_f,U_T(1)]\rightarrow   \tilde{B}H^{2}(G_b,\mathbb Z_2) \rightarrow 0$ \\
\hline
\end{tabular}
\caption{Classifying FSPT phases up to spacial dimension $d_{sp}=3$ with a total symmetry $G_f=G_b \times \mathbb Z_2^f$ using a general group super-cohomology class computed from short exact sequences. Note that $\cH^{d_{sp}+1}[G_f,U_T(1)]$ is the so-called special group super-cohomology proposed in \Ref{Gu2014} and that in lower dimensions with $d_{sp}=0,1$, we have $H^{d_{sp}+1}_{f}[G_f,U_T(1)]\equiv\cH^{d_{sp}+1}[G_f,U_T(1)]$. }\label{supercohomology}
\end{table*}

\begin{table}[tb]
 \centering
 \begin{tabular}{ |c|c|c|c|c| }
 \hline
  $G_b\ \backslash d_{sp}$ & $0$ & $1$ & $2$ & $3$   \\
\hline
$\m Z_2$& $\m Z_2^2$ & $\m Z_2$ & $\m Z_8$ & $\m Z_1$\\
\hline 
$\m Z_{2k+1}$
& $\m Z_{4k+2}$ & $\m Z_1$ & $\m Z_{2k+1}$ & $\m Z_1$  \\
\hline
$\m Z_{2k}$
& $\m Z_{2k}\times \m Z_2$ & $\m Z_2$ &
$\begin{cases}
\m Z_{4k}\times\m Z_2,\ \ k\text{ even}\\
\m Z_{8k},\quad\quad\quad k\text{ odd}
\end{cases}$
& $\m Z_1$   \\
\hline
$\m Z_{2}\times \m Z_2$& $(\m Z_{2})^3$ & $(\m Z_{2})^3$ & $(\m Z_8)^2 \times \m Z_4$ & $(\m Z_{2})^2 $   \\
\hline
$\m Z_{2}\times \m Z_4$& $\m Z_4\times (\m Z_{2})^2 $ & $(\m Z_{2})^3$ & $(\m Z_8)^2 \times (\m Z_2)^3$ & $\m Z_4 \times \m Z_2$   \\
\hline
$\m Z_{4}\times \m Z_4$& $(\m Z_{4})^2\times \m Z_2 $ & $(\m Z_2)^2 \times \m Z_4$ & $(\m Z_{8})^2 \times \m Z_{4} \times (\m Z_{2})^3 $ & $(\m Z_{4})^2 \times \m Z_{2}$   \\
\hline
$\m Z_{2}\times \m Z_8$& $\m Z_8\times (\m Z_{2})^2$ & $(\m Z_{2})^3$ & $ \m Z_{16} \times \m Z_{8}\times (\m Z_{2})^3 $ & $\m Z_{8} \times \m Z_{2}$   \\
\hline
 \end{tabular}
 \caption{
Classification of FSPT phases with a total symmetry $G_f=G_b \times \mathbb Z_2^f$ in $d_{sp}$-spatial dimensions constructed
using general group super-cohomoloy  for some
simple symmetries (represented by the bosonic symmetry groups $G_b$).   Here $\m Z_1$ means that
our construction only gives rise to the trivial phase.  $\m Z_n$ means that the
constructed non-trivial SPT phases plus the trivial phase are labeled by the
elements in $\m Z_n$.
}
\label{tbF}
\end{table}

Finally, regarding the completeness of general group super-cohomology classification for 3D FSPT phases, we present some physical arguments. Although the decoration of complex fermions on the intersection points of $G_b$-symmetry domain walls and the decoration of Majorana chains on $G_b$-symmetry domain walls give rise to a complete classification of 2D FSPT phases, this does not necessarily imply that this is also true in 3D. In fact, it has been pointed out \cite{wen15} that the decoration of invertible TQFT on the $G_b$-symmetry domain walls may also give rise to new SPT states. For bosonic SPT states, decoration of the so-called $E_8$ state on the $G_b$-symmetry domain walls indeed produces the $efmf$ SPT state beyond group cohomology classification. It has also been pointed out that $H^1(G_b,\mathbb Z)$ classifies these additional bosonic SPT states. As $H^1(G_b,\mathbb Z)$ is trivial for the unitary symmetry group $G_b$ and $H^1(\mathbb Z_2^T,\mathbb Z)=\mathbb Z_2$ for the anti-unitary time-reversal symmetry, we understand why the $efmf$ state is the only non-trivial root state of bosonic SPT states beyond group cohomology classification with time-reversal symmetry. For interacting fermion systems, in principle, we can decorate a $p+ip$ state (the root state of 2D fermionic invertible TQFT) onto the $G_b$-symmetry domain walls. However, as $H^1(G_b,\mathbb Z)$ is trivial for the unitary symmetry group $G_b$, there are no new FSPT states with unitary symmetry group $G_b$. For time-reversal symmetry, it is possible to generate new FSPT states in this way, and we discuss this possibility in our future work. 

\subsection{Organization of the paper}
The remainder of this paper is organized as follows. We begin with the definition of Hilbert space and the basic structure of fixed-point wavefunctions for FSPT states with total symmetry $G_f=G_b\times \mathbb Z_2^f$ in 1D, 2D and 3D in section~\ref{sec:wavefunction}. In section~\ref{sec:2D:spin}, we give a brief review of discrete spin structures and Kasteleyn orientations in 2D. In section~\ref{sec:2D:fSPT}, we derive the fixed-point conditions for FSLU transformations under wavefunction renormalization and re-derive the classifications of 2D FSPT phases. In section~\ref{sec:3D:spin}, we discuss how to generalize the discrete spin structure and \emph{local} Kasteleyn orientation in 3D. In section~\ref{sec:3D:fSPT}, we use the concept of equivalence class of FSLU transformations and wavefunction renormalization to obtain the construction and classification of 3D FSPT phases. Finally, we offer conclusions and discussions for possible future directions.

Readers less interested in the detailed mathematical construction of (local) Kasteleyn orientations are invited to skip some part of section~\ref{sec:2D:spin} and \ref{sec:3D:spin}, and read directly section~\ref{sec:2D:fSPT} and \ref{sec:3D:fSPT} of constructing FSPT states. The only prerequisites are some terminology conventions and the conclusion that we can construct (local) Kasteleyn orientations systematically and rigorously on the resolved dual lattice of arbitrary triangulations of spin manifolds in arbitrary dimensions.

\section{Fixed-point wavefunctions of FSPT phases}
\label{sec:wavefunction}

\subsection{Constructing fixed-point wavefunction and classification for FSPT phases in 1D}

\begin{figure*}[h!]
\centering
\includegraphics[scale=2]{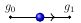}
\caption{Bosonic and fermionic degrees of freedoms for 1D fixed-point FSPT states on a link. The black dots are bosonic degrees of freedom labelled by $g_i\in G$ on sites. The blue ball represents the complex fermion $c_{(ij)}$ at the center of the link $\langle ij\rangle$. The arrow represents the local order of two sites.}
\label{fig:dof1D}
\end{figure*}

As a warm up, let us begin with fixed-point wavefunction in 1D and use FSLU transformation to re-derive the well know classification result of 1D FSPT phases. The building block of bosonic and fermionic degrees of freedom in the 1D FSPT model is shown in \fig{fig:dof1D}. Similar to the bosonic SPT phase, every (locally ordered) vertex $i$ of the 1D lattice has bosonic degrees of freedom labeled by a group element $g_i\in G_b$. (Recall that the FSPT phases have a total symmetry $G_f=G_b\times \mathbb Z_2^f$.) A spinless complex fermion $c_{(ij)}$ is at the center of each link $\langle ij\rangle$ (see the blue ball in \fig{fig:dof1D}), and the fermion occupation number $n_1(g_i,g_j)$ is either $0$ or $1$. Let $|0\rangle$ be the ground state of no fermions on any of the links; then, a generating set of the Fock space is given by $\prod _{(ij) \in l}c^\dagger_{(ij)}|0\rangle$, where $l\subset L$ is a subset of all links $L$, including the empty set. Thus, the full local Hilbert space for our 1D model on a fixed lattice $\cT$ (triangulation of 1D spacial manifold) is:
\begin{equation}
L^{1D}_{\cT}=\bigoplus_{l\subset L}\left(\prod _{(ij) \in l}c^\dagger_{(ij)}|0\rangle \bigotimes\prod _{v \in V(\cT)} \mathbb{C}^{|G_b|}\right).
\end{equation}

Here, $|G_b|$ is the order of the bosonic symmetry group $G_b$. As a vector space, the fermionic Hilbert space on the links is the same as the tensor product $\bigotimes_{L(\cT)}\mathbb{C}^2$; however, the Fock space structure means that a local Hamiltonian for a fermion system is non-local when regarded as one for a boson system. We note that the structure of total bosonic and fermionic Hilbert space on arbitrary triangulations is the same as the 1D case of \Ref{Gu2014}, although the latter is considering the spacetime picture.

Our 1D fixed-point state is a superposition of those basis states with all possible triangulations $\cT$:
\newcommand{\wfconfoneD}{
\includegraphics[scale=1.3]{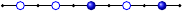}
}
\begin{align}
\label{eq:1Dwf}
|\Psi\rangle = \sum_{\text{all conf.}} \Psi\left(
\vcenter{\hbox{\wfconfoneD}}
\right) \stretchleftright{\Bigg|}{\ 
\vcenter{\hbox{\wfconfoneD}}
\ }{\Big\rangle}.
\end{align}
In the following, we will derive the rules of wavefunction renormalization generated by FSLU transformations for the above wavefunction. We will obtain the conditions for fixed-point wavefunction and show how to construct all FSPT states with total symmetry $G_f=G_b\times \mathbb Z_2^f$ in 1D.

\subsubsection{Fermionic symmetric local unitary transformation}

To obtain a fixed-point wavefunction for \eq{eq:1Dwf}, we need to understand the changes of the wavefunction under renormalization. In 1D, renormalization can be understood as removing some bosonic or fermionic degrees of freedom by reducing the number of vertices. The basic renormalization process is known as (2-1) Pachner move of triangulation of 1D manifold. Since we have a bosonic degree of freedom at each vertex and a fermionic degree of freedom at each link, the (2-1) move effectively reduces the Hilbert space of one bosonic mode and one fermionic mode.

To be more precise, the (2-1) move is a FSLU transformation between the fermionic Fock spaces on two different triangulations:
\begin{align}
\label{eq:1D_F1_a}
\Psi\left(
\vcenter{\hbox{\includegraphics[scale=1.5]{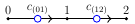}}}
\right)\quad = \quad
F(g_0,g_1,g_2)
\quad
\Psi\left(
\vcenter{\hbox{\includegraphics[scale=1.5]{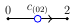}}}
\right),
\end{align}
where the $F$ operator is defined as
\begin{align}
\label{eq:F1D}
F(g_0,g_1,g_2) = 
\frac{1}{|G_b|^{1/2}}\nu_2(g_0,g_1,g_2)
c^{\dagger n_1(g_1,g_2)}_{(12)}
c^{\dagger n_1(g_0,g_1)}_{(01)}
c^{n_1(g_0,g_2)}_{(02)}.
\end{align}
We note that the $|G_b|$ is the order of the group $G_b$ and we introduce the normalization factor $1/|G_b|^{1/2}$ in the above expression due to the change of vertex number. 
Here, $\nu_2(g_0,g_1,g_2)$ is a $U_T(1)$-valued function with variables $g_i\in G_b$ and $c^{\dagger}_{(ij)}$ is the creation operator for $c$ fermions at link $\langle ij\rangle$, etc. $n_1(g_i,g_j)\in \{0,1\}$ is a $\mathbb Z_2$-valued function indicating whether there is a $c$ fermion at link $\langle ij\rangle$ or not. Since we are constructing symmetric state, both $\nu_2$ and $n_1$ should be symmetric under the action of $G_b$(We note that $\nu_2(gg_0,gg_1,gg_2)=\nu_2^*(g_0,g_1,g_2)$ if $g$ is anti-unitary). So they are $U_T(1)$-valued 2-cochain and $\mathbb Z_2$-valued 1-cochain respectively. Because the renormalization process (2-1) move should preserve the fermion parity, we have $\dd n_1(g_0,g_1,g_2) = n_1(g_1,g_2) + n_1(g_0,g_2) + n_1(g_0,g_1) = 0$ (mod 2). Therefore $n_1$ is in fact a $\mathbb Z_2$-valued 1-cocycle.

\subsubsection{Consistent equations and equivalence classes}

Since we are constructing fixed-point wavefunction, \eq{eq:1Dwf} should be invariant under renormalization. For instance, we can use two different sequences of $F$ moves \eq{eq:1D_F1_a} to connect a fixed initial state and a fixed final state. Different approaches should give rise to the same wavefunction. These constraints give us the consistent equations for $\nu_2$.

The simplest example is the following two paths between two fixed states:
\begin{align}
\label{eq:1D_eq1}
\Psi\left(
\vcenter{\hbox{\includegraphics[scale=1.5]{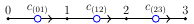}}}
\right) 
&=
F(g_1,g_2,g_3)\ 
\Psi\left(
\vcenter{\hbox{\includegraphics[scale=1.5]{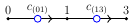}}}
\right),\\\nonumber
&=
F(g_1,g_2,g_3) F(g_0,g_1,g_3) \ 
\Psi\left(
\vcenter{\hbox{\includegraphics[scale=1.5]{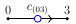}}}
\right),
\end{align}
\begin{align}
\label{eq:1D_eq2}
\Psi\left(
\vcenter{\hbox{\includegraphics[scale=1.5]{1D_fig12.pdf}}}
\right) 
&=
F(g_0,g_1,g_2)\ 
\Psi\left(
\vcenter{\hbox{\includegraphics[scale=1.5]{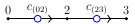}}}
\right),\\\nonumber
&=
F(g_0,g_1,g_2) F(g_0,g_2,g_3) \ 
\Psi\left(
\vcenter{\hbox{\includegraphics[scale=1.5]{1D_fig9.pdf}}}
\right).
\end{align}
The constraint is that the product of $F$ moves for the above two processes equal to each other:
\begin{align}
\label{eq:1Deq}
F(g_0,g_1,g_3) F(g_1,g_2,g_3)=F(g_0,g_2,g_3) F(g_0,g_1,g_2).
\end{align}
Substituting the expression of $F$ move \eq{eq:F1D} into this equation and using the fact $\dd n_1=0$ (mod 2), we find that the above equation for fermionic operators is equivalent to a purely bosonic one without any fermion sign:
\begin{align}
\dd \nu_2(g_0,g_1,g_2,g_3) = \frac{\nu_2(g_1,g_2,g_3) \nu_2(g_0,g_1,g_3)}{\nu_2(g_0,g_2,g_3) \nu_2(g_0,g_1,g_2)} =1.
\end{align}
That means $\nu_2$ should be a $U_T(1)$-valued 2-cocycle, provided that the wavefunction \eq{eq:1Dwf} is a fix-point wavefunction. This $\nu_2$ data is the same as the construction of bosonic SPT states.


Using a FSLU transformation, we can redefine the basis state $|\{g_l\}\rangle$ as
\begin{align}
|\{g_l\}\rangle' = U_{\mu_1,m_0} |\{g_l\}\rangle 
= \prod_{\langle ij \rangle} \mu_1(g_i,g_j) \prod_{\langle i \rangle} \left[ f^{m_0(g_i)}_{iA}f^{m_0(g_i)}_{iB} \right] \prod_{\langle ij\rangle} \left[ f^{\dagger m_0(g_j)}_{jA} f^{\dagger m_0(g_i)}_{iB} \right] |\{g_l\}\rangle,
\end{align}
where we first create two complex fermions $f_{jA}$ and $f_{iB}$ near the two ends of the link $\langle ij\rangle$ ($i<j$), and then annihilate the two fermions $f_{iA}$ and $f_{iB}$ near the vertex $i$ when gluing the two links sharing vertex $i$. To preserve the fermion parity and be symmetric, $m_0$ should be a 0-cocycle (with $\mathbb Z_2$ coefficient): $m_0(gg_i)=m_0(g_i)$ and $\dd m_0(g_i,g_j)=m_0(g_j)+m_0(g_i)=0$. $\mu_1(g_i,g_j)$ is a phase factor associated with link $\langle ij\rangle$. In this new basis, the fermionic  $F$ move is $F' = U_{\mu_1,m_0} F U_{\mu_1,m_0}^\dagger$. After eliminating all $f$ fermions (all the fermion signs are cancelled), one find that the phase factor in \eq{eq:F1D} becomes
\begin{align}
\nu_2'(g_0,g_1,g_2) = \nu_2(g_0,g_1,g_2) \frac{\mu_1(g_1,g_2)\mu_1(g_0,g_1)}{\mu_1(g_0,g_2)}.
\end{align}
Since our gapped phases are defined by FSLU transformations, $\nu_2'$ and $\nu_2$ belong to the same phase. In general, the elements $\nu_2$ in the same group cohomology class in $H^2(G_b, U_T(1))$ correspond to the same 1D FSPT phase. This is consistent with the general result obtained from the path-integral formalism in \Ref{Gu2014} that $\nu_{d+1}$ can be gauge transformed to
\begin{align}\label{eq:fcoboundary}
\nu'_{d+1} = \nu_{d+1}\cdot\dd \mu_{d}\cdot(-1)^{Sq^2(m_{d-1})},
\end{align}
for $Sq^2(m_0)$ is trivial in the 1D FSPT case.

In summary, 1D FSPT is characterized by $n_1\in H^1(G_b,\mathbb Z_2)$ and $\nu_2\in H^2(G_b,U_T(1))$. This is consistent with the previous result \cite{Gu2014}.

\subsection{Constructing fixed-point wavefunction for FSPT phases in 2D and 3D}
The fixed-point wavefucitons for FSPT phases in 2D and 3D are much more complicated. We will describe all the details and explain the corresponding physical meanings below. 

Similar to the wavefunction renormalization scheme for 2D bosonic SET phases, we consider the quantum state defined on an arbitrary triangulation for 2D and 3D FSPT phases. The triangulation admits a branching structure that can be labeled by a set of local arrows on all links (edges) with no oriented loop for any triangle. Mathematically, the branching structure can be regarded as a discrete version of a $\rm spin^c$ structure and can be consistently defined on arbitrary triangulations of 2D and 3D orientable manifolds. 

We begin with the construction of fixed-point wavefunction in 2D; then, the generalization to 3D becomes straightforward. As any 2D FSPT state can be naturally mapped to a 2D bosonic SET state by gauging the fermion parity symmetry, our construction for fixed-point wavefunctions is greatly inspired by such connections. In particular, an FSPT state with total symmetry $G_f=G_b\times \mathbb Z_2^f$ can be mapped to a $G_b$ symmetry-enriched toric code model. As a simple example, fixed-point wavefunctions and commuting projector parent Hamiltonians on an arbitrary trivalent graph (due to triangulation) with $G_b=\mathbb Z_2$ were constructed in \Ref{cheng16}. It has also been shown that all of these SET states can be obtained by gauging fermion parity from FSPT states with total symmetry $G_f=\mathbb Z_2\times \mathbb Z_2^f$.

\begin{figure*}[h!]
\centering
\includegraphics[]{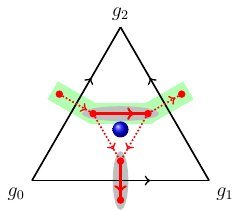}
\caption{Fermionic degrees of freedom in a triangle. The red dots represent Majorana fermions at the two sides of each link. The blue ball represents the complex fermion of the special group super-cohomology model at the center of the triangle. The green strip is the decorated Kitaev's Majorana chain onto the dual lattice $\cP$.}
\label{fig:dof}
\end{figure*}

The building block of bosonic and fermionic degrees of freedom in our 2D FSPT model is shown in \fig{fig:dof}. Exactly as in the bosonic SET phase, every vertex $i$ of the space triangulation has bosonic degrees of freedom labeled by a group element $g_i\in G_b$. (Recall that the FSPT phases have a total symmetry $G_f=G_b\times \mathbb Z_2^f$.) A spinless complex fermion $c$ is at the center of each triangle/face (see the blue ball in \fig{fig:dof}), and the fermion occupation number is either $0$ or $1$. Let $|0\rangle$ be the ground state of no fermions on any of the triangles; then, a generating set of the Fock space is given by $\prod _{(ijk) \in f}c^\dagger_{(ijk)}|0\rangle$, where $f\subset F$ is a subset of all triangles $F$, including the empty set. In addition, each link has two Majorana fermions on its two sides, an arrangement that is equivalent to spinless complex fermion $a$. Similar to the $c$ fermion on each triangle, let $|\tilde{0}\rangle$ be the ground state of no fermions on any of the links; then, a generating set of the Fock space is given by $\prod_{(ij) \in l}a^\dagger_{(ij)}|\tilde{0}\rangle$, where $l\subset L$ is a subset of all links $L$, including the empty set. Thus, the full local Hilbert space for our 2D model on a fixed triangulation $\cT$ is
\begin{equation}
L^{2D}_{\cT}=\bigoplus_{f\subset F} \bigoplus_{l\subset L}\left(\prod _{(ijk) \in f}c^\dagger_{(ijk)}|0\rangle \bigotimes \prod _{(ij) \in l}a^\dagger_{(ij)}|\tilde{0}\rangle \bigotimes\prod _{v \in V(\cT)} \mathbb{C}^{|G_b|}\right).
\end{equation}
Here, $|G_b|$ is the order of the bosonic symmetry group $G_b$. As a vector space, the fermionic Hilbert space on the triangles and links is the same as the tensor product $\bigotimes_{F(\cT)}\mathbb{C}^2 \otimes \bigotimes_{L(\cT)}\mathbb{C}^2$; however, the Fock space structure means that a local Hamiltonian for a fermion system is non-local when regarded as one for a boson system. We note that the structure of total fermionic Hilbert space on arbitrary triangulations is slightly more general than that given in \Ref{Gu2015}; this allows us to construct very general FSET states in 2D. However, the construction of general FSET states is beyond the scope of this paper. 

As mentioned above, for FSPT states, the support space of FSLU transformations must be one-dimensional such that it can adiabatically connect to a product state in the absence of global symmetry. Therefore, the fermionic states of $c$ and $a$ fermions on the triangles and edges are completely fixed by the configuration of group elements $\{g_i\}$ on the vertices. In particular, the equivalence classes of complex fermion occupation number of $c$ fermions are uniquely determined by the elements in $BH^2(G_b, \mathbb Z_2)$ (the obstruction-free subgroup of $H^2(G_b, \mathbb Z_2)$), which was first proposed by the special group super-cohomology construction of FSPT phases. Essentially, the complex fermion $c$ can be regarded as a decoration on the intersection points of $G_b$-symmetry domain walls. \Ref{cheng15, Tarantino2016,Ware2016} pointed out that a Majorana chain can be decorated onto the $G_b$-symmetry domain walls to generate a complete set of FSPT states in 2D.
This layer of decoration is uniquely determined by the elements in $H^1(G_b, \mathbb Z_2)$. The Majorana fermions must be paired (see the gray ellipse in \fig{fig:dof}) to form Kitaev's Majorana chains on the $G_b$ symmetry domain walls (see the green strip in \fig{fig:dof}). This requires a discrete spin structure---the Kasteleyn orientations on the dual trivalent lattice (with proper resolution for the lattice sites, as seen in \fig{fig:dof})---such that the total fermion parity of the $a$ fermion is always even on any closed loop. We review all of the details in section~\ref{sec:2D:spin}. An example of triangulation of the torus and decoration of Kitaev's Majorana chains is given in \fig{fig:torus}. The full details are discussed in section~\ref{sec:2D:fSPT}.

\newcommand{\wfconf}{
\includegraphics[]{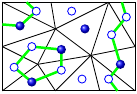}
}

Thus, our 2D fixed-point state is a superposition of those basis states with all possible triangulations $\cT$.
\begin{align}
|\Psi\rangle = \sum_{\text{all conf.}} \Psi\left(
\vcenter{\hbox{\wfconf}}
\right) \stretchleftright{\Bigg|}{\ 
\vcenter{\hbox{\wfconf}}
\ }{\Big\rangle}.
\end{align}
In section~\ref{sec:2D:fSPT}, we derive the rules of wavefunction renormalization generated by FSLU transformations. We also obtain the conditions for fixed-point wavefunctions and show how to construct all FSPT states with total symmetry $G_f=G_b\times \mathbb Z_2^f$ on arbitrary triangulations in 2D.

\begin{figure*}[h!]
\centering
\includegraphics[]{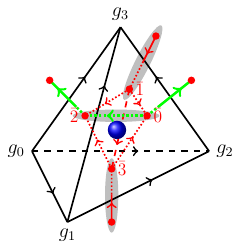}
\caption{Fermionic degrees of freedom in a tetrahedron. The red dots represent Majorana fermions on the two sides of each triangle. The blue ball represents the complex fermion of the (special) group super-cohomology model at the center of the tetrahedron. The green line is the decorated Kitaev's Majorana chain on the dual lattice $\cP$.}
\label{fig:dof3D}
\end{figure*}

In the following, we generalize all of the above constructions to 3D. The building block of bosonic and fermionic degrees of freedom is shown in \fig{fig:dof3D}. Again, every vertex $i$ of the 3D space triangulation has a bosonic degree of freedom labeled by a group element $g_i\in G_b$. However, the spinless complex fermion $c$ introduced by special group super-cohomology theory now resides on each tetrahedron (see the blue ball in \fig{fig:dof3D}). In addition, each triangle of the space tetrahedron has two Majorana fermions on its two sides, which is again equivalent to a spinless complex fermion $a$. Similar to the 2D case, let $|0\rangle$ and $|\tilde{0}\rangle$ be the ground states of no fermions on any tetrahedron and triangle; then, a generating set of the Fock space is given by $\prod _{(ijkl) \in t}c^\dagger_{(ijkl)}|0\rangle \bigotimes \prod _{(ijk) \in f}a^\dagger_{(ijk)}|\tilde{0}\rangle$, where $t\subset T$ is a subset of all tetrahedra $T$, including the empty set, and $f\subset F$ is a subset of all triangles $F$, including the empty set. Thus, the full local Hilbert space of our 3D model on a fixed triangulation $\cT$ is
\begin{equation}
L^{3D}_{\cT}=\bigoplus_{t\subset T} \bigoplus_{f\subset F}\left(\prod _{(ijkl) \in t}c^\dagger_{(ijkl)}|0\rangle \bigotimes \prod _{(ijk) \in f}a^\dagger_{(ijk)}|\tilde{0}\rangle \bigotimes\prod _{v \in V(\cT)} \mathbb{C}^{|G_b|}\right).
\end{equation}

Similar to the 2D case, the fermionic states of $c$ and $a$ fermions on the tetrahedra and triangles are also completely fixed by the configuration of group elements $\{g_i\}$ on the vertices. The equivalence classes of complex fermion occupation number of the $c$ fermion are uniquely determined by the elements in $BH^3(G_b, \mathbb Z_2)$ (the obstruction-free subgroup of $H^3(G_b, \mathbb Z_2)$), which was also first proposed by the special group super-cohomology construction of FSPT states. It is not a surprise that in 3D, the complex fermion $c$ can also be regarded as a decoration (subjected to obstructions) onto the intersection points of $G_b$-symmetry domain walls. The most interesting new feature here is that a Majorana chain can also be decorated onto the intersection lines of $G_b$-symmetry domain walls, and such a construction generates a new set of FSPT states in 3D. As expected, this layer of decoration also requires a discrete spin structure on the dual trivalent lattice (with a proper resolution for the lattice sites as well, as seen in \fig{fig:dof3D}), and the Majorana fermions must also be paired to form Kitaev's Majorana chains (see the green line in \fig{fig:dof3D}). However, such decorations are subjected to a fundamental obstruction on $H^4(G_b, \mathbb Z_2)$ due to fermion parity conservation. We discuss all of the details in section~\ref{sec:3D:spin}. Furthermore, the fixed-point condition of wavefunction renormalization gives rise to a secondary obstruction on $H^5(G_b, U_T(1))$, which is explored in full in section~\ref{sec:3D:fSPT}.

\newcommand{\wfconfB}{
\includegraphics[]{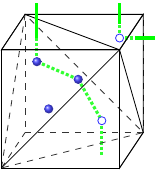}
}

Finally, our 3D fixed-point state is a superposition of those basis states with all possible triangulations $\cT$.
\begin{align}
|\Psi\rangle = \sum_{\text{all conf.}} \Psi\left( 
\vcenter{\hbox{\wfconfB}}
\right) \stretchleftright{\Bigg|}{
\vcenter{\hbox{\wfconfB}}
}{\Big\rangle}.
\end{align}
In section~\ref{sec:3D:fSPT}, we derive the rules of wavefunction renormalization generated by FSLU transformations. We also obtain the conditions for fixed-point wavefunction and show how to construct all FSPT states with total symmetry $G_f=G_b\times \mathbb Z_2^f$ on arbitrary triangulations in 3D.

\section{Constructions and classifications for FSPT states in 2D}

\begin{figure*}[h!]
\centering
\includegraphics[]{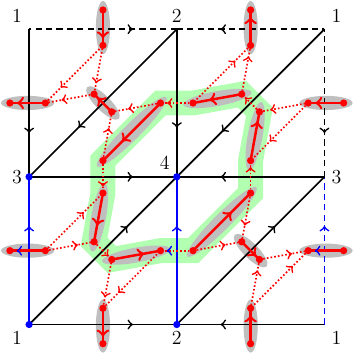}
\caption{Example of triangulation $\cT$ of torus and Kitaev chain decoration. All vertices $\langle1\rangle$, $\langle2\rangle$, $\langle3\rangle$, and $\langle4\rangle$ (blue dots) are singular vertices, i.e., $w_0=\langle1\rangle+\langle2\rangle+\langle3\rangle+\langle4\rangle$. We choose link $\langle 13\rangle$ and $\langle 24\rangle$ (blue lines) to be singular lines, i.e., $w_0=\partial(\langle 13\rangle+\langle 24\rangle)$. The direction of the red links dual to $\langle 13\rangle$ and $\langle 24\rangle$ are changed. Vertices $i=1,2,3$, and $4$ of $\cT$ are labelled by group elements $g_i\in G$. Majorana fermions (red dots) reside on the vertices of the resolved dual lattice $\tcP$ (solid and dashed red links). The solid red links and gray ellipses indicate that the two Majorana fermions at their two ends are paired with respect to the link direction. The green strip is the $\mathbb Z_2$ domain wall of the ``spin'' configuration $\{g_i\}$ and is decorated by a Kitaev's Majorana chain (Majorana fermions along the domain wall are paired differently from the ``vacuum'').}
\label{fig:torus}
\end{figure*}

\Ref{Tarantino2016} pointed out that discrete spin structures and Kasteleyn orientation played an essential role in constructing FSPT phases decorated with Kitaev's Majorana chains on $G_b$-symmetry domain walls. In this section, we give a brief review of the essential idea and generalize the construction to arbitrary triangulations in 2D (see \fig{fig:torus}). In particular, we use Poincar\'e dual to show how to implement discrete spin structures and Kasteleyn orientation for an arbitrary triangulation with a branching structure in 2D. Essentially, the Poincar\'e dual enables us to define discrete spin structures in arbitrary dimensions and gives rise to the notion of \textit{local} Kasteleyn orientation, which serves as the key step toward decorating Kitaev's Majorana chains onto the intersection lines of $G_b$-symmetry domain walls in 3D. 

In the following, we start by defining spin structures in terms of the second Stiefel--Whitney class. Then, we clarify the relation between the discrete spin structures and the Kasteleyn orientation in 2D. Finally, we make use of the novel concept of equivalence classes of FSLU transformations (with a one-dimensional support space) to obtain the full classification of FSPT states with total symmetry $G_f=G_b\times \mathbb Z_2^f$ in 2D.

\subsection{Discrete spin structure and Kasteleyn orientations}
\label{sec:2D:spin}

In this subsection, we construct the zeroth Stiefel--Whitney homology class on arbitrary 2D triangulation lattice and relate it to the Kasteleyn orientations on the resolved dual lattice. The general procedures of constructing Kasteleyn orientations are summarized as follows:
\begin{enumerate}
\item
Given a (black) triangulation lattice $\cT$ with branching structure for a 2D spin manifold;
\item
Construct the (red) resolved dual lattice $\tcP$ and (red) link orientations using convention \fig{fig:direction}. At this stage, some of the vertices in $\cT$ are non-Kasteleyn-oriented;
\item
Find the expression of $w_0$ in \eq{eq:w_0} as a formal summation of singular vertices of $\cT$ (i.e., non-Kasteleyn-oriented vertices in step-2);
\item
Connect singular vertices in $\cT$ by (blue) lines $S$ (i.e., $\partial S=w_0$);
\item
Using convention \fig{fig:convention:2}, reverse the orientations of (red) links dual to (blue) links belonging to $S$;
\item
Now all the vertices in $\cT$ are Kasteleyn-oriented.
\end{enumerate}
After all the above steps, the resolved dual lattice $\tcP$ now has Kasteleyn orientations. In this way, any decorations of Kitaev's Majorana chains will have the same fermion parity. We will use them to construct generic FSPT states in section~\ref{sec:2D:fSPT}.

\subsubsection{Discrete Stiefel--Whitney homology class $w_0$}
\label{sec:2D:w_0}

It is well known that an oriented manifold $M$ (with dimension $n$) admits spin structures if and only if its second Stiefel--Whitney class $[w^2]\in H^2(M,\mathbb Z_2)$ vanishes. In the construction of the lattice models upon triangulation of $M$, we found it more convenient to use the $(n-2)$-th Stiefel--Whitney homology class $[w_{n-2}]$, which is the Poincar\'e dual of $[w^2]$.

\begin{figure}[h!]
\centering
\begin{subfigure}[h]{.2\textwidth}
\centering
\includegraphics[]{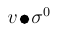}
\caption{$\sigma^0=v$ is a $0$-simplex.}
\end{subfigure}
\begin{subfigure}[h]{.2\textwidth}
\centering
\includegraphics[]{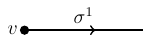}
\caption{$\sigma^1$ is an $1$-simplex.}
\label{fig:regular1}
\end{subfigure}
\begin{subfigure}[h]{.2\textwidth}
\centering
\includegraphics[]{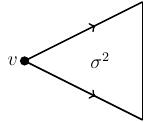}
\caption{$\sigma^2$ is a $2$-simplex.}
\label{fig:regular2}
\end{subfigure}
\caption{Regular pair $v\subseteq\sigma^i$ ($i=0,1,2$) for vertex $v$.}
\label{fig:regular}
\end{figure}

In this subsection, we consider only the 2D case. For a spatial manifold $M$ ($n=2$) with triangulation $\cT$, the Stiefel--Whitney homology class $[w_0]$ has a representative that is the summation of all vertices $v$ with some (mod 2) coefficients as follows \cite{Goldstein1976} \footnote{Note that the barycentric subdivision triangulation illustrated in \Ref{Gaiotto2016} is a special case of this general construction of the Stiefel--Whitney homology class \cite{Goldstein1976}.}
\begin{align}\label{eq:w0}
w_0 = \sum_{v\in \cT} \# \{\sigma | v \subseteq \sigma \text{ is regular} \} \cdot v.
\end{align}
Here, $v \subseteq \sigma$ means that $v$ is a sub-simplex of simplex $\sigma$. $v \subseteq \sigma$ is called regular if $v$ and $\sigma$ have one of the three relative positions shown in \fig{fig:regular}. $\# \{\sigma | v \subseteq \sigma \text{ is regular} \} \cdot v$ denotes the formal product of the (mod 2) number of regular pairs $v \subseteq \sigma$ and the vertex $v$.
We call vertex $v$ singular if $\# \{\sigma | v \subseteq \sigma \text{ is regular} \}$ is odd. In this language, $w_0$ in \eq{eq:w0} is the formal summation of all singular vertices. $w_0$ is a vector (0-th singular chain) in the vector space (of 0-th singular chains) spanned by the formal bases of all vertices with $\mathbb Z_2$ coefficients.

On the other hand, it is known that all oriented 2D surfaces admit spin structures. Thus, the second Stiefel--Whitney class $[w^2]$ or the zeroth Stiefel--Whitney homology class $[w_0]$ of any oriented surface is (co)homologically trivial. As a result, the collection of singular vertices $w_0$ in \eq{eq:w0} can be viewed as a boundary $\partial S$ for some lines $S$ (we call them singular lines). The singular lines $S$ are colored blue in the following figures.

For a fixed collection of singular vertices (a fixed representative $w_0$ for $[w_0]$), different inequivalent choices of singular lines $S$ correspond to different spin structures that are isomorphic to $H^1(M,\mathbb Z_2)$ non-canonically. This can be seen as follows. We first choose arbitrary fixed $S_0$ such that $w_0=\partial S_0$. Then for any other choice $S$ with also $w_0=\partial S$, we can add $S_0$ and $S$ formally. The summation $S_0+S$ is a collection of closed loops on the manifold (recall that lines in $S_0$ and $S$ have the same end points). We can ask whether $S_0+S$ is in the trivial class of $H_1(M,\mathbb Z_2)$ or not. If it is, we say $S_0$ and $S$ are equivalent. In this way, with the fixed $S_0$, we have a one-to-one correspondence between equivalence classes of singular lines $S$ and $H_1(M,\mathbb Z_2)$. In other words, the set of equivalence classes of singular lines $S$ is an affine $H_1(M,\mathbb Z_2)$-space. This is also one of the most important properties of spin structures of a manifold. As a result, we have a one-to-one correspondence between $2^{2g}$ equivalence classes of singular lines $S$ and $2^{2g}$ spin structures of the manifold with genus $g$.

\subsubsection{Kasteleyn orientations and gauge transformations}

To decorate Kitaev chains onto domain walls of a 2D spin model, it is useful to determine the Kasteleyn orientation \cite{Kasteleyn1963} for the edges of the lattice \cite{Tarantino2016,Ware2016}. In this section, we relate the existence of discrete spin structures (the vanishing of $[w_0]$) of a triangulation $\cT$ to the existence of Kasteleyn orientation of the resolved dual lattice. Then, in the next section, we use FSLU transformations \cite{Chen2010,Gu2015} to classify FSPT states and define exactly solvable models on arbitrary triangulations in 2D.

\begin{figure*}[h!]
\centering
\begin{subfigure}[h!]{.3\textwidth}
\centering
\includegraphics[]{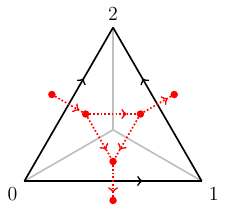}
\caption{Positive oriented triangle}
\end{subfigure}
~
\begin{subfigure}[h!]{.3\textwidth}
\centering
\includegraphics[]{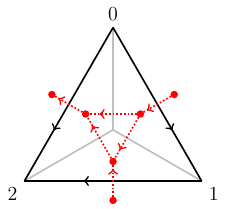}
\caption{Negative oriented triangle}
\end{subfigure}
\caption{Triangulation $\cT$ (black line), resolved triangulation $\tcT$ (black and gray line), and resolved dual lattice $\tcP$ (dashed red line). The resolved triangulation $\tcT$ was obtained from the original $\cT$ by adding a new vertex to the center of each triangle. The links of $\cP$ have orientations induced from the link orientations of $\cT$ according to the conventions shown in \fig{fig:convention:1}. Red dots on the vertices of $\tcP$ represent Majorana fermions that split from the complex fermions on each link of $\cT$ (see the discussion in section~\ref{sec:2D:fSPT}).}
\label{fig:direction}
\end{figure*}

Our set up begins with a fixed triangulation $\cT$ of the surface $M$. The first step is to construct a polyhedral decomposition $\cP$ of $M$ that is a trivalent graph dual to $\cT$. We add a spinless fermionic degree of freedom to every link of $\cT$ and split it into two Majorana fermions on the two sides of this link for convenience. Equivalently, we can resolve the triangulation $\cT$ by adding a new vertex to each triangle center and obtain a new triangulation $\tcT$. The Majorana fermions reside on the vertices of the resolved dual lattice $\tcP$, which is a trivalent graph dual to $\tcT$ (see \fig{fig:direction} and \fig{fig:torus} for this construction on a torus).

\begin{figure}[h!]
\centering
\begin{subfigure}[h!]{.3\textwidth}
\centering
\includegraphics[]{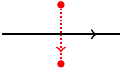}
\caption{Orientation convention for (red) link dual to (black) link $l\notin S$.}
\label{fig:convention:1}
\end{subfigure}
\begin{subfigure}[h!]{.3\textwidth}
\centering
\includegraphics[]{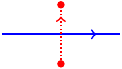}
\caption{Orientation convention for (red) link dual to (blue) singular link $l\in S$.}
\label{fig:convention:2}
\end{subfigure}
\caption{Conventions for the orientation of links in $\tilde{\cP}$ (dotted red line) from branching structure of triangulation $\cT$ (solid black line). Non-singular (singular) black (blue) links $l\notin S$ ($l\in S$) induce orientation conventions for the dual link in $\tilde{\cP}$. We introduce a spinless fermion on each (black/blue) link in $\cT$ and split it into two Majorana fermions on the two sides of this link or vertices of $\tcP$ (red dots).}
\label{fig:convention}
\end{figure}

The second step is adding directions to links in $\cT$ and $\tcP$. We order all of the vertices in $\cT$ and use the convention that all links are from vertices of smaller number to vertices of larger number. This is a branching structure of $\cT$ such that there is no cycle for any triangle. The dual-link direction in $\cP$ is obtained from $\cT$ using the convention shown in \fig{fig:convention:1}. The directions of the new links in $\tcP$ are also obtained from triangulation $\cT$ by using the conventions in \fig{fig:direction}.

The essential point of the aforementioned link orientation conventions can be explained as follows. When traveling along the smallest red loop in $\tcP$ around vertex $v\in\cT$ counterclockwise, we encounter even numbers of red links (due to the resolvation) with the direction along or opposite to our direction (for example, the red loop inside the green strip around vertex 4 in \fig{fig:torus}). Using the conventions in \fig{fig:direction}, the red link direction is opposite the counterclockwise direction if and only if (1) the red link is dual to a black link in $\cT$ such that $v$ is the initial point of this black link (this corresponds to the case in \fig{fig:regular1}) or (2) the red link is resolved to a new link inside a triangle in $\cT$ such that $v$ is the first point of this triangle, i.e., the $0$ point of triangle $\langle 012\rangle$ (this is the case in \fig{fig:regular2}). If the total number of red links with opposite directions is odd, the vertex $v$ is considered Kasteleyn-oriented. As the smallest loop in $\tcP$ around $v$ has an even number of red links, it does not matter whether we use counterclockwise or clockwise conventions. Under the above construction, we relate the zeroth Stiefel--Whitney homology class $w_0$ in \eq{eq:w0} and the orientation of links in $\tcP$, i.e., $w_0$ is the summation of all non-Kasteleyn-oriented vertices.
\begin{align}\label{eq:w_0}
w_0 &= \sum_{v\in \cT} \left( 1+ \#\{\text{$\sigma^1|v\subseteq\sigma^1$ is regular}\} + \#\{\text{$\sigma^2|v\subseteq\sigma^2$ is regular}\} \right) \cdot v \\\nonumber
&= \sum_{v\in \cT} v\ \text{($v$ is non-Kasteleyn-oriented)}.
\end{align}

As the zeroth Stiefel--Whitney homology class $[w_0]$ for any oriented surface is trivial, we have $w_0=\partial S$ for some singular line $S$. If we further reverse the direction of the links in $\tcP$, thus crossing the singular lines $S$ as shown in \fig{fig:convention:2}, then all of the vertices in $\cT$ are Kasteleyn-oriented, as this operation only changes the Kasteleyn property of the singular vertices in $w_0$ while preserving this property for all other vertices, including those in the interval of $S$. After completing the procedure above, we relate the vanishing of the zeroth Stiefel--Whitney homology class $[w_0]$ to the property of Kasteleyn orientation of the smallest loop around each vertex.

\begin{figure*}[h!]
\centering
\begin{subfigure}[h!]{.2\textwidth}
\centering
\includegraphics[]{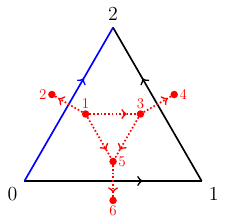}
\end{subfigure}
$\quad\quad\longrightarrow\quad\quad$
\begin{subfigure}[h!]{.2\textwidth}
\centering
\includegraphics[]{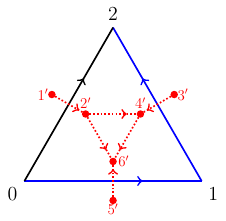}
\end{subfigure}
\caption{Shape changing of singular lines $S$ of $\cT$ and ``gauge transformation'' of Kasteleyn orientations of $\tcP$. We perform ``gauge transformations'' on the three red vertices inside a black triangle, which effectively change the shape of $S$ and the directions of the three outreaching red links. The Majorana fermions are mapped from $n$ to $n'$ $(n = 1,2,\cdots,6)$ with respect to the directions of red links dual to black links under this FSLU.}
\label{fig:gauge}
\end{figure*}

Note that the construction of link direction in $\tcP$ depends on the choice of the singular line $S$. On the one hand, the local shape of $S$ is not important as long as $\partial S$ is fixed. In fact, if we change the shape of $S$ locally, the change in link direction in $\tcP$ can be obtained by several ``gauge transformations'' of Kasteleyn orientation, which relate two different but equivalent Kasteleyn orientations (simultaneously changing the directions of links sharing a common vertex in $\tcP$) \cite{Cimasoni2007}. An example of the basic shape changes of singular lines on $\cT$ and gauge transformation of Kasteleyn orientation on $\tcP$ is shown in \fig{fig:gauge}. Note that the Majorana degrees of freedom on the vertices of $\tcP$ are mapped from one lattice to another according to the link direction under the gauge transformation of Kasteleyn orientation ($n$ to $n'$ in \fig{fig:gauge}). In this way, the vacuum state without fermions (without Kitaev chain) on the left lattice is mapped to the vacuum state on the right lattice without changing the fermion parity.

On the other hand, the homology class of $S$ matters. Different choices of topological classes of $S$ (fixed $w_0=\partial S$) correspond to different spin structures on $M$. Our constructions make sure that, for arbitrary choices of $S$, the \emph{local} Kasteleyn properties along the smallest loop around every vertex is satisfied. However, the \emph{global} Kasteleyn property along non-trivial cycles of $M$ can be either preserved or broken. They correspond to $2^{2g}$ different spin structures on closed oriented surface $M$ with genus $g$ \cite{Cimasoni2007,Cimasoni2008}. Different choices of $S$ induce different global Kasteleyn properties and thus correspond to different spin structures.

\subsubsection{Kasteleyn orientations under retriangulations}
 
In the above, we focus only on a fixed triangulation $\cT$ of $M$ and relate its discrete Stiefel--Whitney homology class $[w_0]$ to the Kasteleyn orientations and spin structures. To use FSLU transformations to classify FSPT phases, we must understand the relation of Kasteleyn orientations for different triangulations. In fact, we only have to determine the changes of Kasteleyn orientations under Pachner moves, which are basic moves of retriangulation \cite{Pachner1991}.

Ordinary Pachner moves for a two-dimensional manifold consist of a (2-2) move and a (1-3) move. With branching structures, there are three types of (2-2) move and four types of (1-3) move in total. (We do not consider the mirror images of these moves; otherwise, the number of moves would double). Only two types of (2-2) move and two types of (1-3) move have branching structures that can be induced by global ordering \cite{Gu2014}. Examples of these moves are as follows:
\begin{figure*}[h!]
\centering
\begin{subfigure}[h!]{.2\textwidth}
\centering
\includegraphics[]{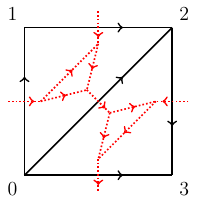}
\end{subfigure}
$\quad \longrightarrow \quad$
\begin{subfigure}[h!]{.2\textwidth}
\centering
\includegraphics[]{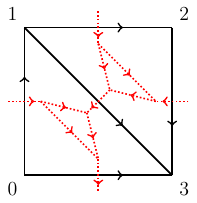}
\end{subfigure}
\caption{Standard (2-2) move.}
\label{fig:(2-2)standard}
\end{figure*}


\begin{figure*}[h!]
\centering
\begin{subfigure}[h!]{.25\textwidth}
\centering
\includegraphics[scale=1.1]{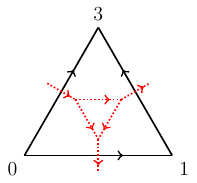}
\end{subfigure}
$\longrightarrow$
\begin{subfigure}[h!]{.25\textwidth}
\centering
\includegraphics[scale=1.1]{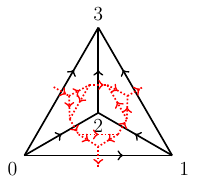}
\end{subfigure}
\caption{One of the (1-3) moves.}
\label{fig:(1-3)a}
\end{figure*}

Other types of (2-2) and (1-3) moves are shown in Supplementary Material. 
For Pachner moves that are not induced by global ordering, the representative $w_0$ of Stiefel--Whitney class $[w_0]$ in \eq{eq:w0} may be changed. For Pachner moves that are induced by a global ordering, the representative $w_0$ of Stiefel--Whitney class $[w_0]$ is unchanged. This makes the 2D case much easier than the 3D case.

\subsection{FSLU transformations and consistent conditions for fixed-point states}
\label{sec:2D:fSPT}

In the above subsection, we discuss discrete spin structures and Kasteleyn orientation construction on arbitrary 2D triangulation lattices. We can now decorate Kitaev's Majorana chains using these rules and systematically classify 2D FSPT states by using FSLU transformations.

\subsubsection{Decoration of Kitaev's Majorana chains}

As discussed in section \ref{sec:wavefunction}, our model has two types of fermionic degrees of freedom. The first type is the complex fermion $c_{(ijk)}$, which resides at the center of triangle $\langle ijk\rangle$ of space manifold triangulation $\cT$. We use $n_2(g_i,g_j,g_k)=0,1$ to denote the number of $c$ fermions at triangle $\langle ijk\rangle$. In fact, the parity conservation constraint for $c$ fermions under retriangulation is $\dd n_2=0$ (mod 2). Therefore $n_2$ is an element of $H^2(G_b,\mathbb Z_2)$.

The second type of (complex) fermion, $a_{(ij)}$, resides on the link $\langle ij\rangle$ of $\cT$. To describe Kitaev's Majorana chain more conveniently, we separate fermion $a_{(ij)}$ to two Majorana fermions.
\begin{align}
\gamma_{ijA} &= a_{(ij)} + a_{(ij)}^\dagger,\\
\gamma_{ijB} &= \frac{1}{i} \left( a_{(ij)} - a_{(ij)}^\dagger \right).
\end{align}
The Majorana fermions $\gamma_{ijA}$ and $\gamma_{ijB}$ reside on the two sides of link $\langle ij\rangle$. They also reside on the two ends of the link in $\tcP$ dual to link $\langle ij\rangle$. Our convention is that the dual link has direction from vertex $\langle ijA\rangle$ to vertex $\langle ijB\rangle$. The fermion parity operator of $a$ fermions or $\gamma$ fermions at link $\langle ij\rangle$ is $P_f^\gamma=-i\gamma_{ijA}\gamma_{ijB}$.

Now we decorate Kitaev's Majorana chains onto the dual lattice. We use a $\mathbb Z_2$-valued 1-cochain $\tilde n_1(g_i,g_j)$ to indicate whether there is a domain wall between vertices $i$ and $j$. $\tilde n_1\in H^1(G_b,\mathbb Z_2)$ is a cocycle because we are constructing SPT states without deconfined Majorana fermions (the Kitaev chains should form closed loops). Depending on the configurations of $\{g_i\}$ and the choices of $\tilde n_1$, the domain wall configuration in a particular lattice is different. We pair the Majorana fermions depending on the domain wall configuration as follows. If there is no domain wall on link $\langle ij\rangle$, then the Majorana fermions $\gamma_{ijA}$ and $\gamma_{ijB}$ on the two sides of this link are paired (vacuum pair) with respect to the direction of the dual red link (we use a solid blue line and gray ellipse to indicate this pairing). If there is a domain wall on link $\langle ij\rangle$, then the Majorana fermion of this link is paired with another Majorana fermion belonging to another link with a domain wall within the same triangle. 

After fermion decoration, the (2-2) move becomes a fermionic unitary transformation between the fermionic Fock spaces on two different triangulation lattices $\cT$ and $\cT'$. An example of this $F$ move (the standard $F$ move) is presented as follows (there are $\mathbb Z_2$ domain walls on links $\langle 01\rangle$, $\langle 02\rangle$, $\langle 03\rangle$, and no domain wall on the other links):
\begin{align}
\label{eq:F1_a}
\Psi\left(
\vcenter{\hbox{\includegraphics[]{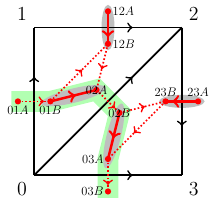}}}
\right)\quad = \quad
F(g_0,g_1,g_2,g_3)
\quad
\Psi\left(
\vcenter{\hbox{\includegraphics[]{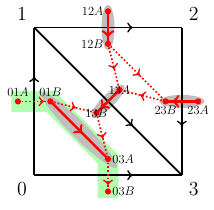}}}
\right),
\end{align}
where the $F$ operator is defined as
\begin{align}
\label{fig:F2D}
F(g_0,g_1,g_2,g_3) = 
\nu_3(g_0,g_1,g_2,g_3)
c^{\dagger n_2(g_0,g_1,g_2)}_{(012)}
c^{\dagger n_2(g_0,g_2,g_3)}_{(023)}
c^{n_2(g_0,g_1,g_3)}_{(013)}
c^{n_2(g_1,g_2,g_3)}_{(123)}
X[\tilde n_1(g_i,g_j)].
\end{align}
Here, $\nu_3(g_0,g_1,g_2,g_3)$ is a $U_T(1)$-valued 3-cochain and $c^{\dagger}_{(012)}$ is the creation operator for $c$ fermions at triangle $\langle 012\rangle$, etc. $X[\tilde n_1(g_i,g_j)]$ is a projection operator changing the Majorana fermion configurations. In the above example, the $X$ operator has an explicit form as follows:
\begin{align}
\label{eq:X20}
X[\tilde n_1] = 2^{1/2} \left(P_{01B,02A} P_{02B,03A}\right) P_{13A,13B},
\end{align}
where $P_{a,b}=(1-i\gamma_a\gamma_b)/2$ is the projection operator for Majorana pairs $\langle a,b\rangle$ (the direction is from vertex $a$ to vertex $b$). The first two projection operators in the above equation project the state to the Majorana dimer configuration in the left figure. Note that the Majorana fermions, which do not appear explicitly on one lattice, are considered to be in vacuum pairs. For example, the two Majorana fermions $\gamma_{13A}$ and $\gamma_{13B}$ appear only in the right figure. They are considered to be paired from $\gamma_{13A}$ to $\gamma_{13B}$ in the left figure. Therefore, we have a third projection operator in \eq{eq:X20} to put the two Majorana fermions $\gamma_{13A}$ and $\gamma_{13B}$ into a vacuum state ($a_{(13)}^\dagger a_{(13)}=0$) in the left figure. All other Majorana fermions that are not shown in \eq{eq:X20} are unchanged under the aforementioned $F$ move.

In order to make $X$ an unitary operator acting on the Hilbert space of Majorana fermions, we introduce a normalization factor in the front of $X$. By directly calculating the norm of the final state after the action of $X$ operator, we can obtain a general expression of the normalization factor
\begin{align}
\label{eq:factor}
\prod_{\mathrm{loop\;}i\mathrm{\;in\; (\tcP,\tcP')}}2^{(L_i-1)/2},
\end{align}
with $2L_i$ being the length of the $i$-th loop in the transition graph of dimer configurations in $\tcP$ and $\tcP'$. For example, the transition graph of the two states in \eq{eq:F1_a} has only one loop with length bigger than two: $01B$-$02A$-$02B$-$03A$-$01B$. So the factor is $2^{(4/2-1)/2}=2^{1/2}$, according to \eq{eq:factor}.

As we are constructing FSPT states, the fermionic local unitary transformation $F$ should be $G_b$ symmetric in the sense that
\begin{align}
F(g_0,g_1,g_2,g_3) = F(gg_0,gg_1,gg_2,gg_3),
\end{align}
for all $g\in G_b$ if $G_b$ is a unitary symmetry group. That is why $\nu_3(g_0,g_1,g_2,g_3)$, $n_2(g_0,g_1,g_2)$, and $\tilde n_1(g_0,g_1)$ are all cochains that are invariant under unitary $g$ action. (We note that $\nu_3(gg_0,gg_1,gg_2,gg_3)=\nu_3^*(g_0,g_1,g_2,g_3)$ for anti-unitary $g$ action.)

In general, there are eight kinds of domain wall configuration in the above $F$ move. One can show that for all configurations, the fermion parities of Majorana fermions are the same in the initial and final wavefunctions (this comes from the Kasteleyn orientation property of retriangulations; see section~\ref{sec:2D:spin} and Supplementary Material). Therefore, the fermion parities of $c$ fermions and $\gamma$ fermions should also be conserved separately, and both $n_2$ and $\tilde n_1$ are cocycles.


Similar to the 1D case, we can use FSLU to redefine the basis state $|\{g_l\}\rangle$ as
\begin{align}
|\{g_l\}\rangle' = U_{\mu_2,m_1} |\{g_l\}\rangle 
= \prod_{\langle ijk \rangle} \mu_2(g_i,g_j,g_k)^{s_{\langle ijk \rangle}} \prod_{\langle ij \rangle} \left[ f^{m_1(g_i,g_j)}_{ijA}f^{m_1(g_i,g_j)}_{ijB} \right] \prod_{\langle ijk\rangle} \left[ f^{\dagger m_1(g_j,g_k)}_{jkA} f^{\dagger m_1(g_i,g_j)}_{ijA} f^{\dagger m_1(g_i,g_k)}_{ikB} \right] |\{g_l\}\rangle,
\end{align}
where we first create three complex fermions $f_{jkA}$, $f_{ijA}$ and $f_{ikB}$ near the three links of the triangle $\langle ijk\rangle$ ($i<j<k$), and then annihilate the two fermions $f_{ijA}$ and $f_{ijB}$ on the two sides of link $\langle ij\rangle$ when gluing the two triangles sharing link $\langle ij\rangle$. To preserve the fermion parity and be symmetric, $m_1$ should be a 1-cocycle (with $\mathbb Z_2$ coefficient): $m_1(gg_i,gg_j)=m_1(g_i,g_j)$ and $\dd m_1(g_i,g_j,g_k)=0$. $\mu_2(g_i,g_j,g_k)$ is a phase factor associated with triangle $\langle ijk\rangle$, and $s_{\langle ijk \rangle}=\pm 1$ denotes the orientation of the triangle. After eliminating all $f$ fermions in the new $F$ move operator $F' = U_{\mu_2,m_1} F U_{\mu_2,m_1}^\dagger$ (all fermion signs are cancelled again), one find that the phase factor in \eq{fig:F2D} becomes
\begin{align}\label{eq:nu3p}
\nu_3'(g_0,g_1,g_2,g_3) = \nu_3(g_0,g_1,g_2,g_3) \frac{\mu_2(g_1,g_2,g_3)\mu_2(g_0,g_1,g_3)}{\mu_2(g_0,g_2,g_3)\mu_2(g_0,g_1,g_2)},
\end{align}
So the elements $\nu_3$ in the same group cohomology class in $H^3(G_b, U_T(1))$ correspond to the same 2D FSPT phase. This is also consistent with the general result in \eq{eq:fcoboundary} \cite{Gu2014}, since $Sq^2(m_1)$ is also trivial in the 2D FSPT case.

Apart from the (2-2) move, there is another (2-0) move that can change the total number of vertices for triangulations. An example of domain wall configurations for the (2-0) move is
\begin{align}
\label{fig:(2-0)}
\Psi\left(
\vcenter{\hbox{\includegraphics[]{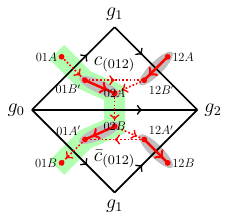}}}
\right)
= \frac{1}{|G_b|^{1/2}}  c_{(012)}^{\dagger n_2(g_0,g_1,g_2)} \bar c_{(012)}^{\dagger n_2(g_0,g_{1},g_2)} X[\tilde n_1(g_i,g_j)]\ 
\Psi\left(
\vcenter{\hbox{\includegraphics[]{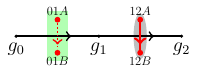}}}
\right),
\end{align}
where $c_{(012)}$ and $\bar c_{(012)}$ are the annihilation operators of the $c$ fermions at the center of two triangles with opposite orientations in the left figure ($c_{(012)}$ and $\bar c_{(012)}$ are at the centers of the upper and lower triangles in the left figure respectively). The Hilbert dimension of the bosonic degrees of freedom on the vertices of a fixed triangulation is $|G_b|^{N_v}$, where  $|G_b|$ is the order of the group $G_b$ and $N_v$ is the number of vertices. Therefore we add a normalization factor $|G_b|^{-1/2}$ in the front of the (2-0) move operator, for the vertex number is reduced by one from the left state to the right state. $X[\tilde n_1]$ is also the projection operator from the state of Majorana dimer pairs in the right figure to the state of the left figure. Note that there are six Majorana fermions ($\gamma_{02A}, \gamma_{02B}$, $\gamma_{01A'}, \gamma_{01B'}, \gamma_{12A'}$, and $\gamma_{12B'}$) that do not appear explicitly in the right figure. Similar to the case of the (2-2) move, these fermions should also be considered to be in vacuum pairs in the right figure state, such that $-i\gamma_{02A}\gamma_{02B}=-i\gamma_{01B'}\gamma_{01A‘}=-i\gamma_{12B'}\gamma_{12A'}=1$ when acting on the right figure state. This choice is possible because the dimer loop $01A$-$01B'$-$01A'$-$01B$-$01A$ is Kasteleyn oriented. Therefore, one can also use the convention that the two Majorana fermions on the two sides of a link are paired up by regarding the projection operators $-i\gamma_{01A}\gamma_{01B'},-i\gamma_{01A'}\gamma_{01B},-i\gamma_{12A}\gamma_{12B'},and -i\gamma_{12A'}\gamma_{12B}$ as $1$ when acting on the vacuum state of the left figure. The $X$ operator then projects the state to the Majorana dimer configuration state in the left figure. The fermion parities of the left and right states are always the same. An explicit expression of $X$ for this particular(2-0) move is
\begin{align}
\label{eq:20X}
X[\tilde n_1] = 2P_{01B',02A} P_{02B,01A'} P_{12A,12B'} P_{12A',12B}.
\end{align}
Using the (2-0) moves, we can deduce all (3-1) moves and other (2-2) moves form the standard (2-2) $F$ move in \eq{eq:F1_a}. The normalization factor is obtained from \eq{eq:factor}. there are two loops in the transition graph of the Majorana dimer states with length bigger than two: $01B'$-$02A$-$02B$-$01A'$-$01B'$ and $12A$-$12B'$-$12A'$-$12B$-$12A$. So the normalization factor is $2^{(4/2-1)/2}\times 2^{(4/2-1)/2}=2$.

\subsubsection{Fermionic pentagon equations}
\label{sec:2D:pentagon}

\begin{figure*}[h!]
\centering
\includegraphics[]{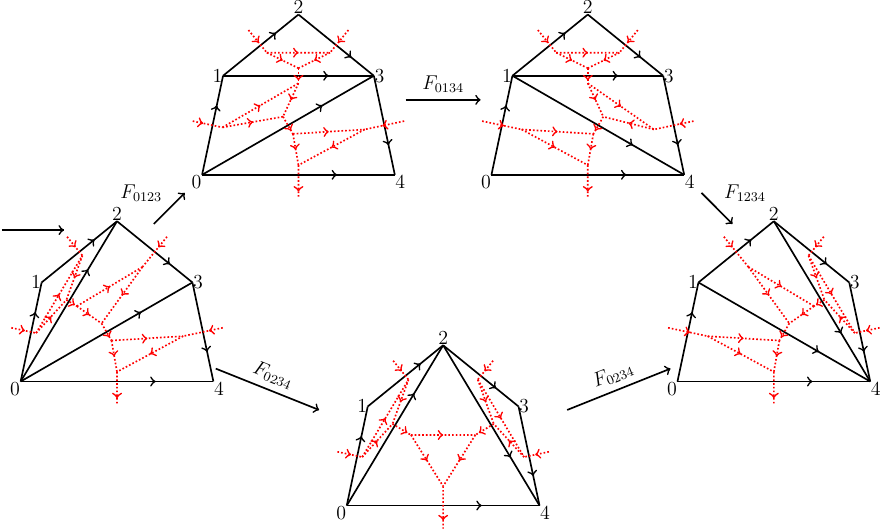}
\caption{Fermionic pentagon equation.}
\label{fig:pentagon}
\end{figure*}

In the above, we discuss the FSLU moves. The most important one is the standard $F$ move in \eq{eq:F1_a}. Similar to the bosonic pentagon equation for the bosonic $F$ move, we have a fermionic pentagon equation as a consistent equation for FSLU transformations (see \fig{fig:pentagon}). This fermionic pentagon equation involves only the standard $F$ move. Using the unitary conditions, one can also derive other pentagon equations, and they essentially give the same constraint for $\nu_3$.

We now calculate the constraint for $\nu_3$ from the pentagon equation in \fig{fig:pentagon}. As the $c$ fermions and Majorana fermions are decoupled in the $F$ move (the $c$ fermion part and the Majorana fermion part of $X$ in \eq{fig:F2D} commute), only the $c$ fermion twists the cocycle condition for $\nu_3$. The $X$ operators are merely projection operators that do not introduce any nontrivial phases in two different paths of pentagon equation. The final result of the equation for $\nu_3$ is the same as (special) group super-cohomology theory \cite{Gu2014,Gu2015}, i.e.,
\begin{align}
(\dd\nu_3)(g_0,g_1,g_2,g_3,g_4) = (-1)^{Sq^2(n_2)(g_0,g_1,g_2,g_3,g_4)} = (-1)^{n_2(g_0,g_1,g_2)n_2(g_2,g_3,g_4)}.
\end{align}

Now, we see that only $BH^2(G_b,\mathbb Z_2)$, the obstruction-free subgroup of $H^2(G_b,\mathbb Z_2)$ formed by elements $n_{2} \in H^2(G_b,\mathbb Z_2)$ that satisfy $Sq^2(n_{2})=0$ in $H^{4}(G_b,U_T(1))$, can give rise to solutions for $\nu_3$, and inequivalent solutions of $\nu_3$ are still given by $H^3(G_b,U_T(1))$ according to the gauge transformations of $\nu_3$. Thus, the mathematical objects that classify 2D FSPT phases with a total symmetry $G_f=G_b\times \mathbb Z_2^f$ can be summarized as three group cohomologies of the symmetry group $G_b$ \cite{cheng15,gaiotto16}: $H^1 (G_b,\mathbb Z_2 )$, $BH^2(G_b,\mathbb Z_2)$ and $H^3(G_b, U_T(1))$.
Finally, by using the method proposed in \Ref {Gu2015}, one can derive the commuting projector parent Hamiltonian of all of these FSPT states on arbitrary 2D triangulations with a branching structure.

\section{Constructions and classifications for FSPT states in 3D}

In this section, we construct and classify the 3D FSPT states parallel to the discussions of 2D FSPT states. Compared with the 2D case, the most nontrivial part of 3D phases is the fermion parity mixing of the $c$ fermions and Majorana fermions. In section~\ref{sec:3D:spin}, we find that there are in general no Kasteleyn orientations on a 3D lattice. The existence of a spin structure only implies \emph{local} Kasteleyn orientations. If we decorate Kitaev's Majorana chains onto a 3D lattice, the shape-changing process of the Majorana chain may also change the fermion parity of the corresponding Majorana chain. In this case, we should use the $c$ fermion to compensate the fermion parity changes. As a result of this fermion parity mixing, the cocycle equation for $\nu_4$ is much more complicated than special group super-cohomology theory.

\subsection{Discrete spin structure in 3D and local Kasteleyn orientations}
\label{sec:3D:spin}

In this subsection, we discuss the first Stiefel--Whitney homology class on a discrete lattice and relate it to the \emph{local Kasteleyn orientations} on the dual lattice. The overall constructions are parallel to the 2D case. The difference is that Kasteleyn orientations are only satisfied for the smallest loops in 3D, not for the general large loops. The fermion parity of a Kitaev chain decorated onto a fluctuating loop is therefore not conserved.

\subsubsection{Discrete Stiefel--Whitney homology class $w_1$}

\begin{figure}[h!]
\centering
\begin{subfigure}[h!]{.2\textwidth}
\centering
\includegraphics[]{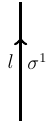}
\caption{$\sigma^1=l$ is an $1$-simplex.}
\end{subfigure}
\begin{subfigure}[h!]{.2\textwidth}
\centering
\includegraphics[]{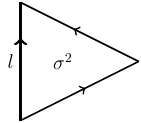}
\caption{$\sigma^2$ is a $2$-simplex.}
\label{fig:regular3D1}
\end{subfigure}
\begin{subfigure}[h!]{.2\textwidth}
\centering
\includegraphics[]{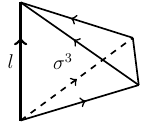}
\caption{$\sigma^3$ is a $3$-simplex.}
\label{fig:regular3D2}
\end{subfigure}
\caption{Regular pairs $l\subseteq\sigma^i$ ($i=1,2,3$) for link $l$.}
\label{fig:regular3D}
\end{figure}

Similar to the oriented 2D manifolds, all oriented 3D manifolds admit spin structures. The second Stiefel--Whitney cohomology class $[w^2]$ is always trivial. We can consider the first discrete Stiefel--Whitney homology class for a triangulation $\cT$ with a branching structure of 3D spatial manifold $M$ \cite{Goldstein1976}:
\begin{align}\label{eq:w1}
w_1 = \sum_{l\in \mathcal T} \# \{\sigma | l \subseteq \sigma \text{ is regular} \} \cdot l.
\end{align}
$w_1$ is the summation of all links in $\cT$ with some $\mathbb Z_2$ coefficients. Again, $l \subseteq \sigma$ means that link $l$ is a sub-simplex of simplex $\sigma$. $l \subseteq \sigma$ is called regular if $l$ and $\sigma$ have one of the three relative positions shown in \fig{fig:regular3D}. If $\# \{\sigma | l \subseteq \sigma \text{ is regular} \}$ is odd, we call the link $l$ singular. Thus, $w_1$ in \eq{eq:w1} is the formal summation of all singular lines. $w_1$ is a vector (1-th singular chain) in the vector space (of 1-th singular chains) spanned by formal bases of all links with $\mathbb Z_2$ coefficients.

As the second Stiefel--Whitney cohomology class $[w^2]$ of any oriented 3D manifold is trivial, we can find some surface $S$ such that $w_1=\partial S$. For a fixed collection of singular links (a fixed representative $w_1$ for $[w_1]$), different inequivalent choices of $S$ correspond to different spin structures (see the discussions at the end of section~\ref{sec:2D:w_0}).

\subsubsection{Local Kasteleyn orientations and gauge transformations}

In 3D, we also want to decorate Kitaev chains onto the intersection lines of $G_b$-symmetry domain walls. A natural question inherited from 2D is whether there are Kasteleyn properties for all even-link loops in 3D. This question is related to the fermion parity of the Kitaev chain. The answer is that the existence of discrete spin structures (the vanishing of $[w_1]$) is related to the existence of \emph{local} Kasteleyn orientations of the resolved dual lattice. In other words, Kasteleyn properties are satisfied for the smallest loops but generally broken for large loops in 3D.

\begin{figure*}[h!]
\centering
\begin{subfigure}[h!]{.3\textwidth}
\centering
\includegraphics[]{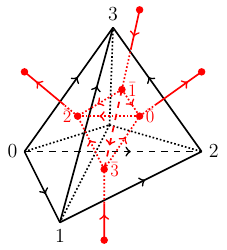}
\caption{Positive oriented tetrahedron}
\end{subfigure}
~
\begin{subfigure}[h!]{.3\textwidth}
\centering
\includegraphics[]{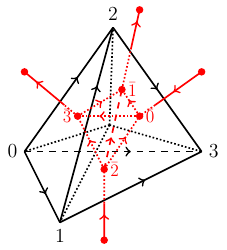}
\caption{Negative oriented tetrahedron}
\end{subfigure}
\caption{Triangulation $\cT$ (solid black line), resolved triangulation $\tcT$ (solid and dashed black lines), and resolved dual lattice $\tcP$ (red line). The resolved triangulation $\tcT$ is obtained from the original $\cT$ by adding a new vertex to the center of each tetrahedron. The links of $\cP$ have orientations induced from the link orientations of $\cT$ according to the conventions shown in \fig{fig:convention3D:1}. Red dots on the vertices of $\tcP$ represent Majorana fermions, which are split from complex fermions on each face of $\cT$.}
\label{fig:direction3D}
\end{figure*}

We now consider the construction that is similar to the 2D case. For a fixed triangulation $\cT$ of 3D manifold $M$, the first step is to construct a polyhedral decomposition $\cP$ of $M$, which is a 4-valent graph dual to $\cT$. We now add a spinless fermionic degree of freedom to every face (triangle) of $\cT$ and split it into two Majorana fermions on the two sides of this face for convenience. Equivalently, we resolve the triangulation $\cT$ by adding a new vertex to each tetrahedron center and obtain a new resolved triangulation $\tcT$. The Majorana fermions reside on the vertices of the resolved dual lattice $\tcP$, which is a 4-valent graph dual to $\tcT$ (see \fig{fig:direction3D}).

\begin{figure}[h!]
\centering
\begin{subfigure}[h!]{.3\textwidth}
\centering
\includegraphics[]{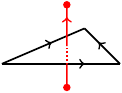}
\caption{Orientation convention for (red) link dual to (black) triangle $f\notin S$.}
\label{fig:convention3D:1}
\end{subfigure}
\begin{subfigure}[h!]{.3\textwidth}
\centering
\includegraphics[]{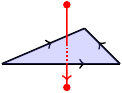}
\caption{Orientation convention for (red) link dual to (blue) singular triangle $f\in S$.}
\label{fig:convention3D:2}
\end{subfigure}
\caption{Conventions for orientations of links in $\tilde{\cP}$ (red line) from branching structure of triangulation $\cT$ (black line). Non-singular (singular) black (blue) triangle $f\notin S$ ($f\in S$) induces orientation for the dual link in $\tilde{\cP}$. We introduce a spinless fermion on each (black/blue) triangle in $\cT$ and split it into two Majorana fermions on two sides of this triangle or vertices of $\tcP$ (red dots).}
\label{fig:convention3D}
\end{figure}

The second step is again adding directions to links in $\cT$ and $\tcP$. The directions of the links in $\cT$ are given by the branching structure. The dual link direction in $\cP$ is obtained from $\cT$ using the convention shown in \fig{fig:convention3D:1}. The directions of the new links in $\tcP$ are obtained from the triangulation $\cT$ by using the conventions in \fig{fig:direction3D}.

The above link direction construction has the following properties. Consider a fixed link $l\in \cT$. When going along the smallest red loop in $\tcP$ around this link $l$ along the right-hand rule direction, we encounter an even number of red links with directions along or opposite to our direction. Using the conventions in \fig{fig:direction3D}, the red link direction is opposite to our direction if and only if (1) the red link is dual to a black triangle in $\cT$ such that the initial and final vertices of $l$ are the first and last vertices of this black triangle (this corresponds to the case in \fig{fig:regular3D1}) and (2) the red link is a resolved new link inside a tetrahedron in $\cT$ such that the initial and final vertices of $l$ are the first and the last vertices of this tetrahedron (this is the case in \fig{fig:regular3D2}). If the total number of red links with opposite directions is odd, we call the link $l$ Kasteleyn-oriented. Under this construction, we relate the first Stiefel--Whitney homology class $w_1$ in \eq{eq:w1} to the orientations of links in $\tcP$, i.e., $w_1$ is the summation of all non-Kasteleyn-oriented links.
\begin{align}
w_1 &= \sum_{l\in \cT} \left( 1+ \#\{\text{$\sigma^2|l\subseteq\sigma^2$ is regular}\} + \#\{\text{$\sigma^3|l\subseteq\sigma^3$ is regular}\} \right) \cdot l \\\nonumber
&= \sum_{l\in \cT} l\ \text{($l$ is non-Kasteleyn-oriented)}.
\end{align}

As discussed above, the first Stiefel--Whitney homology class $[w_1]$ for any oriented 3D manifold is trivial. Therefore, we have some singular surface $S$ such that $w_1=\partial S$. Now, if we reverse the direction of the links in $\tcP$ crossing the singular surface $S$, as shown in \fig{fig:convention3D:2}, then all of the links in $\cT$ are Kasteleyn-oriented. After following the procedures above, we relate the vanishing of the zeroth Stiefel--Whitney homology class $[w_0]$ to the property of \emph{local} Kasteleyn orientation of the smallest loops around all of the links in $\cT$. Here, ``local'' means that only the smallest loops in $\tcP$ around links in $\cT$ are Kasteleyn-oriented. Larger loops with an even number of links do not have Kasteleyn properties in general.

\begin{figure*}[h!]
\centering
\begin{subfigure}[h!]{.2\textwidth}
\centering
\includegraphics[]{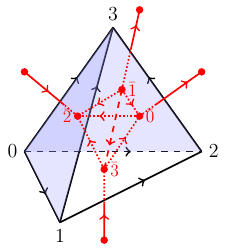}
\end{subfigure}
$\quad\longrightarrow\quad$
\begin{subfigure}[h!]{.2\textwidth}
\centering
\includegraphics[]{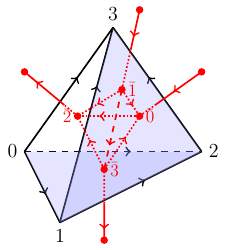}
\end{subfigure}
\caption{Shape changing of singular surfaces $S$ of $\cT$ and ``gauge transformation'' of Kasteleyn orientations of $\tcP$. We perform ``gauge transformations'' on the four red vertices inside a black tetrahedron, which effectively changes the shape of $S$ and the directions of the four outreaching red links. The Majorana fermions are mapped with respect to the directions of red links dual to black triangles under this FSLU transformation.}
\label{fig:gauge3D}
\end{figure*}

The above construction of link directions in $\tcP$ depends on the choice of singular surface $S$. The shape of $S$ can also be changed with fixed $\partial S$. If we change the shape of $S$ locally, the changes of link directions in $\tcP$ can be obtained by several ``gauge transformations'' of Kasteleyn orientations. We define this by simultaneously changing the directions of links sharing a common vertex in $\tcP$, similar to the 2D case \cite{Cimasoni2007}. Different Kasteleyn orientations related by these ``gauge transformations'' are said to be equivalent. An example of shape changes of singular surfaces on $\cT$ and gauge transformation of Kasteleyn orientations on $\tcP$ are shown in \fig{fig:gauge3D}. The Majorana degrees of freedom on the vertices of $\tcP$ are also mapped from one lattice to another according to the link directions (similar to the 2D case in \fig{fig:gauge}). This ensures that the vacuum state (without Kitaev chain) on one lattice is mapped to the vacuum state on another lattice, without fermion parity changing (no fermion on either lattice).

With fixed $w_0=\partial S$, the choices of $S$ are not unique. Different choices of topological classes of $S$ correspond to different spin structures on $M$. The global Kasteleyn properties along non-trivial cycles of $M$ can also be either preserved or broken depending on the choices of $S$. Our construction generalizes the relation of Kasteleyn orientations and discrete spin structures from 2D \cite{Cimasoni2007} to 3D.

\subsubsection{Local Kasteleyn orientations under retriangulations}

\begin{figure*}[h!]
\centering
\begin{subfigure}[h!]{.49\textwidth}
\centering
$\vcenter{\hbox{\includegraphics[]{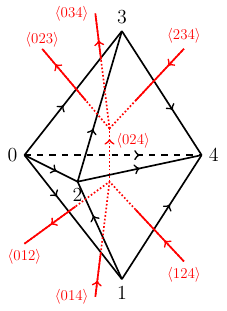}}}
\longrightarrow
\vcenter{\hbox{\includegraphics[]{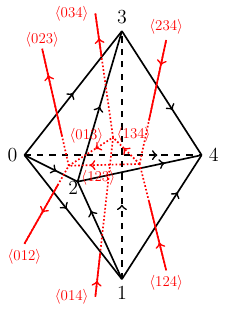}}}$
\caption{For triangulation $\mathcal T$ and dual lattice $\mathcal P$.}
\end{subfigure}
~
\begin{subfigure}[h!]{.49\textwidth}
\centering
$\vcenter{\hbox{\includegraphics[]{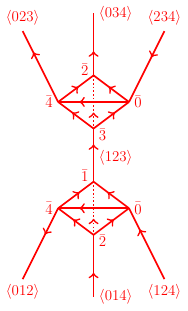}}}
\longrightarrow
\vcenter{\hbox{\includegraphics[]{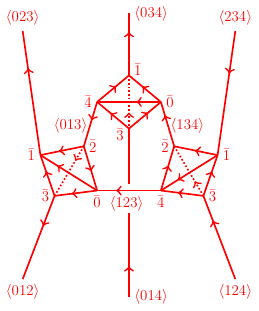}}}$
\caption{For resolved dual lattice $\tilde{\mathcal P}$.}
\end{subfigure}
\caption{Standard (2-3) move.}
\label{fig:23F}
\end{figure*}

To perform FSLU transformations, we now consider that the Kasteleyn orientation changes under retriangulation of $M$. Pachner moves for the 3D manifold consist of a (2-3) move and a (1-4) move \cite{Pachner1991}. When introducing branching structures, there are 10 types of (2-3) move and 5 types of (1-4) move (again, we do not consider the mirror images of these moves, otherwise the number would double) \cite{Gu2014}. Eight types of (2-3) move and three types of (1-4) move have branching structures that can be induced by global ordering \cite{Costantino2005,Gu2014}. The standard (2-3) move is given in \fig{fig:23F}, which does not involve singular surfaces. Two examples of moves involving singular surfaces (the representative $w_1$ of Stiefel--Whitney class $[w_1]$ in \eq{eq:w1} is changed) are presented in Figs. \ref{fig:2-3move} and \ref{fig:1-4move}. Other types of (2-3) and (1-4) moves are shown in Supplementary Material.

\begin{figure*}[h!]
\centering
\begin{subfigure}[h!]{.49\textwidth}
\centering
$\vcenter{\hbox{\includegraphics[]{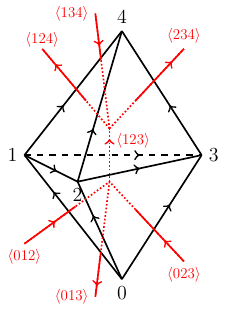}}}
\longrightarrow
\vcenter{\hbox{\includegraphics[]{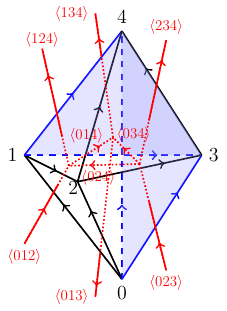}}}$
\caption{(2-3) move for triangulation $\mathcal T$ and dual lattice $\mathcal P$. Singular links $w_1=\langle 14 \rangle + \langle 34 \rangle + \langle 03 \rangle + \langle 04 \rangle$ and surfaces $S=\langle 034 \rangle+\langle 134 \rangle$ are represented by blue lines and blue triangles, respectively.}
\end{subfigure}
~
\begin{subfigure}[h!]{.49\textwidth}
\centering
$\vcenter{\hbox{\includegraphics[]{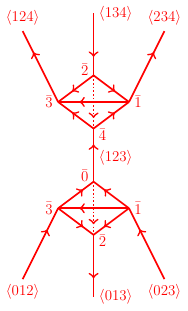}}}
\longrightarrow
\vcenter{\hbox{\includegraphics[]{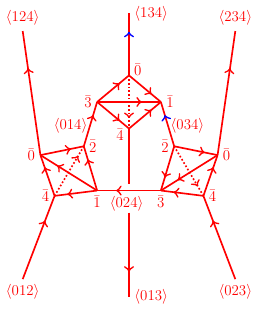}}}$
\caption{(2-3) move for resolved dual lattice $\tilde{\mathcal P}$. Blue arrows on dual (red) links $\langle 034 \rangle$ and $\langle 134 \rangle$ indicate that the directions are reversed by singular surfaces $S$.}
\end{subfigure}
\caption{A (2-3) move that involves singular surfaces.}
\label{fig:2-3move}
\end{figure*}

\begin{figure*}[h!]
\centering
\begin{subfigure}[h!]{.45\textwidth}
\centering
$\vcenter{\hbox{\includegraphics[]{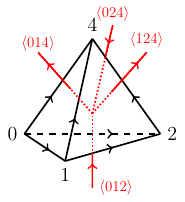}}}
\longrightarrow
\vcenter{\hbox{\includegraphics[]{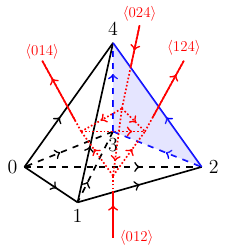}}}$
\caption{(1-4) move for triangulation $\mathcal T$ and dual lattice $\mathcal P$. Singular links $w_1=\langle 23 \rangle + \langle 34 \rangle + \langle 24 \rangle$ and surfaces $S=\langle 234 \rangle$ are represented by blue lines and blue triangles, respectively.}
\end{subfigure}
~
\begin{subfigure}[h!]{.45\textwidth}
\centering
$\vcenter{\hbox{\includegraphics[]{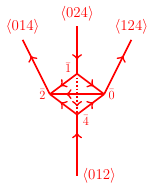}}}
\longrightarrow
\vcenter{\hbox{\includegraphics[]{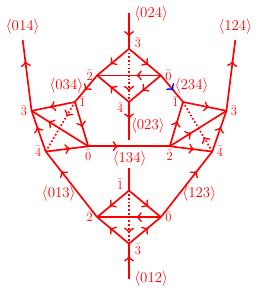}}}$
\caption{(1-4) move for resolved dual lattice $\tilde{\mathcal P}$. Blue arrow on dual (red) links $\langle 234 \rangle$ indicates that the directions are reversed by singular surfaces $S$.}
\end{subfigure}
\caption{A (1-4) move.}
\label{fig:1-4move}
\end{figure*}

Until now, the construction in 3D has been very similar to the 2D case. However, there is a very crucial difference. Although both the lattices before and after the Pachner move have \emph{local} Kasteleyn properties, the Kasteleyn orientation for larger loops may be broken. Consider for example the case in \fig{fig:2-3move}. If we consider that the large loop consists of links $\langle012\rangle$, $\langle123\rangle$, and $\langle124\rangle$ (and the newly resolved links between them) on the left lattice $\tcP$ and links $\langle012\rangle$, and $\langle124\rangle$ (and the newly resolved links between them) on the right lattice $\tcP'$, the Kasteleyn properties are the same (all five links on $\tcP$ and three links on $\tcP'$ have an up direction). On the other hand, if we consider that the large loop consists of links $\langle012\rangle$, $\langle123\rangle$, and $\langle234\rangle$ on the left lattice $\tcP$ and links $\langle012\rangle$, $\langle024\rangle$, and $\langle234\rangle$ on the right lattice $\tcP'$, the Kasteleyn properties are changed (all five links on $\tcP$ have an up direction, but only two of five links on $\tcP'$ have an up direction). In the next subsection, we systematically analyze the Kasteleyn properties of loops under Pachner moves. After decorating Majorana fermions, we see that the Majorana fermion parity changes, which is crucial in constructing legitimate FSPT states.


\subsection{FSLU transformations and consistent conditions for fixed-point states}
\label{sec:3D:fSPT}

With the above setup of discrete spin structures and Kasteleyn orientation construction on a 3D lattice, we can now use the FSLU transformation to classify 3D FSPT states systematically.

\subsubsection{Fermion parity conservation and the obstruction of Kitaev's Majorana chain decoration}

Similar to the 2D case, our 3D model has two types of fermionic degrees of freedom. The first type is the complex fermion $c_{(ijkl)}$, which resides at the center of tetrahedron $\langle ijkl\rangle$ of triangulation $\cT$ of the space manifold. In the special group super-cohomology wavefunction, the $c$ fermion parity $P_f^c$ is unchanged under (2-3) and (1-4) moves. If we use $n_3(g_i,g_j,g_k,g_l)= 0,1$ to denote the number of $c$ fermions at tetrahedron $\langle ijkl\rangle$, then the parity conserved condition becomes $\dd n_3=0$ (mod 2). Therefore, $n_3$ is an element of $H^3(G_b,\mathbb Z_2)$. This is not true if we introduce the second type of fermion.

The second type of fermion, complex fermion $a_{(ijk)}$, resides on triangle $\langle ijk\rangle$ of $\cT$. Similar to the 2D case, we also separate fermion $a_{(ijk)}$ into two Majorana fermions:
\begin{align}
\gamma_{ijkA} &= a_{(ijk)} + a_{(ijk)}^\dagger,\\
\gamma_{ijkB} &= \frac{1}{i} \left( a_{(ijk)} - a_{(ijk)}^\dagger \right).
\end{align}
The Majorana fermions $\gamma_{ijkA}$ and $\gamma_{ijkB}$ reside on the two sides of triangle $\langle ijk\rangle$, or dually, on two ends of the link in $\tcP$ dual to triangle $\langle ijk\rangle$. Our convention is that the dual link (we also use $\langle ijk\rangle$ to denote the dual link) has direction from vertex $\langle ijkA\rangle$ to $\langle ijkB\rangle$. As such, the fermion parity operator of $a$ fermion or $\gamma$ fermion at triangle $\langle ijk\rangle$ is $P_f^\gamma=-i\gamma_{ijkA}\gamma_{ijkB}$.

Now we decorate Kitaev's Majorana chains onto the loops in dual lattice $\cP$. We introduce a $\mathbb Z_2$ cochain $\tilde n_2(g_i,g_j,g_k)=0,1$ to specify the decoration configuration of Kitaev's Majorana chain. If there is a Kitaev chain that goes though link $\langle ijk \rangle$ in $\cP$ (see the green links in \fig{fig:dof3D} and figures in \eq{eq:Fmove}), then we set $\tilde n_2(g_i,g_j,g_k)=1$. On the other hand, $\tilde n_2(g_i,g_j,g_k)=0$ means there is no Kitaev chain. The Kitaev chain decorations can be translated to dimer configurations of Majorana pairs in the resolved dual lattice $\tcP$. $\tilde n_2(g_i,g_j,g_k)=0$ indicates vacuum pairing, i.e., the two Majorana fermions at triangle $\langle ijk\rangle$ are paired up from $\langle ijkA\rangle$ to $\langle ijkB\rangle$. If $\tilde n_2(g_i,g_j,g_k)=1$, then $\gamma_{ijkA}$ and $\gamma_{ijkB}$ are paired up with other nearby Majorana fermions separately, similar to the construction of Kitaev's Majorana chain (see figures of \eq{eq:Fmove2}, where gray ellipses indicate paired Majorana fermions).

\begin{figure*}[h!]
\centering
\begin{subfigure}[h!]{.3\textwidth}
\centering
\includegraphics[]{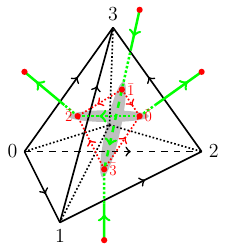}
\caption{Positive oriented tetrahedron}
\end{subfigure}
~
\begin{subfigure}[h!]{.3\textwidth}
\centering
\includegraphics[]{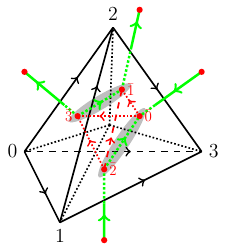}
\caption{Negative oriented tetrahedron}
\end{subfigure}
\caption{Resolvation of four strings of Kitaev's Majorana chains meeting at one tetrahedron. If four strings meet at one tetrahedron, we should pair the Majorana fermions $\bar 0$ to $\bar 2$ and $\bar 1$ to $\bar 3$, respectively (gray ellipses). The green lines on $\tilde{\mathcal P}$ indicate that these lines are decorated by Kitaev chains.}
\label{fig:4string}
\end{figure*}

As we are constructing an SPT state without intrinsic anyonic excitations, the decorated Kitaev chain should form a closed loop without ends. Therefore, similar to the 2D case, we have the equation $\dd \tilde n_2=0$ (mod 2), which means that the cochain $\tilde n_2$ is an element of $H^2(G_b,\mathbb Z_2)$. It is possible that all four faces of a tetrahedron $\langle 0123\rangle$ in $\cT$ are decorated with Kitaev chains, i.e., $\dd \tilde n_2(g_0,g_1,g_2,g_3)=4$. There are ambiguities in pairing four Majorana fermions inside the tetrahedron. For the total three possible pairings, we use the convention that the Majorana fermion $\bar 0$ is paired to $\bar 2$ and $\bar 1$ is paired to $\bar 3$ (see \fig{fig:4string}). One can also choose other conventions which essentially produce the same results \footnote{If we choose the convention of pairing \unexpanded{$\langle \bar 0 \bar 3\rangle$} and \unexpanded{$\langle \bar 1 \bar 2\rangle$} of the four Majorana fermions at the center of tetrahedron \unexpanded{$\langle 0123 \rangle$}, all of the results are the same as those of the conventional \unexpanded{$\langle \bar 0 \bar 2 \rangle$} and \unexpanded{$\langle \bar 1 \bar 3 \rangle$} used in the main text of our paper. \unexpanded{$\gamma_{012A}$} and \unexpanded{$\gamma_{234B}$} can also be chosen consistently in the definition of $X$ operators (see the discussion below \eq{eq:X}). The last choice of pairing \unexpanded{$\langle \bar0\bar1\rangle$} and \unexpanded{$\langle\bar2\bar3\rangle$} is different from the above two conventions because the Majorana fermion parity is changed when four Kitaev strings meet at one tetrahedron. However, we can shift the $c$ fermion number $n_3(g_0,g_1,g_2,g_3)$ by one (mod 2) to absorb this fermion parity change and then obtain the same obstruction results.}.

Now we turn to the Pachner moves for different triangulations. We find that the Majorana fermion parity is changed under a (2-3) move if and only if
\begin{align}
Sq^2(\tilde n_2)(g_0,g_1,g_2,g_3,g_4) = \tilde n_2(g_0,g_1,g_2) \tilde n_2(g_2,g_3,g_4) = 1.
\end{align}
We use the notation $Sq^2(\tilde n_2) = \tilde n_2^2 = \tilde n_2 \smile \tilde n_2$ later. This can be obtained by directly checking the property of Kasteleyn orientation for the loops in the transition graph of two Majorana dimer states for $64$ kinds of string configurations of each (2-3) move (see Supplementary Material for details). To compensate for the Majorana fermion parity changes, the fermion parity of the fermions at the center of tetrahedron should also be changed, leading to the (mod 2) equation
\begin{align}
\label{eq:fparity}
\dd n_3 = \tilde n_2 \smile \tilde n_2,
\end{align}
or more explicitly
\begin{align}
n_3(g_1,g_2,g_3,g_4) + n_3(g_0,g_2,g_3,g_4) + n_3(g_0,g_1,g_3,g_4) + n_3(g_0,g_1,g_2,g_4) + n_3(g_0,g_1,g_2,g_3) = \tilde n_2(g_0,g_1,g_2) \tilde n_2(g_2,g_3,g_4).
\end{align}
The above equation shows that the cocycle equation of $n_3$ is twisted by $\tilde n_2^2$, which is different from $\dd n_3=0$ in the special group super-cohomology model \cite{Gu2014}. The above equation for $n_3$ has solutions if and only if $Sq^2(\tilde n_2)$ is the trivial element in $H^4(G_b,\mathbb Z_2)$.

\subsubsection{Fermionic symmetric local unitary transformations}

After fermion decoration, the standard (2-3) move \fig{fig:23F} becomes a fermionic unitary transformation between the fermionic Fock spaces on two different triangulation lattices $\cT$ (left) and $\cT'$ (right). An example of this standard $F$ move, which changes the fermion parity of Majorana fermions, is [on lattice $\cT$, $\cP$ in \eq{eq:Fmove} and on $\tcP$ in \eq{eq:Fmove2}]:
\begin{align}
\label{eq:Fmove}
\Psi\left(
\vcenter{\hbox{\includegraphics[]{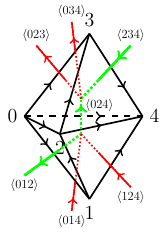}}}
\right)\quad = \quad
F(g_0,g_1,g_2,g_3,g_4)
\quad
\Psi\left(
\vcenter{\hbox{\includegraphics[]{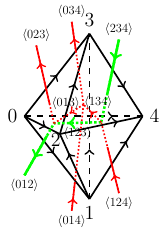}}}
\right),
\end{align}

\begin{align}
\label{eq:Fmove2}
\Psi\left(
\vcenter{\hbox{\includegraphics[]{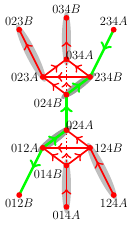}}}
\right)\quad = \quad
F(g_0,g_1,g_2,g_3,g_4)
\quad
\Psi\left(
\vcenter{\hbox{\includegraphics[]{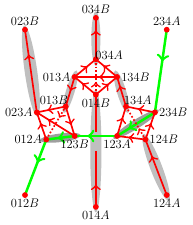}}}
\right),
\end{align}
where the $F$ operator is given by
\begin{align}\label{eq:3DF}
F(g_0,g_1,g_2,g_3,g_4) = \nu_4(g_0,g_1,g_2,g_3,g_4)
c^{\dagger n_3(0124)}_{(0124)} c^{\dagger n_3(0234)}_{(0234)} c^{n_3(0123)}_{(0123)} c^{n_3(0134)}_{(0134)} c^{n_3(1234)}_{(1234)} X[\tilde n_2(g_i,g_j,g_k)].
\end{align}
Here, $c^{\dagger n_3(0124)}_{(0124)}$ is the abbreviation of $c^{\dagger n_3(g_0,g_1,g_2,g_4)}_{(0124)}$, which is the creation operator for $c$ fermions at tetrahedron $\langle0124\rangle$, etc. The $X[\tilde n_2]$ operator in the above $F$ move from the resolved dual lattice $\tcP$ to $\tcP'$ has the following general expression:
\begin{align}
\label{eq:Xgeneral}\nonumber
X[\tilde n_2(g_i,g_j,g_k)]
& = P_{\tcP} \cdot \gamma_{012A}^{\tilde n_2^2(g_0,g_1,g_2,g_3,g_4)},\\
P_{\tcP} &= \left(\prod_{\mathrm{loop\;}i\mathrm{\;in\; (\tcP,\tcP')}}2^{(L_i-1)/2}\right) \left( \prod_{\mathrm{Majorana\; pairs\;} \langle a,b\rangle \mathrm{\;in\;}\tcP} P_{a,b} \right) \left( \prod_{\mathrm{link\;}\langle ijk\rangle \notin \tcP} P_{ijkA,ijkB} \right),
\end{align}
where $2L_i$ is the length of the $i$-th loop in the transition graph of dimer configurations in $\tcP$ and $\tcP'$ and $P_{a,b}=(1-i\gamma_a\gamma_b)/2$ is the projection operator for Majorana pairs $\langle a,b\rangle$ (from vertex $a$ to vertex $b$). The projection operators in the second parenthesis project the state in the right figure to the Majorana dimer configuration states in the left figure. The projection operators in the third parenthesis are the vacuum projection operators for those Majorana fermions that do not appear in the left figure explicitly (this is similar to the projection $P_{02A,02B}$ in \eq{eq:X20} of the 2D case). An explicit expression of $X$ for the Majorana pair configuration in Eqs.~(\ref{eq:Fmove}) and (\ref{eq:Fmove2}) is
\begin{align}
\label{eq:X}
X[\tilde n_2] = 2 \left( P_{024B,234B} P_{012A,024A} \right) \left( P_{013A,013B} P_{123A,123B} P_{134A,134B} \right) \gamma_{012A}.
\end{align}
Note that the normalization factor in the front of a general $X$ operator is the same as \eq{eq:factor}. For the $F$ move in Eqs.~(\ref{eq:Fmove}) and (\ref{eq:Fmove2}), there are only one loop in the Majorana dimer transition graph with length bigger than two: $012A$-$024A$-$024B$-$234B$-$123A$-$123B$-$012A$. So the normalization factor is $2^{(6/2-1)/2}=2$, as shown in \eq{eq:X}.

When the $F$ move changes the Majorana fermion parity, the last term of the $X$ operator is the Majorana fermion operator $\gamma_{012A}=a_{(012)}+a_{(012)}^\dagger$. The $X$ operator is now an operator with an odd number of $a$ fermion creation or annihilation operators, which changes the fermion parity of the state. We check that, for all possible Kitaev's Majorana chain configurations, the loop-breaking Kasteleyn orientation in the transition graph of the two Majorana dimer states always contains vertex $012A$. Therefore, the $X$ operator with $\gamma_{012A}$ should indeed project the state to the desired Majorana configuration state (not $0$). In fact, $\gamma_{234B}$ is also an allowed choice. We calculate the consistent equation of $\nu_4$ for both choices later.

The fermionic local unitary transformation $F$ should also be $G_b$ symmetric.
\begin{align}
F(g_0,g_1,g_2,g_3,g_4) = F(gg_0,gg_1,gg_2,gg_3,gg_4),
\end{align}
for all $g\in G_b$ if $G_b$ is a unitary symmetry group. So, $\nu_4(g_0,g_1,g_2,g_3,g_4)$, $n_3(g_0,g_1,g_2,g_3)$, and $\tilde n_2(g_0,g_1,g_2)$ are all cochains that are invariant under unitary $g$ action. (We note that $\nu_4(gg_0,gg_1,gg_2,gg_3,gg_4)=\nu_4^*(g_0,g_1,g_2,g_3,g_4)$ for anti-unitary $g$ action.)


Similar to the 1D and 2D cases, we can use FSLU to redefine the basis state $|\{g_n\}\rangle$ as
\begin{align}\nonumber
|\{g_n\}\rangle' = U_{\mu_3,m_2} |\{g_n\}\rangle
&= \prod_{\langle ijkl \rangle} \mu_3(g_i,g_j,g_k,g_l)^{s_{\langle ijkl \rangle}} \prod_{\langle ijk \rangle} \left[ f^{m_2(g_i,g_j,g_k)}_{ijkB}f^{m_2(g_i,g_j,g_k)}_{ijkA} \right]\\
&\quad\cdot \prod_{\langle ijkl\rangle} \left[ f^{\dagger m_2(g_j,g_k,g_l)}_{jklA} f^{\dagger m_2(g_i,g_j,g_l)}_{ijlA} f^{\dagger m_2(g_i,g_k,g_l)}_{iklB} f^{\dagger m_2(g_i,g_j,g_k)}_{ijkB} \right] |\{g_n\}\rangle,
\end{align}
where we first create four complex fermions $f_{jklA}$, $f_{ijlA}$, $f_{iklB}$ and $f_{ijkB}$ near the four triangles of the tetrahedron $\langle ijkl\rangle$ ($i<j<k<l$), and then annihilate the two fermions $f_{ijkA}$ and $f_{ijkB}$ on the two sides of triangle $\langle ijk\rangle$ when gluing the two tetrahedra sharing triangle $\langle ijk\rangle$. To preserve the fermion parity and be symmetric, $m_2$ should be a 2-cocycle (with $\mathbb Z_2$ coefficient): $m_2(gg_i,gg_j,gg_k)=m_2(g_i,g_j,g_k)$ and $\dd m_2(g_i,g_j,g_k,g_l)=0$. $\mu_3(g_i,g_j,g_k,g_l)$ is a phase factor associated with tetrahedron $\langle ijkl\rangle$, and $s_{\langle ijkl \rangle}=\pm 1$ denotes the orientation of the tetrahedron. After tedious calculation of eliminating all $f$ fermions in the new $F$ move operator $F' = U_{\mu_3,m_2} F U_{\mu_3,m_2}^\dagger$, one find that the phase factor in \eq{eq:3DF} becomes
\begin{align}
\nu_4'(g_0,g_1,g_2,g_3,g_4) = \nu_4(g_0,g_1,g_2,g_3,g_4) \frac{\mu_3(g_1,g_2,g_3,g_4)\mu_3(g_0,g_1,g_3,g_4)\mu_3(g_0,g_1,g_2,g_3)}{\mu_3(g_0,g_2,g_3,g_4)\mu_3(g_0,g_1,g_2,g_4)}(-1)^{Sq^2(m_2)(g_0,g_1,g_2,g_3,g_4)},\label{3Dboundary}
\end{align}
where $Sq^2(m_2)(g_0,g_1,g_2,g_3,g_4)=(m_2\smile m_2)(g_0,g_1,g_2,g_3,g_4)=m_2(g_0,g_1,g_2)m_2(g_2,g_3,g_4)$ is the Steenrod square of $m_2$. If $\nu_4$ and $\nu_4'$ can be related by the above equation, then they correspond to the same 3D FSPT phase. This is also consistent with the general result in \eq{eq:fcoboundary} \cite{Gu2014}.

In addition to the standard (2-3) move, there are (2-0) moves as FSLU that can change the vertex number. One example of the Majorana fermion configuration for a (2-0) move is
\begin{align}
\label{fig:(2-0)3D}
\Psi\left(
\vcenter{\hbox{\includegraphics[]{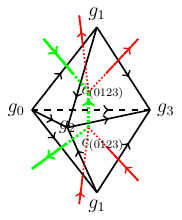}}}
\right)
= \frac{1}{|G_b|^{1/2}}  c_{(0123)}^{\dagger n_3(g_0,g_1,g_2,g_3)} \bar c_{(0123)}^{\dagger n_3(g_0,g_1,g_2,g_3)} X[\tilde n_2(g_i,g_j,g_k)]\ 
\Psi\left(
\vcenter{\hbox{\includegraphics[]{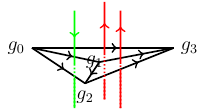}}}
\right),
\end{align}
\begin{align}
\label{fig:(2-0)3D-b}
\Psi\left(
\vcenter{\hbox{\includegraphics[]{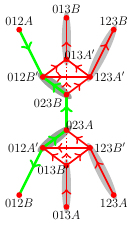}}}
\right)
= \frac{1}{|G_b|^{1/2}}  c_{(0123)}^{\dagger n_3(g_0,g_1,g_2,g_3)} \bar c_{(0123)}^{\dagger n_3(g_0,g_1,g_2,g_3)} X[\tilde n_2(g_i,g_j,g_k)]\ 
\Psi\left(
\vcenter{\hbox{\includegraphics[]{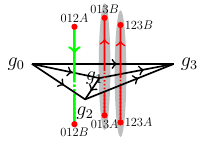}}}
\right).
\end{align}
The notations and conventions of vacuum pairs are similar to those in the 2D case in the discussion following \eq{fig:(2-0)}. $c_{(0123)}$ and $\bar c_{(0123)}$ fermions are at the centers of the lower and upper tetrahedra respectively. An explicit expression for the $X$ operator is
\begin{align}
\label{eq:20X3D}
X[\tilde n_2] = 2^{3/2}P_{023B,012B'} P_{012A',023A} P_{013A',013B} P_{123A',123B}P_{013A,013B'}P_{123A,123B'}.
\end{align}
Note that the normalization factor in the front of $X$ operator is $(2^{(4/2-1)/2})^3=2^{3/2}$, according to \eq{eq:factor}. Because there are three loops in the Majorana dimer transition graph with length bigger than two: 
$012B'$-$023B$-$023A$-$012A'$-$012B'$, $013B$-$013A'$-$013B'$-$013A$-$013B$, and $123B$-$123A'$-$123B'$-$123A$-$123B$.
Using two kinds of (2-0) moves (the other one is shown in Supplementary Material) and the standard (2-3) move, we can deduce all (1-4) moves and other (2-3) moves from the standard $F$ move in Eqs.~(\ref{eq:Fmove}) and (\ref{eq:Fmove2}) (see Supplementary Material for all details).

\subsubsection{Fermionic hexagon equations}

\begin{figure*}[t]
\centering
\includegraphics[]{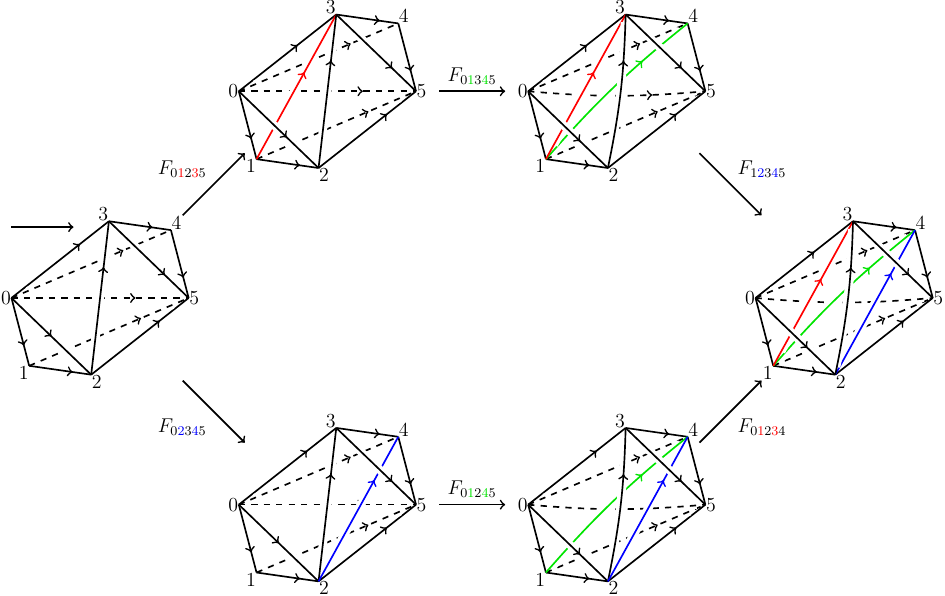}
\caption{Fermionic fusion hexagon equation. All of the $F$ moves are of the standard (2-3) move (see \fig{eq:Fmove} and \fig{eq:Fmove2}). Colored numbers $i$ and $j$ in the subscript of $F$ indicate that the link $\langle ij\rangle$ with the same color is added after this $F$ move. All six $F$ moves do not introduce new singular lines and surfaces. There is a global direction from left to right such that the vertex with the smaller number has the earlier time.}
\label{fig:hexagon}
\end{figure*}

The consistent equation for the moves defined in the above subsection is the fermionic hexagon equation (see \fig{fig:hexagon}), which is a higher dimensional version of the fermionic pentagon equation for (2+1)D fermionic topological phases \cite{Gu2015}.
Reordering the fermionic operators in the fermionic hexagon equation will twist the bosonic cocycle equation for $\nu_4$:
\begin{align}
(\dd\nu_4)(g_0,g_1,g_2,g_3,g_4,g_5) = \mathcal O (g_0,g_1,g_2,g_3,g_4,g_5).
\end{align}
The phase twist or obstruction on the left-hand side has three terms
\begin{align}
\mathcal O = \mathcal O_c \mathcal O_{c\gamma} \mathcal O_\gamma,
\end{align}
resulting from three different fermion phase factors: (1) $\mathcal O_c$ from reordering $c$ fermions, (2) $\mathcal O_{c\gamma}$ from reordering $c$ fermions and $\gamma$ Majorana fermions, and (3) $\mathcal O_\gamma$ from reordering $\gamma$ Majorana fermions. The final result of the phase twist is
\begin{align}
\label{eq:O}
\mathcal O (012345) &= (-1)^{(n_3 \smile_1 n_3) (012345) + (n_3 \smile_2 \dd n_3) (012345) + \dd n_3(02345) \dd n_3(01245) \dd n_3(01234)}\\\nonumber
&\quad\cdot i^{\dd n_3(01235) \dd n_3(02345) + \dd n_3(01345) \dd n_3(12345)}
\cdot (-i)^{\dd n_3(02345) \dd n_3(01245) + \dd n_3(02345) \dd n_3(01234)}.
\end{align}
Here, $(012345)$ is the abbreviation of $(g_0,g_1,g_2,g_3,g_4,g_5)$, etc.
Note that the obstruction depends only on $n_3$ and $\dd n_3$ (or $\tilde n_2^2$ through the fermion parity equation $\dd n_3=\tilde n_2^2$). In the following, we derive the above obstruction equation in detail.

The first part $\mathcal O_c$ contains the special group super-cohomology results \cite{Gu2014}, which are obtained by reordering $c$ fermion operators.
\begin{align}
\label{eq:Oc}
\mathcal O_c (012345) &= (-1)^{\left[Sq^2(n_3)+\dd n_3\smile_2 n_3\right](012345)}
=(-1)^{\left[n_3\smile_1 n_3 + \dd n_3\smile_2 n_3\right](012345)}\\\nonumber
&= (-1)^{n_3(0345)n_3(0123) + n_3(0145)n_3(1234) + n_3(0125)n_3(2345)}\\\nonumber
&\quad\cdot (-1)^{\dd n_3(01234)n_3(0145) +\dd n_3(02345)n_3(0125) +\dd n_3(01234)n_3(1245) +\dd n_3(01345)n_3(1235) +\dd n_3(01234)n_3(2345) +\dd n_3(01245)n_3(2345)}.
\end{align}
Here, $\smile_i$ is Steenrod's $i$-th cup product \cite{Steenrod1947}. Apart from the term $Sq^2(n_3)$, there is an additional term $\dd n_3\smile_2 n_3$ because of $\dd n_3=\tilde n_2^2 \neq 0$.

The second part $\mathcal O_{c\gamma}$ comes from reordering $c$ fermions and $\gamma$ Majorana fermions (we always put $c$ fermions in front of $\gamma$ fermions). For example, the composition of $F_{01345}$ and $F_{01235}$ gives a sign $(-1)^{\dd n_3(01345) \dd n_3(01235)}$. In total, the upper path of three $F$ moves gives the power $\dd n_3(01345)\dd n_3(01235)+\dd n_3(12345)\dd n_3(01345)+\dd n_3(12345)\dd n_3(01235)$, whereas the lower path gives $\dd n_3(01245)\dd n_3(01234)+\dd n_3(02345)\dd n_3(01234)+\dd n_3(02345)\dd n_3(01245)$. The final result is then
\begin{align}
&\quad\mathcal O_{c\gamma} (012345) = (-1)^{\left(\dd n_3 \smile_3 \dd n_3\right) (g_0,g_1,g_2,g_3,g_4,g_5)}\\\nonumber
&= (-1)^{\dd n_3(01245)\dd n_3(01234)+\dd n_3(01235)\dd n_3(01345)+\dd n_3(02345)\dd n_3(01234) + \dd n_3(02345)\dd n_3(01245)+\dd n_3(01235)\dd n_3(12345)+\dd n_3(01345)\dd n_3(12345)}.
\end{align}
After adding coboundary $(-1)^{\dd (n_3\smile_3 \dd n_3)} = (-1)^{\dd n_3\smile_3 \dd n_3 + n_3\smile_2\dd n_3 + \dd n_3\smile_2 n_3}$ to the combination of the phase factor $\mathcal O_c$ and $\mathcal O_{c\gamma}$, we obtain a simpler expression.
\begin{align}
\label{eq:Ocg}
\mathcal O_c \mathcal O_{c\gamma} = (-1)^{n_3\smile_1 n_3 + n_3 \smile_2 \dd n_3}.
\end{align}
Note that adding the coboundary changes the $U(1)$-cochain $\nu_4 \rightarrow (-1)^{n_3\smile_3 \dd n_3} \nu_4$.

\begin{figure*}[t]
\centering
\includegraphics[]{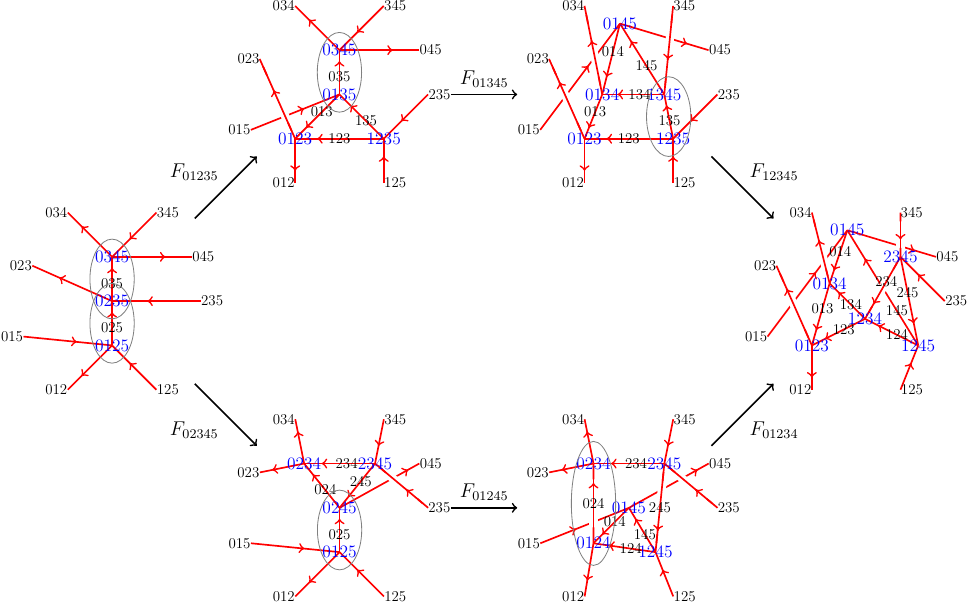}
\caption{Fermionic fusion hexagon equation on dual lattice $\cP$.}
\label{fig:hexagon_dual}
\end{figure*}

We now turn to the subtlest part $\mathcal O_\gamma$ coming from a decorated Majorana chain. In addition to $\pm 1$, $\mathcal O_\gamma$ can take values in $\pm i$(See Supplementary Material for the physical origin of this purely imaginary phase factor). If all six $F$ moves do not change the Majorana fermion parity, then $X_\gamma$ operators in $F$ moves are merely projections with an even number of $\gamma$ Majorana operators. The $c$ fermions and Majorana fermions are decoupled, and both $\mathcal O_{c\gamma}$ and $\mathcal O_\gamma$ are trivial. The obstruction $\mathcal O$ is the same as the special group super-cohomology result. Therefore, we only need to check the case in which some of the six $F$ moves in \fig{fig:hexagon} change the Majorana fermion parity, i.e., some of $\tilde n_2^2(01234)$, $\tilde n_2^2(01245)$, $\tilde n_2^2(02345)$, $\tilde n_2^2(01345)$, $\tilde n_2^2(01235)$, and $\tilde n_2^2(12345)$ are equal to one. We denote the six $X$ operators in $F$ moves as $X_{01235}=P_1 \gamma_{012A}^{\tilde n_2^2(01235)}$, $X_{01345}=P_2 \gamma_{013A}^{\tilde n_2^2(01345)}$, $X_{12345}=P_3 \gamma_{123A}^{\tilde n_2^2(12345)}$, $X_{02345}=P_1 \gamma_{023A}^{\tilde n_2^2(02345)}$, $X_{01245}=P_4 \gamma_{012A}^{\tilde n_2^2(01245)}$, and $X_{01234}=P_5 \gamma_{012A}^{\tilde n_2^2(01234)}$. Here, $P_i$ means the projection operator onto the Majorana chain configuration of the $i$-th figure in the hexagon equation. $\mathcal O_\gamma$ is defined as the Majorana chain phase difference of the upper and lower paths in hexagon equation
\begin{align}
X_{02345} X_{01245} X_{01234} |\mathrm{final}\rangle = \mathcal O_\gamma(012345) X_{01235} X_{01345} X_{12345} |\mathrm{final}\rangle,
\end{align}
where $|\mathrm{final}\rangle$ is the state of the Majorana chain configuration in the rightmost figure of the hexagon equation \fig{fig:hexagon_dual}. We can calculate $\mathcal O_\gamma$ from the expression
\begin{align}
\label{eq:f_gamma}
\nonumber
\mathcal O_\gamma (012345) &= \langle \mathrm{final}| X_{12345}^\dagger X_{01345}^\dagger X_{01235}^\dagger X_{02345} X_{01245} X_{01234} |\mathrm{final}\rangle\\
&= \langle \mathrm{final}| \gamma_{123A}^{\tilde n_2^2(12345)} P_3 \gamma_{013A}^{\tilde n_2^2(01345)} P_2 \gamma_{012A}^{\tilde n_2^2(01235)} P_1 P_1 \gamma_{023A}^{\tilde n_2^2(02345)} P_4 \gamma_{012A}^{\tilde n_2^2(01245)} P_5 \gamma_{012A}^{\tilde n_2^2(01234)} |\mathrm{final}\rangle.
\end{align}
The above equation suggests that $\mathcal O_\gamma$ depends only on the values of $\tilde n_2^2(01235)$, $\tilde n_2^2(01345)$, $\tilde n_2^2(12345)$, $\tilde n_2^2(02345)$, $\tilde n_2^2(01245)$, and $\tilde n_2^2(01234)$, i.e., the Majorana parity changes of the six $F$ moves.

\begin{figure*}[t]
\centering
\includegraphics[]{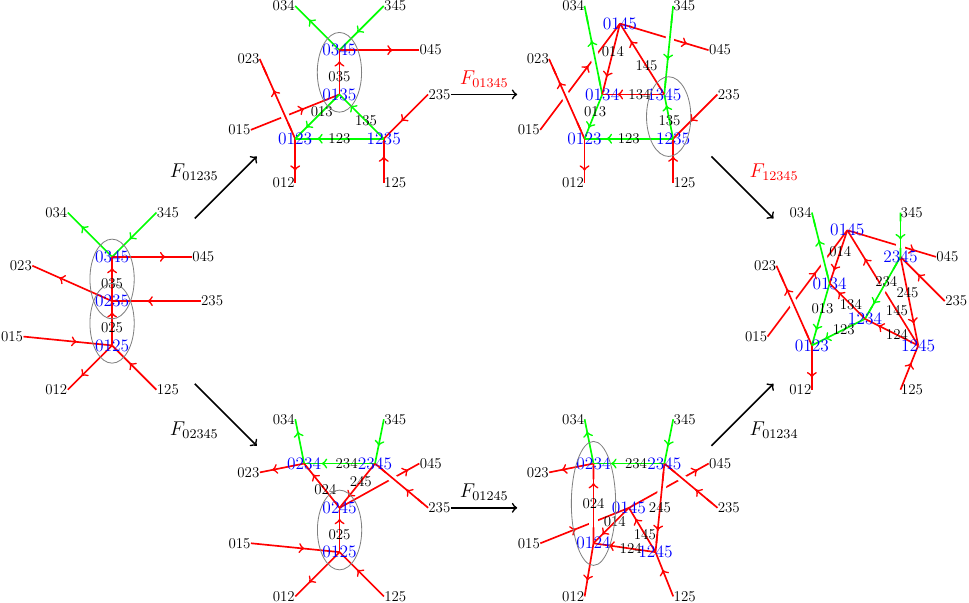}
\caption{One example of Kitaev's Majorana chain decoration configuration of fermionic fusion hexagon equation. Green lines indicate that these dual links are decorated by Kitaev's Majorana chains. Among the six $F$ moves for this configuration, only $F_{01345}$ and $F_{12345}$ (red color in the figure) change the Majorana fermion parity. Therefore, this choice of $\tilde n_2(g_i,g_j,g_k)$ belongs to the $(0,1,1,0,0,0)$ row in \tab{tab:Ogamma} and has obstruction $\mathcal O_\gamma=i$.}
\label{fig:hexagon_dual2}
\end{figure*}

Consider, for example, the case in which only $F_{01345}$ and $F_{12345}$ change the Majorana fermion parity [$(0,1,1,0,0,0)$ in the eighth row of \tab{tab:Ogamma}. See \fig{fig:hexagon_dual2} for an example of $\tilde n_2(g_i,g_j,g_k)$ satisfying this condition. We can expand projection operators $P_i$ to Majorana fermion operators. For simplicity, we can consider the term with only contributions $-i\gamma_{013A}\gamma_{013B}$ and $-i\gamma_{123A}\gamma_{123B}$ from $P_1$. We can also add $-i\gamma_{013B}\gamma_{123B}$, which equals 1 when acting on $|\mathrm{final}\rangle$ in front of this state. The result is then
\begin{align}
\mathcal O_\gamma(012345)|_{(0,1,1,0,0,0)} &= \langle \mathrm{final}| \gamma_{123A} P_3 \gamma_{013A} P_2 P_1 P_4 P_5 P_6 |\mathrm{final}\rangle \\\nonumber
&= \langle \mathrm{final}| \gamma_{123A} \gamma_{013A} (-i\gamma_{013A}\gamma_{013B}) (-i\gamma_{123A}\gamma_{123B}) (-i\gamma_{013B}\gamma_{123B}) |\mathrm{final}\rangle \\\nonumber
&= i.
\end{align}
We assume that there are only two Kitaev strings meeting at tetrahedron $\langle0123\rangle$. (this is always true for all possible choices of $\tilde n_2(g_i,g_j,g_k)$ that belong to the eighth row of \tab{tab:Ogamma}.) Similarly, one can calculate other choices of $\tilde n_2$, and at last obtain the results listed in \tab{tab:Ogamma}. The final results can be summarized to the expression
\begin{align}\label{eq:Og_tab}
\mathcal O_\gamma (012345) = i^{\tilde n_2^2(01235) \tilde n_2^2(02345) + \tilde n_2^2(01345) \tilde n_2^2(12345)} (-i)^{\tilde n_2^2(02345) \tilde n_2^2(01245) + \tilde n_2^2(02345) \tilde n_2^2(01234)} (-1)^{\tilde n_2^2(02345) \tilde n_2^2(01245) \tilde n_2^2(01234)}.
\end{align}
Combining this $\mathcal O_\gamma$ result with $\mathcal O_c \mathcal O_{c\gamma}$ in \eq{eq:Ocg}, we obtain the obstruction claimed in \eq{eq:O}.

%

\begin{table*}[h]
\centering
\caption{Calculations of $\mathcal O_\gamma$ from all possible Kitaev chain configurations in the hexagon equation. The first column has entries $\left(\tilde n_2^2(01235),\tilde n_2^2(01345),\tilde n_2^2(12345),\tilde n_2^2(02345),\tilde n_2^2(01245), \tilde n_2^2(01234)\right)$, indicating whether the six $F$ moves in the hexagon equation change the Majorana fermion parity. The second column shows the $\gamma$ operators appearing in \eq{eq:f_gamma}. The third and fourth columns are lines of Majorana dimer pairs ($\emptyset$ means there are no $\tilde n_2(g_i,g_j,g_k)$ that have two or four strings at tetrahedron $\langle 0123\rangle$). The last two columns are the values of $\mathcal O_\gamma$ for $\tilde n_2(g_i,g_j,g_k)$ belonging to this row (the last column but one uses the convention $\gamma_{012A}$ in \eq{eq:Xgeneral}, and the last column uses the convention $\gamma_{234B}$). Nontrivial results are labelled in red or blue. The results in this table can be summarized to the expression shown in \eq{eq:Og_tab} for convention $\gamma_{012A}$ and in \eq{eq:Og_tab2} for convention $\gamma_{234B}$.}
\label{tab:Ogamma}
\begin{tabular}{|c||c|c|c|c||c|}
\hline
fermion parity changes & $\gamma$ operators in \eq{eq:f_gamma} & line (2 strings at $\langle0123\rangle$) & line (4 strings at $\langle0123\rangle$) & $\mathcal O_\gamma|_{012A}$ & $\mathcal O_\gamma|_{234B}$ \\
\hline\hline
$(0,0,0,0,0,0)$ & $-$ & $-$ & $-$ & $1$ & $1$ \\
\hline
$(1,1,0,0,0,0)$ & $\gamma_{012A},\gamma_{013A}$ & $012A$-$013B$-$013A$ & $\emptyset$ & $1$ & {\color{blue}$-i$} \\
$(1,0,1,0,0,0)$ & $\gamma_{012A},\gamma_{123A}$ & $012A$-$123B$-$123A$ & $\emptyset$ & $1$ & {\color{blue}$-i$} \\
${\color{red}(1,0,0,1,0,0)}$ & $\gamma_{012A},\gamma_{023A}$ & $012A$-$023A$ & $\emptyset$ & {\color{red}$i$} & {\color{blue}$i$} \\
$(1,0,0,0,1,0)$ & $\gamma_{012A}^2=1$ & $-$ & $-$ & $1$ & $1$ \\
$(1,0,0,0,0,1)$ & $\gamma_{012A}^2=1$ & $-$ & $-$ & $1$ & $1$ \\
${\color{red}(0,1,1,0,0,0)}$ & $\gamma_{013A},\gamma_{123A}$ & $013A$-$013B$-$123B$-$123A$ & $\emptyset$ & {\color{red}$i$} & $1$ \\
$(0,1,0,1,0,0)$ & $\gamma_{013A},\gamma_{023A}$ & $013A$-$013B$-$023A$ & $\emptyset$ & $1$ & $1$ \\
$(0,1,0,0,1,0)$ & $\gamma_{013A},\gamma_{012A}$ & $013A$-$013B$-$012A$ & $\emptyset$ & $1$ & $1$ \\
$(0,1,0,0,0,1)$ & $\gamma_{013A},\gamma_{012A}$ & $013A$-$013B$-$012A$ & $\emptyset$ & $1$ & $1$ \\
$(0,0,1,1,0,0)$ & $\gamma_{123A},\gamma_{023A}$ & $123A$-$123B$-$023A$ & $\emptyset$ & $1$ & $1$ \\
$(0,0,1,0,1,0)$ & $\gamma_{123A},\gamma_{012A}$ & $123A$-$123B$-$012A$ & $\emptyset$ & $1$ & $1$ \\
$(0,0,1,0,0,1)$ & $\gamma_{123A},\gamma_{012A}$ & $123A$-$123B$-$012A$ & $\emptyset$ & $1$ & $1$ \\
${\color{red}(0,0,0,1,1,0)}$ & $\gamma_{023A},\gamma_{012A}$ & $023A$-$012A$ & $\emptyset$ & {\color{red}$-i$} & $1$ \\
${\color{red}(0,0,0,1,0,1)}$ & $\gamma_{023A},\gamma_{012A}$ & $023A$-$012A$ & $\emptyset$ & {\color{red}$-i$} & $1$ \\
$(0,0,0,0,1,1)$ & $\gamma_{012A}^2=1$ & $-$ & $-$ & $1$ & {\color{blue}$i$} \\
\hline
$(1,1,0,0,1,1)$ & $\gamma_{013A},\gamma_{012A}^3$ &$012A$-$013B$-$013A$, $012A$ & $\emptyset$ & $1$ & $1$ \\
$(1,0,1,0,1,1)$ & $\gamma_{123A},\gamma_{012A}^3$ & $012A$-$123B$-$123A$, $012A$ & $\emptyset$ & $1$ & $1$ \\
${\color{red}(1,0,0,1,1,1)}$ & $\gamma_{023A},\gamma_{012A}^3$ & $012A$-$023A$, $012A$ & $\emptyset$ & {\color{red}$i$} & {\color{blue}$-1$} \\
${\color{red}(1,1,1,1,0,0)}$ & $\gamma_{012A},\gamma_{013A},\gamma_{123A},\gamma_{023A}$ & $\emptyset$ &
$\begin{cases}
123A\text{-}123B\text{-}013B\text{-}013A\\
\quad\quad\ \ 023A\text{-}012A
\end{cases}$
& {\color{red}$-1$} & {\color{blue}$i$} \\
$(0,1,1,1,1,0)$ & $\gamma_{013A},\gamma_{123A},\gamma_{023A},\gamma_{012A}$ & $\emptyset$ &
$\begin{cases}
123A\text{-}123B\text{-}013B\text{-}013A\\
\quad\quad\ \ 023A\text{-}012A
\end{cases}$
& $1$ & $1$ \\
$(0,1,1,1,0,1)$ & $\gamma_{013A},\gamma_{123A},\gamma_{023A},\gamma_{012A}$ & $\emptyset$ &
$\begin{cases}
123A\text{-}123B\text{-}013B\text{-}013A\\
\quad\quad\ \ 023A\text{-}012A
\end{cases}$
& $1$ & $1$ \\
\hline
${\color{red}(1,1,1,1,1,1)}$ & $\gamma_{013A},\gamma_{123A},\gamma_{023A},\gamma_{012A}^3$ & $\emptyset$ &
$\begin{cases}
123A\text{-}123B\text{-}013B\text{-}013A\\
\quad\quad\ \ 023A\text{-}012A
\end{cases}$
& {\color{red}$-1$} & {\color{blue}$-1$} \\
\hline
\end{tabular}
\end{table*}


As discussed above, we can also use the convention $X_{01234}=P_\cP\gamma_{234B}^{\tilde n_2^2(01234)}$ in the definition of a standard $F$ move. The results using different conventions may differ from each other by some coboundaries. The obstruction from Majorana fermions in the convention $\gamma_{234B}$ is
\begin{align}\label{eq:Og_tab2}
\mathcal O_\gamma (012345) = i^{\tilde n_2^2(01235) \tilde n_2^2(02345) + \tilde n_2^2(01245) \tilde n_2^2(01234)} (-i)^{\tilde n_2^2(01235) \tilde n_2^2(01345) + \tilde n_2^2(01235) \tilde n_2^2(12345)} (-1)^{\tilde n_2^2(01235) \tilde n_2^2(01345) \tilde n_2^2(12345)}.
\end{align}
By checking all possible choices of $\tilde n_2$ numerically, we find that the above expression of $\mathcal O_\gamma$ equals exactly to
\begin{align}
\mathcal O_\gamma (012345) = (-i)^{[\tilde n_2\smile(\tilde n_2 \smile_1 \tilde n_2)](012345)} (-1)^{\tilde n_2^2(01235) \tilde n_2^2(02345)}.
\end{align}
This can be obtained from the Pontrjagin dual of the four-dimensional spin bordism \cite{Morgan}.

We note that although $\mathcal O$ takes values in $\mathbb Z_4$, the additivity property actually requires $\mathcal O$ to be a cohomology map on $\mathbb Z_8$; see Supplementary Material for details. Thus, to find a solution for $\nu_4$, we must define an obstruction-free subgroup of $H^2 (G_b,\mathbb Z_2 )$, which is formed by elements $\tilde  n_{2} \in H^2(G_b, \mathbb Z_2)$ that simultaneously satisfy $Sq^2(\tilde n_{2})=0$ in
$H^{4}(G_b,\mathbb Z_2)$ and $\mathcal O(\tilde n_{2})=0$ in $H^{5}(G_b,U_T(1))$, where $\mathcal O$ is some unknown cohomology operation (to the best of our knowledge) that maps $\tilde n_{2}$ satisfying $Sq^2(\tilde n_{2})=0$ in $H^{2}(G_b,\mathbb Z_2)$ into an element in $H^{5}(G_b,\mathbb Z_8) \subset H^{5}[G_b,U_T(1)]$. We note that $n_3$ is completely determined by $\tilde n_{2}$ and that any solution of $\dd n_3=\tilde n_{2}^2$ can be used in the above definition.  
Together with the special group super-cohomolgy results, we conclude that the precise mathematical objects that classify 3D FSPT phases with a total symmetry $G_f=G_b\times \mathbb Z_2^f$ can also be summarized as three group cohomologies of the symmetry group $G_b$, $\tilde{B}H^2 (G_b,\mathbb Z_2 )$, $BH^3(G_b,\mathbb Z_2)$, and $H^4_{\rm rigid}(G_b, U_T(1))$. Commuting projector parent Hamiltonians for all of these FSPT states can also be constructed on arbitrary 3D triangulations with a branching structure.

Finally, let us provide some physical arguments to support our classification scheme for 3D FSPT phases with a total symmetry $G_f=G_b\times \mathbb Z_2^f$ when $G_b$ is a unitary symmetry group. (a) From the concept of equivalence classes of FSLU transformations, if two fixed point wavefunctions corresponded to the same phase, there would be
a finite depth FSLU circuit connecting the two. Thus, it is obvious that the fixed point wavefunctions have the same
algebraic data, e.g., cocycle solutions (up to coboundary transformation) from the the above three group cohomologies of the symmetry group $G_b$, since these algebraic data only depends on ``long
distance" physics which can't be changed by a finite depth FSLU circuit. In fact, finite depth FSLU circuit can at most generate some coboundary transformations, e.g. the transformation defined in Eq.(\ref{3Dboundary}). (b) Our constructions are consistent with the spin-cobordism classifications for 3D FSPT phases. (c) Similar to the bosonic 3D SPT states, if we turn on background gauge field $G_b$ and couple it to the 3D FSPT states constructed here, the corresponding $G_b$ flux lines will carry new types of three-loop braiding statistics. Some examples from the $BH^3(G_b,\mathbb Z_2)$ layer are studied in a recent work \cite{ChenjieFSPT}. We believe that nontrivial solutions from the layer $\tilde{B}H^2 (G_b,\mathbb Z_2 )$ will give rise to non-Abelian three-loop braiding statistics, and full details will be studied in our future work.   

\section{Discussion and conclusions }
We have constructed fixed-point wavefunctions for FSPT phases in two and three dimensions based on the novel concept of FSLU transformations. All of these FSPT states admit parent Hamiltonians consisting of commuting projectors on arbitrary triangulations with a branching structure. We believe that our construction will give rise to a complete classification for FSPT states with total symmetry $G_f=G_b \times \mathbb Z_2^f$ when $G_b$ is a unitary symmetry group. Mathematically, our constructions naturally define a general group super-cohomology theory that generalizes the so-called special 
group super-cohomology theory proposed in \Ref{Gu2014}.

In particular, one can start with a spin manifold in an arbitrary spacial dimension $d_{sp}$ and define the corresponding discrete spin structure via Poincare dual. Then, one can decorate Kitaev's Majorana chain onto the intersection lines of the $G_b$ symmetry domain walls if the first obstruction vanishes for elements $\tilde n_{d_{sp}-1} \in H^{d_{sp}-1}(G_b, \mathbb Z_2)$. That is, $Sq^2(\tilde n_{d_{sp}+1})=0$ in $H^{{d_{sp}}+1}(G_b,\mathbb Z_2)$, and the obstruction-free elements $\tilde n_{d_{sp}-1} \in H^{d_{sp}-1}(G_b, \mathbb Z_2)$ will give rise to all inequivalent patterns of Mjorana chain decoration. Finally, by applying the wavefunction renormalization on arbitrary triangulations, one may derive twisted cocycle equations where the twisted factors define some unknown cohomologoy map $\mathcal O$ that maps elements in $H^{d_{sp}-1}(G_b,\mathbb Z_2)$ satisfying $Sq^2(\tilde n_{d_{sp}-1})=0$ in $H^{d_{sp}+1}(G_b,\mathbb Z_2)$ into elements in $H^{d_{sp}+2}(G_b,\mathbb Z_8)\subset H^{d_{sp}+2}[G_b,U_T(1)]$, and the second obstruction-free condition requires $\mathcal O(\tilde n_{d_{sp}-1})=0$ in $H^{d_{sp}+2}(G_b,U_T(1))$. Elements $\tilde n_{d_{sp}-1} \in H^{d_{sp}-1}(G_b, \mathbb Z_2)$ satisfying both the first and second obstruction-free conditions may define a subgroup $\tilde BH^{d_{sp}-1}(G_b, \mathbb Z_2)\in H^{d_{sp}-1}(G_b, \mathbb Z_2)$, which allows us to write down another short exact sequence $0\rightarrow \cH^{d_{sp}+1}[G_f,U_T(1)] \rightarrow H^{d_{sp}+1}_{f}[G_f,U_T(1)]\rightarrow   \tilde{B}H^{d_{sp}-1}(G_b,\mathbb Z_2) \rightarrow 0$ to define a general group super-cohomology theory. We note that here $\cH^{d_{sp}+1}[G_f,U_T(1)]$ is the special group super-cohomology class defined by \Ref{Gu2014}.

In future, it would be of great importance to understand the physical properties of 3D FSPT phases classified by general group super-cohomology theory, e.g., understanding the braiding statistics of $G_b$-flux lines. Of course, constructing time-reversal symmetry-protected topological states with both $T^2=1$ and $T^2=P^f$ (where $P^f$ is the total fermion parity) is another interesting direction. It should also be interesting to investigate the phase transition theory among FSPT phases in arbitrary dimensions.

\begin{acknowledgements}
We thank Zheng-Xin Liu and Jianwei Li for helpful discussions on solving the system of linear Diophantine equations. We also thank Shuo Yang for helping to calculate the equations numerically. ZCG acknowledges start-up support via Direct Grant nos. 4053163 and 3132745 from The Chinese University of Hong Kong and funding from Hong Kong’s Research Grants Council (ECS no.2191110, GRF no.14306714). We also would like to thank Tianhe-1A platform at the National Supercomputer Center in Tianjin and Tianhe-2 platform at the National Supercomputer Center in Guangzhou for computational support.
\end{acknowledgements}

\textit{Note added} -- During the preparation of this work, we noticed a relevant work \Ref{kapustin17} considering similar problems. However, our expression for $\mathcal O_\gamma$
 is slightly different from \Ref{kapustin17}.

\appendix
 
\section{Kitaev's Majorana chain, fermion parity and Kasteleyn orientation}
\label{Kitaevchain}
In this Supplementary Material, we present some basic preliminaries of Kitaev's Majorana chain. The relation of fermion parity and Kasteleyn orientation will also be discussed.

For a $p$-wave superconductor with zero chemical potential on a chain with $N$ sites, the BCS Hamiltonian is
\begin{align}
H=-\sum_{j}(c_j^\dagger c_{j+1}+\mathrm{h.c.})-\sum_{j}(c_jc_{j+1}+\mathrm{h.c.}),
\end{align}
where the summation is from $j=1$ to $N-1$ (N) for open (periodic) boundary condition. $c_{N+1}$ is identified with $c_1$ in periodic boundary condition. After introducing two Majorana fermions $\gamma_{2j-1}$ and $\gamma_{2j}$ for each complex fermion $c_j$ at site $j$, i.e.,
\begin{align}
\gamma_{2j-1} &= c_j + c_j^\dagger,\\
\gamma_{2j} &= \frac{1}{i} (c_j - c_j^\dagger),
\end{align}
the BCS Hamiltonian becomes a summation of mutually commuting terms of Majorana fermion bilinear form:
\begin{align}\label{eq:H_K}
H=i\sum_{j} \gamma_{2j}\gamma_{2j+1}.
\end{align}
Since $(i\gamma_{2j}\gamma_{2j+1})^2=1$, each term of the above Hamiltonian is a projection operator (up to some constant and scaling). Therefore, the full spectrum of the model can be obtained by specifying $i\gamma_{2j}\gamma_{2j+1} = \pm 1$ for all $j$.

\subsection{Ground states}
\begin{figure*}[h!]
\centering
\begin{subfigure}[h!]{.6\textwidth}
\centering
\includegraphics[width=\linewidth]{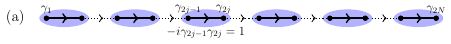}
\end{subfigure}
\begin{subfigure}[h!]{.6\textwidth}
\centering
\includegraphics[width=\linewidth]{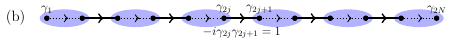}
\end{subfigure}
\begin{subfigure}[h!]{.6\textwidth}
\centering
\includegraphics[width=\linewidth]{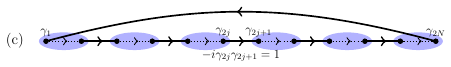}
\end{subfigure}
\begin{subfigure}[h!]{.6\textwidth}
\centering
\includegraphics[width=\linewidth]{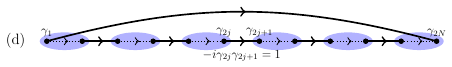}
\end{subfigure}
\caption{Several states of the Majorana chain. Gray ellipses represent physical sites. The complex fermion $c_{j}$ at site $j$ is decomposed into two Majorana fermions as $c_j=(\gamma_{2j-1}+i\gamma_{2j})/2$. Solid oriented line from $j$ to $k$ indicates that $\gamma_j$ is paired to $\gamma_k$, i.e., $-i\gamma_{j}\gamma_k=1$ when acting on this state. (a) Trivial state $\otimes_j|n_{c_j}=0\rangle$. In terms of Majorana fermions, $\gamma_{2j-1}$ is paired to $\gamma_{2j}$ ($j=1,2,\cdots,N$), i.e., $-i\gamma_{2j-1}\gamma_{2j}=1$ when acting on this state. (b) Ground state of the Kitaev's Majorana chain \eq{eq:H_K} in open boundary condition. The first and the last Majorana fermions can be paired either as (c) or as (d), resulting in two fold degeneracy of the ground states. (c) Ground state of the Kitaev's Majorana chain \eq{eq:H_K} in periodic boundary condition. The graph is not Kasteleyn oriented, and the state has odd fermion parity. (d) Ground state of the Kitaev's Majorana chain \eq{eq:H_K} in anti-periodic boundary condition. The graph has Kasteleyn oriented, and the state has even fermion parity.}
\label{fig:K}
\end{figure*}

The ground state $|GS\rangle$ of the Kitaev chain is defined such that
\begin{align}\label{eq:MGS}
-i\gamma_{2j}\gamma_{2j+1} |GS\rangle = |GS\rangle, \quad \forall j.
\end{align}
One can imagine the two Majorana fermion $\gamma_{2j}$ and $\gamma_{2j+1}$ are paired up from the former to the later. Graphically, we can use an oriented link from $\gamma_{2j}$ to $\gamma_{2j+1}$ to illustrate the pairing. If we combine these two Majorana fermions to a complex fermion
\begin{align}
c_{2j,2j+1}=\frac{1}{2}(\gamma_{2j}+i\gamma_{2j+1}),
\end{align}
then the term in the Hamiltonian is exactly the fermion parity operator of the new complex fermion
\begin{align}
P_f^{2j,2j+1} = -i\gamma_{2j}\gamma_{2j+1} = 1-2c_{2j,2j+1}^\dagger c_{2j,2j+1}.
\end{align}
In this language, the $\gamma_{2j}$ and $\gamma_{2j+1}$ Majorana fermion paired dimer state is the unoccupied vacuum state $|n_{c_{2j,2j+1}}=0\rangle$ ($n_{c_{2j,2j+1}}=c_{2j,2j+1}^\dagger c_{2j,2j+1}$) for the new complex fermion $c_{2j,2j+1}$. The ground state $|GS\rangle=\otimes_j |n_{c_{2j,2j+1}}=0\rangle$ is a Majorana fermion dimer state with $\gamma_{2j}$ paired to $\gamma_{2j+1}$ for all $j$ (see \fig{fig:K}).

In periodic boundary condition, all Majorana fermions are paired up from $\gamma_{2j}$ to $\gamma_{2j+1}$ (in particular, $\gamma_{2N}$ is paired to $\gamma_1$). So the ground state is unique (see \fig{fig:K}b).

In open boundary condition, however, the first and the last Majorana fermions $\gamma_1$ and $\gamma_{2N}$ are dangling. We can pair them either from $\gamma_1$ to $\gamma_{2N}$, or from $\gamma_{2N}$ to $\gamma_{1}$. This gives rise to the two-fold-degenerate ground states (see \fig{fig:K}c and \fig{fig:K}d).

\subsection{Excited states and the appearance of phase factor $\pm i$}

Since the Hamiltonian \eq{eq:H_K} is a summation of commuting projectors, we can easily construct all excited states of the model. In the Fock space of the new complex fermions $c_{2j,2j+1}$, ground state is merely $\otimes_j |n_{c_{2j,2j+1}}=0\rangle$, and the excited states are states that some of these fermion modes are occupied.

In terms of Majorana fermion operators, we can construct excited states as
\begin{align}
\label{eq:ex}
|k\rangle = \gamma_k |GS\rangle,
\end{align}
where $k=1,2,\cdots,2N$ is some site of the Majorana fermions.
Using the commutation relations of Majorana fermions, one can easily show
\begin{align}
(-i\gamma_{2j}\gamma_{2j+1})|k\rangle=
\begin{cases}
-|k\rangle,\quad k= 2j,2j+1\\
|k\rangle,\quad k\ne 2j,2j+1
\end{cases}
\end{align}
Therefore, $|k\rangle$ is indeed an excited state of the Kitaev chain with energy $\Delta E = 2$ above the ground state energy. All other excited states can be constructed similarly.

Since the Majorana fermion $\gamma_k$ is a superposition of creation and annihilation operators of the original $c_i$ fermions, the fermion parity of $|k\rangle$ is always different from the ground state $|GS\rangle$. 
Moreover, the excited states $|k\rangle$ constructed above are not linearly independent. In fact, using the definition of the ground state \eq{eq:MGS} and the algebraic relations of Majorana fermions, we have
\begin{align}
\gamma_{2j+1}|GS\rangle=i\gamma_{2j}|GS\rangle.
\end{align}
This equation shows that the Majorana fermion will pick up a phase factor $\pm i$ when hopping one lattice site along the chain. 

Similar to \eq{eq:ex} to change the fermion parity, we introduced a Majorana fermion operator $\gamma_{012A}$ (or $\gamma_{234B}$) in the 3D $F$ move when the fermion parity of the Kitaev chain in this $F$ move is changed. It indicates that the state of the final Majorana chain in the $F$ move is the excited state of the original one with one Majorana fermion $\gamma_{012A}$ (or $\gamma_{234B}$) excited. In the fermionic hexagon equation, if some of the six $F$ moves changes Majorana fermion parity, then the excited Majorana fermion will hope along the Majorana chain. This is the reason why the obstruction $\mathcal O_\gamma$ may have phase factor $\pm i$.

\subsection{Fermion parity and Kasteleyn orientation}

The two ground states of Kitaev chain in open boundary condition have different fermion parity (see \fig{fig:K}c and \fig{fig:K}d). One can check that the fermion parity of the ground state in periodic boundary condition (see \fig{fig:K}c) is odd:
\begin{align}
P_f |GS_{PBC}\rangle &= (-i\gamma_1\gamma_2)(-i\gamma_3\gamma_4)\cdots(-i\gamma_{2N-1}\gamma_{2N}) |GS_{PBC}\rangle \\\nonumber
&= (-i\gamma_2\gamma_3)(-i\gamma_4\gamma_5)\cdots(-i\gamma_{2N-2}\gamma_{2N-1})(i\gamma_{2N}\gamma_{1}) |GS_{PBC}\rangle \\\nonumber
&= -|GS_{PBC}\rangle.
\end{align}
where we have moved $\gamma_1$ from the leftmost to the rightmost position, and used the definition of ground state \eq{eq:MGS}.

Since the other ground state in \fig{fig:K}d reverse the link direction connecting the end points of the state in \fig{fig:K}c, their fermion parities are different. So the state in \fig{fig:K}d has even fermion parity. When looking at the link directions of the pictures, one can easily distinguish the difference of the two states. \fig{fig:K}d is Kasteleyn oriented in the sense that the number of arrows pointing backwards along the chain loop is odd (since the number of links is even, the walking direction is not important). Then the fermion parities for the state in \fig{fig:K}a and \fig{fig:K}d are the same. On the other hand, \fig{fig:K}c is not Kasteleyn oriented. And this state has different fermion parity. For states with other choices of link orientation, one can flip the directions of several links. If we flip odd (even) number of link directions, the fermion parity is (not) changed. Therefore the Kasteleyn oriented is a necessary and sufficient condition for the fermion parity changes.

The same is true for Majorana dimer states in 2D and 3D. One can put two Majorana dimer states together. For the transition graph of the two (oriented) dimer states, if a loop is Kasteleyn oriented, then the two Majorana dimer states in this patch have the same fermion parity. If a loop is not Kasteleyn oriented, then the two states have different fermion parity in this loop.

We use this criteria for 3D lattice, and find that the Kasteleyn orientation property may change under the $F$ move. This is why we add $\gamma_{012A}$ (or $\gamma_{234B}$) to make the $F$ operator really transform one state to another with different Majorana fermion pairty.

\section{2D Pachner moves}
\label{Appen:2D}

\subsection{(2-2) moves}
\label{Appen:2-2}

With a branching structure, there are three different types of (2-2) Pachner moves. Only two of them can be induced by a global ordering [see Eqs.~(\ref{eq:22_1}) and (\ref{eq:22_3})].

\begin{align}
\label{eq:22_1}
\vcenter{\hbox{\includegraphics[]{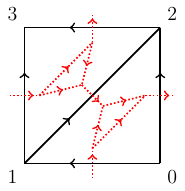}}}
\quad \longrightarrow \quad
\vcenter{\hbox{\includegraphics[]{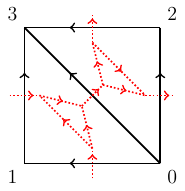}}}
\end{align}



\begin{align}
\label{eq:22_3}
\vcenter{\hbox{\includegraphics[scale=.92]{2D_fig11.pdf}}}
\quad \longrightarrow \quad
\vcenter{\hbox{\includegraphics[scale=.92]{2D_fig12.pdf}}}
\end{align}

Note that the (2-2) move in \eq{eq:22_3} is the standard move. The other (2-2) move in \eq{eq:22_1} can be derive from a sequence of the standard move and the (2-0) moves, as shown below:
\begin{align}\nonumber
\vcenter{\hbox{\includegraphics[scale=.7]{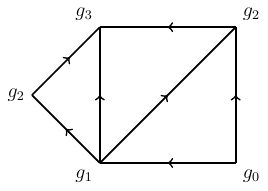}}}
\xrightarrow{\text{(2-0)}}
\vcenter{\hbox{\includegraphics[scale=.7]{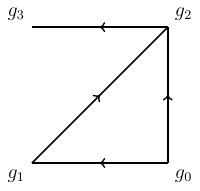}}}
=
\vcenter{\hbox{\includegraphics[scale=.7]{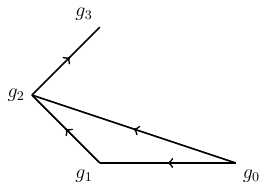}}}
\xrightarrow{\text{(0-2)}}
\vcenter{\hbox{\includegraphics[scale=.7]{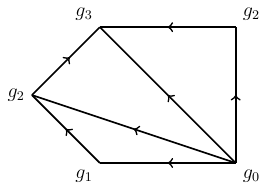}}}
\xrightarrow{\text{ $F$ }}
\vcenter{\hbox{\includegraphics[scale=.7]{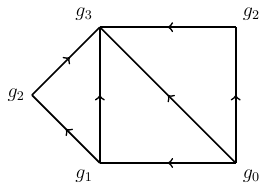}}}
\end{align}
To derive the (2-2) move in \eq{eq:22_1}, we first add a vertex $g_2$ and a triangle labelled by $g_1,g_2,g_3$ on the left of the figure. After a (2-0) move, a (0-2) move and a standard (2-2) $F$ move, we finally obtain the desired triangulation.
The whole process gives the expression for the $F$ move in \eq{eq:22_1} as:
\begin{align}\nonumber
F(0123)|_{\mathrm{Eq.~}(\ref{eq:22_1})} &= \frac{1}{|G_b|^{1/2}}  c_{(123)}^{\dagger n_2(123)} \bar c_{(123)}^{\dagger n_2(123)}
\cdot |G_b|^{1/2} \bar c_{(023)}^{ n_2(023)}  c_{(023)}^{ n_2(023)}
\cdot \nu_3(0123) c_{(012)}^{\dagger n_2(012)} c_{(023)}^{\dagger n_2(023)} c_{(013)}^{n_2(013)} c_{(123)}^{n_2(123)} X[\tilde n_1],\\\nonumber
&= \nu_3(0123) (-1)^{n_2(123)} c_{(012)}^{\dagger n_2(012)} \bar c_{(023)}^{ n_2(023)} c_{(013)}^{n_2(013)} \bar c_{(123)}^{\dagger n_2(123)} X[\tilde n_1]\\
&= \nu_3(0123) \bar c_{(123)}^{\dagger n_2(123)} c_{(012)}^{\dagger n_2(012)} \bar c_{(023)}^{ n_2(023)} c_{(013)}^{n_2(013)} X[\tilde n_1],
\end{align}
where we have used $c_{(023)}^{ n_2(023)} c_{(023)}^{\dagger n_2(023)} = c_{(123)}^{\dagger n_2(123)} c_{(123)}^{n_2(123)}=1$ when acting on appropriate states, and $\dd n_2(0123)=0$ (mod 2). Recall that $c_{(123)}$ and $\bar c_{(123)}$ are annihilation operators of the $c$ fermions at the center of triangles $(123)$ with opposite orientations.

When decorating Kitaev's Majorana chains to the dual lattice, there are in total $2^3=8$ kinds of different decoration configurations (see \tab{tab:2-2}) for each (2-2) move. One can check that, for any decoration configuration, all the loops in the transition graph of the left and right Majorana dimer states have Kasteleyn orientation property. Therefore the Majorana fermion parity is conserved in all (2-2) moves.

\begin{table}[h!]
\centering
\begin{tabular}{ |c|c| } 
\hline
out-leg string $\#$ & configuration\\
\hline\hline
0 &
$\vcenter{\hbox{\includegraphics[]{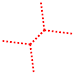}}}
\rightarrow
\vcenter{\hbox{\includegraphics[]{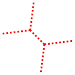}}}$
\\
\hline
&
$\vcenter{\hbox{\includegraphics[]{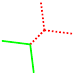}}}
\rightarrow
\vcenter{\hbox{\includegraphics[]{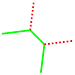}}}$
\\
&
$\vcenter{\hbox{\includegraphics[]{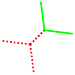}}}
\rightarrow
\vcenter{\hbox{\includegraphics[]{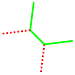}}}$
\\
\multirow{5}{*}{2}&
$\vcenter{\hbox{\includegraphics[]{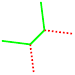}}}
\rightarrow
\vcenter{\hbox{\includegraphics[]{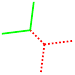}}}$
\\
&
$\vcenter{\hbox{\includegraphics[]{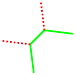}}}
\rightarrow
\vcenter{\hbox{\includegraphics[]{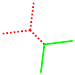}}}$
\\
&
$\vcenter{\hbox{\includegraphics[]{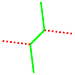}}}
\rightarrow
\vcenter{\hbox{\includegraphics[]{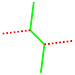}}}$
\\
&
$\vcenter{\hbox{\includegraphics[]{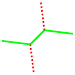}}}
\rightarrow
\vcenter{\hbox{\includegraphics[]{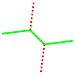}}}$
\\
\hline
4&
$\vcenter{\hbox{\includegraphics[]{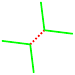}}}
\rightarrow
\vcenter{\hbox{\includegraphics[]{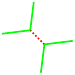}}}$
\\
\hline
\end{tabular}
\caption{$2^3=8$ kinds of configurations of Kitaev's Majorana chain decorations (green line) on the dual lattice $\mathcal P$ for each (2-2) move.}
\label{tab:2-2}
\end{table}

\subsection{(1-3) moves}
\label{Appen:1-3}

There are in total four different types of (1-3) Pachner moves for triangulation with branching structure. Two of them have branching structure that can be induced by a global ordering [see Eqs.~(\ref{fig:(1-3)c})-(\ref{fig:(1-3)b})]. These (1-3) moves can also be derived from a sequence of the standard (2-2) move and (2-0) moves.

\begin{align}
\label{fig:(1-3)c}
\vcenter{\hbox{\includegraphics[]{Appendix_2D_fig23.pdf}}}
\longrightarrow
\vcenter{\hbox{\includegraphics[]{Appendix_2D_fig24.pdf}}}
\end{align}

\begin{align}
\label{fig:(1-3)b}
\vcenter{\hbox{\includegraphics[]{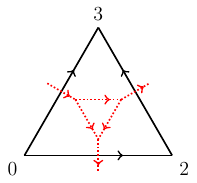}}}
\longrightarrow
\vcenter{\hbox{\includegraphics[]{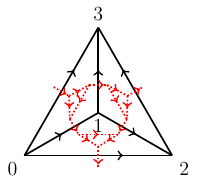}}}
\end{align}

For instance, using the standard $F$ move and (2-0) move, we can derive the (1-3) move \eq{fig:(1-3)b}:
\begin{align}
\label{eq:other13}
\vcenter{\hbox{\includegraphics[scale=.7]{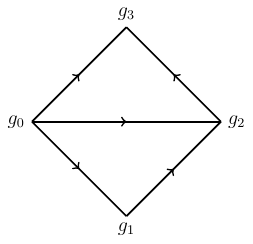}}}
\xrightarrow{\text{ $F$ }}
\vcenter{\hbox{\includegraphics[scale=.7]{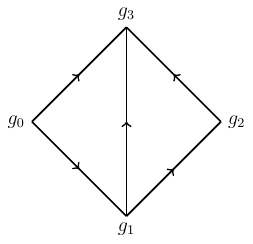}}}
=
\vcenter{\hbox{\includegraphics[scale=.7]{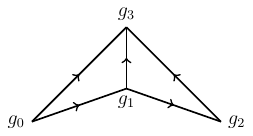}}}
\xrightarrow{\text{(0-2)}}
\vcenter{\hbox{\includegraphics[scale=.7]{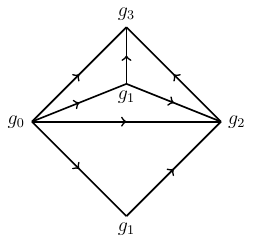}}}.
\end{align}
Note that we have added a triangle labelled by $g_0$, $g_1$ and $g_2$ below on both figure of \eq{fig:(1-3)b}. The above process can be summarized as a FSLU transformation:
\begin{align}
\label{eq:2D:1-3}\nonumber
F(0123)|_{\mathrm{Eq.~}(\ref{fig:(1-3)b})} &= \nu_3^{-1}(0123)
\bar c^{\dagger n_2(012)}_{(012)}
\bar c^{\dagger n_2(023)}_{(023)}
\bar c^{n_2(013)}_{(013)}
\bar c^{n_2(123)}_{(123)}
\cdot |G_b|^{1/2} \bar c_{(012)}^{n_2(012)}  c_{(012)}^{n_2(012)}
X[\tilde n_1]\\
&= \nu_3^{-1}(0123) |G_b|^{1/2} 
c^{ n_2(012)}_{(012)}
\bar c^{\dagger n_2(023)}_{(023)}
\bar c^{n_2(013)}_{(013)}
\bar c^{n_2(123)}_{(123)}
X[\tilde n_1].
\end{align}

Note that, the $F$ move used in the above sequence is slightly different from the standard one. The differences are merely the orientations of all four triangles. There are two consequences of the orientation changes. The first one is that $c$'s are changed to $\bar c$'s by definition. The second consequence is that the $U(1)$ phase factor $\nu_2$ is changed to $\nu_2^{-1}$. This can be obtained by using several steps of (2-0) moves. It is also consistent with the path integral formalism of bosonic SPT states, for $\nu_2^{s(ijkl)}(g_i,g_j,g_k,g_l)$ is assigned to the spacetime tetrahedron $\langle ijkl\rangle$ with orientation $s(ijkl)=\pm 1$.

In total $2^3=8$ kinds of Kitaev's Majorana chain decoration configurations are summarized in \tab{tab:1-3}. Similar to the (2-2) moves, one can check that none of the (1-3) moves change the Majorana fermion parity for arbitrary decoration configurations.

%
%

\begin{table}[h!]
\centering
\begin{tabular}{ |c|c|c| } 
\hline
out-leg string $\#$ & configuration-1 & configuration-2\\
\hline\hline
0 &
$\vcenter{\hbox{\includegraphics[]{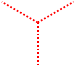}}}
\rightarrow
\vcenter{\hbox{\includegraphics[]{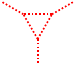}}}$
&
$\vcenter{\hbox{\includegraphics[]{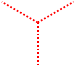}}}
\rightarrow
\vcenter{\hbox{\includegraphics[]{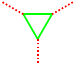}}}$
\\
\hline
&
$\vcenter{\hbox{\includegraphics[]{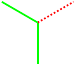}}}
\rightarrow
\vcenter{\hbox{\includegraphics[]{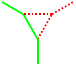}}}$
&
$\vcenter{\hbox{\includegraphics[]{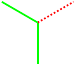}}}
\rightarrow
\vcenter{\hbox{\includegraphics[]{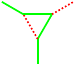}}}$
\\
2 &
$\vcenter{\hbox{\includegraphics[]{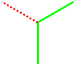}}}
\rightarrow
\vcenter{\hbox{\includegraphics[]{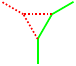}}}$
&
$\vcenter{\hbox{\includegraphics[]{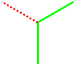}}}
\rightarrow
\vcenter{\hbox{\includegraphics[]{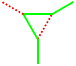}}}$
\\
&
$\vcenter{\hbox{\includegraphics[]{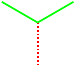}}}
\rightarrow
\vcenter{\hbox{\includegraphics[]{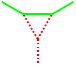}}}$
&
$\vcenter{\hbox{\includegraphics[]{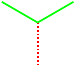}}}
\rightarrow
\vcenter{\hbox{\includegraphics[]{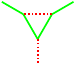}}}$
\\
\hline
\end{tabular}
\caption{$2^3=8$ kinds of configurations of Kitaev's Majorana chain decorations (green line) on the dual lattice $\mathcal P$ for each (1-3) move.}
\label{tab:1-3}
\end{table}

\section{3D Pachner moves and fermion parity changes}
\label{Appen:3D}

\subsection{(2-0) moves}
\label{Appen:3D:2-0}
Besides the (2-0) move shown in main text, there is another (2-0) move which has a branching structure that can be induced by global ordering. This (2-0) move corresponds to the one with vertex labelled by $g_2$ at the center of triangle $013$. The expression for this move is
\begin{align}
\label{fig:(2-0)3D_}
\Psi\left(
\vcenter{\hbox{\includegraphics[]{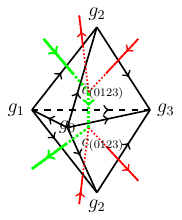}}}
\right)
= \frac{1}{|G_b|^{1/2}}  c_{(0123)}^{\dagger n_3(g_0,g_1,g_2,g_3)} \bar c_{(0123)}^{\dagger n_3(g_0,g_1,g_2,g_3)} X[\tilde n_2(g_i,g_j,g_k)]\ 
\Psi\left(
\vcenter{\hbox{\includegraphics[]{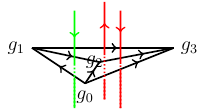}}}
\right),
\end{align}
\begin{align}
\label{fig:(2-0)3D-b_}
\Psi\left(
\vcenter{\hbox{\includegraphics[]{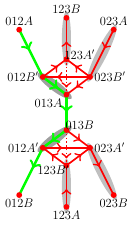}}}
\right)
= \frac{1}{|G_b|^{1/2}}  c_{(0123)}^{\dagger n_3(g_0,g_1,g_2,g_3)} \bar c_{(0123)}^{\dagger n_3(g_0,g_1,g_2,g_3)} X[\tilde n_2(g_i,g_j,g_k)]\ 
\Psi\left(
\vcenter{\hbox{\includegraphics[]{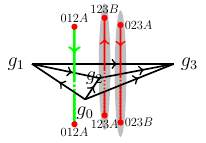}}}
\right).
\end{align}
Using these two (2-0) moves and the standard $F$ move, we can deduce all (1-4) and (2-3) moves with different branching structures.

\subsection{(1-4) moves}

\begin{figure*}[h!]
\centering
\begin{subfigure}[h!]{.45\textwidth}
\centering
$\vcenter{\hbox{\includegraphics[]{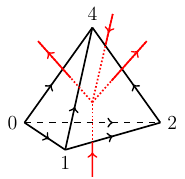}}}
\longrightarrow
\vcenter{\hbox{\includegraphics[]{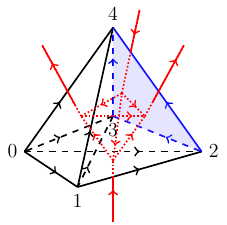}}}$
\caption{For triangulation $\mathcal T$ and dual lattice $\mathcal P$.
}
\end{subfigure}
~
\begin{subfigure}[h!]{.45\textwidth}
\centering
$\vcenter{\hbox{\includegraphics[]{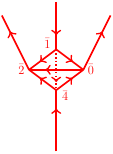}}}
\longrightarrow
\vcenter{\hbox{\includegraphics[]{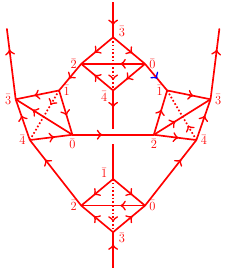}}}$
\caption{For resolved dual lattice $\tilde{\mathcal P}$.
}
\end{subfigure}
\caption{(1-4)-A move.}
\label{fig:(1-4)-A}
\end{figure*}

\begin{figure*}[h!]
\centering
\begin{subfigure}[h!]{.45\textwidth}
\centering
$\vcenter{\hbox{\includegraphics[]{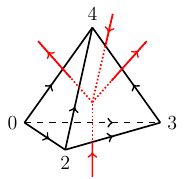}}}
\longrightarrow
\vcenter{\hbox{\includegraphics[]{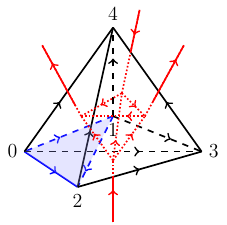}}}$
\caption{For triangulation $\mathcal T$ and dual lattice $\mathcal P$.
}
\end{subfigure}
~
\begin{subfigure}[h!]{.45\textwidth}
\centering
$\vcenter{\hbox{\includegraphics[]{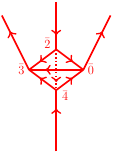}}}
\longrightarrow
\vcenter{\hbox{\includegraphics[]{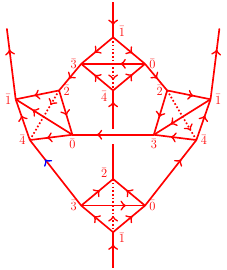}}}$
\caption{For resolved dual lattice $\tilde{\mathcal P}$.
}
\end{subfigure}
\caption{(1-4)-B move.}
\label{fig:14B}
\end{figure*}

\begin{figure*}[h!]
\centering
\begin{subfigure}[h!]{.45\textwidth}
\centering
$\vcenter{\hbox{\includegraphics[]{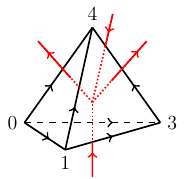}}}
\longrightarrow
\vcenter{\hbox{\includegraphics[]{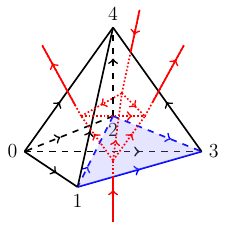}}}$
\caption{For triangulation $\mathcal T$ and dual lattice $\mathcal P$.
}
\end{subfigure}
~
\begin{subfigure}[h!]{.45\textwidth}
\centering
$\vcenter{\hbox{\includegraphics[]{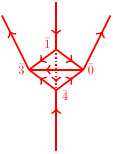}}}
\longrightarrow
\vcenter{\hbox{\includegraphics[]{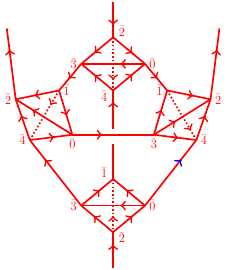}}}$
\caption{For resolved dual lattice $\tilde{\mathcal P}$.
}
\end{subfigure}
\caption{(1-4)-C move.}
\label{fig:1-4C}
\end{figure*}

There are five different kinds of (1-4) 3D Pachner move. Three have branching structures that can be induced by global ordering (see Figs.~\ref{fig:(1-4)-A}--\ref{fig:1-4C}). They can be obtained from a sequence of standard $F$ moves and (2-0) moves. For example, the move of adding vertex $1$ in \fig{fig:14B} can be obtained as follows:
\begin{align}\label{fig:other14}
\vcenter{\hbox{\includegraphics[scale=1]{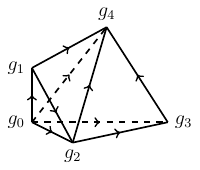}}}
\xrightarrow{\text{ $F$ }}
\vcenter{\hbox{\includegraphics[scale=1]{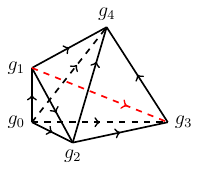}}}
=
\vcenter{\hbox{\includegraphics[scale=1]{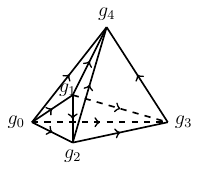}}}
\xrightarrow{\text{ (0-2) }}
\vcenter{\hbox{\includegraphics[scale=1]{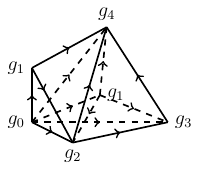}}}
\end{align}
We first add a tetrahedron (0124) to the left-hand-side state of \fig{fig:14B}. Then after a standard $F$ move and a (0-2) move, we obtain the desired (1-4) move in \fig{fig:14B}.
The expression for this sequence of FSLU transformation is
\begin{align}\nonumber\label{eq:other14}
F(0123)|_{\mathrm{Fig.~}(\ref{fig:14B})} &=
\nu_4^{-1}(01234)
\bar c^{\dagger n_3(0124)}_{(0124)} \bar c^{\dagger n_3(0234)}_{(0234)} \bar c^{n_3(0123)}_{(0123)} \bar c^{n_3(0134)}_{(0134)} \bar c^{n_3(1234)}_{(1234)}
\cdot
|G_b|^{1/2} \bar c_{(0124)}^{ n_3(0124)}  c_{(0124)}^{ n_3(0124)}
X[\tilde n_2]\\
&=\nu_4^{-1}(01234) |G_b|^{1/2} 
c^{ n_3(0124)}_{(0124)} \bar c^{\dagger n_3(0234)}_{(0234)} \bar c^{n_3(0123)}_{(0123)} \bar c^{n_3(0134)}_{(0134)} \bar c^{n_3(1234)}_{(1234)} X[\tilde n_2].
\end{align}
Note that we also changed $c$'s to $\bar c$'s and $\nu_3$ to $\nu_3^{-1}$ in the above $F$ move compared to the standard $F$ move, for the orientations of the five tetrahedra are changed. The reason is exactly the same as the 2D case discussed below \eq{eq:2D:1-3}.

For all five kinds of (1-4) moves, the move changes the Majorana fermion parity if and only if both link $012$ and link $234$ in lattice $\cP$ are decorated by Kitaev's Majorana chains. Therefore we also have $\dd n_3 = \tilde n_2\smile\tilde n_2$. $2^6=64$ kinds of Kitaev's Majorana chain decoration configurations are summarized in \tab{tab:1-4}.

\begin{table}[h!]
\centering
\begin{tabular}{ |c|c|c| } 
\hline
out-leg string $\#$ & configuration $\#$ & configuration\\
\hline\hline
\multirow{3}{*}{} & $1$ & 
$\vcenter{\hbox{\includegraphics[]{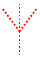}}}
\rightarrow
\vcenter{\hbox{\includegraphics[]{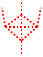}}}$
\\
0 & 4 & 
$\vcenter{\hbox{\includegraphics[]{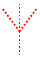}}}
\rightarrow
\vcenter{\hbox{\includegraphics[]{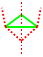}}}$
 \\
 & 3 &
$\vcenter{\hbox{\includegraphics[]{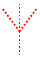}}}
\rightarrow
\vcenter{\hbox{\includegraphics[]{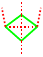}}}$
 \\
\hline
\multirow{4}{*}{} & $6\times1$ & 
$\vcenter{\hbox{\includegraphics[]{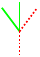}}}
\rightarrow
\vcenter{\hbox{\includegraphics[]{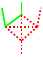}}}$
 \\ 
 & $6\times2$ &
$\vcenter{\hbox{\includegraphics[]{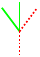}}}
\rightarrow
\vcenter{\hbox{\includegraphics[]{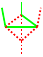}}}$
 \\
2 & $6\times2$ &
$\vcenter{\hbox{\includegraphics[]{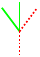}}}
\rightarrow
\vcenter{\hbox{\includegraphics[]{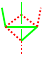}}}$
 \\
 & $6\times2$ &
$\vcenter{\hbox{\includegraphics[]{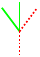}}}
\rightarrow
\vcenter{\hbox{\includegraphics[]{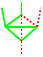}}}$
 \\
 & $6\times1$ &
$\vcenter{\hbox{\includegraphics[]{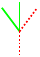}}}
\rightarrow
\vcenter{\hbox{\includegraphics[]{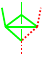}}}$
 \\
\hline
\multirow{3}{*}{} & 3 &
$\vcenter{\hbox{\includegraphics[]{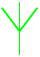}}}
\rightarrow
\vcenter{\hbox{\includegraphics[]{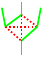}}}$
\\
4 & 4 &
$\vcenter{\hbox{\includegraphics[]{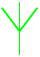}}}
\rightarrow
\vcenter{\hbox{\includegraphics[]{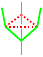}}}$
\\
 & 1 &
$\vcenter{\hbox{\includegraphics[]{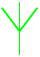}}}
\rightarrow
\vcenter{\hbox{\includegraphics[]{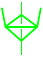}}}$
 \\
\hline
\end{tabular}
\caption{$2^6=64$ kinds of configurations of Kitaev's Majorana chain decorations (green line) on the dual lattice $\mathcal P$ for each (1-4) move. The second column counts the number of configurations of the same symmetry type. When four green lines meet at one point, we should resolve the decoration according to the conventions in main text.}
\label{tab:1-4}
\end{table}

\subsection{(2-3) moves}

\begin{figure*}[h!]
\centering
\begin{subfigure}[h!]{.45\textwidth}
\centering
$\vcenter{\hbox{\includegraphics[]{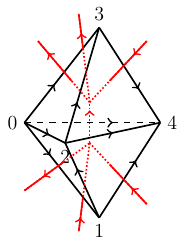}}}
\ \longrightarrow\ 
\vcenter{\hbox{\includegraphics[]{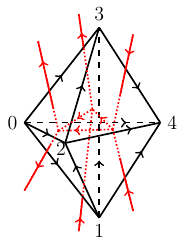}}}$
\caption{For triangulation $\mathcal T$ and dual lattice $\mathcal P$.}
\end{subfigure}
~
\begin{subfigure}[h!]{.45\textwidth}
\centering
$\vcenter{\hbox{\includegraphics[]{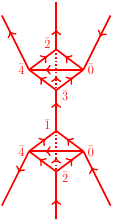}}}
\ \longrightarrow\ 
\vcenter{\hbox{\includegraphics[]{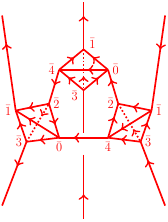}}}$
\caption{For resolved dual lattice $\tilde{\mathcal P}$.}
\end{subfigure}
\caption{(2-3)-A move.}
\label{fig:23A}
\end{figure*}

\begin{figure*}[h!]
\centering
\begin{subfigure}[h!]{.45\textwidth}
\centering
$\vcenter{\hbox{\includegraphics[]{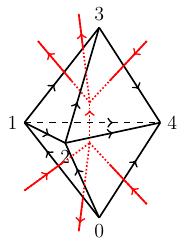}}}
\ \longrightarrow\ 
\vcenter{\hbox{\includegraphics[]{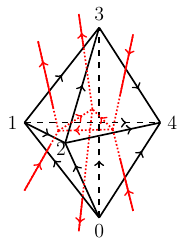}}}$
\caption{For triangulation $\mathcal T$ and dual lattice $\mathcal P$.}
\end{subfigure}
~
\begin{subfigure}[h!]{.45\textwidth}
\centering
$\vcenter{\hbox{\includegraphics[]{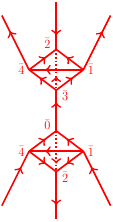}}}
\ \longrightarrow\ 
\vcenter{\hbox{\includegraphics[]{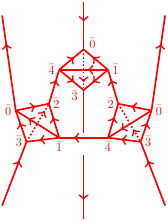}}}$
\caption{For resolved dual lattice $\tilde{\mathcal P}$.}
\end{subfigure}
\caption{(2-3)-B move.}
\label{fig:23B}
\end{figure*}

\begin{figure*}[h!]
\centering
\begin{subfigure}[h!]{.45\textwidth}
\centering
$\vcenter{\hbox{\includegraphics[]{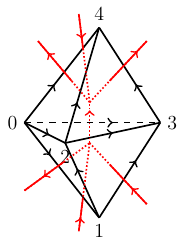}}}
\ \longrightarrow\ 
\vcenter{\hbox{\includegraphics[]{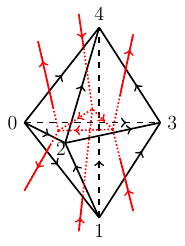}}}$
\caption{For triangulation $\mathcal T$ and dual lattice $\mathcal P$.}
\end{subfigure}
~
\begin{subfigure}[h!]{.45\textwidth}
\centering
$\vcenter{\hbox{\includegraphics[]{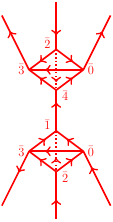}}}
\ \longrightarrow\ 
\vcenter{\hbox{\includegraphics[]{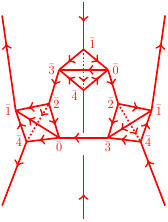}}}$
\caption{For resolved dual lattice $\tilde{\mathcal P}$.}
\end{subfigure}
\caption{(2-3)-C move.}
\label{fig:23C}
\end{figure*}

\begin{figure*}[h!]
\centering
\begin{subfigure}[h!]{.45\textwidth}
\centering
$\vcenter{\hbox{\includegraphics[]{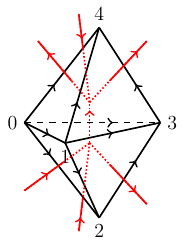}}}
\ \longrightarrow\ 
\vcenter{\hbox{\includegraphics[]{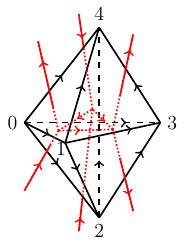}}}$
\caption{For triangulation $\mathcal T$ and dual lattice $\mathcal P$.}
\end{subfigure}
~
\begin{subfigure}[h!]{.45\textwidth}
\centering
$\vcenter{\hbox{\includegraphics[]{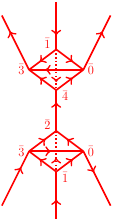}}}
\ \longrightarrow\ 
\vcenter{\hbox{\includegraphics[]{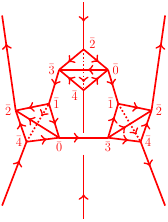}}}$
\caption{For resolved dual lattice $\tilde{\mathcal P}$.}
\end{subfigure}
\caption{(2-3)-E move.}
\label{fig:23I}
\end{figure*}

\begin{figure*}[h!]
\centering
\begin{subfigure}[h!]{.45\textwidth}
\centering
$\vcenter{\hbox{\includegraphics[]{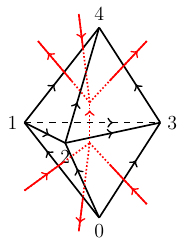}}}
\ \longrightarrow\ 
\vcenter{\hbox{\includegraphics[]{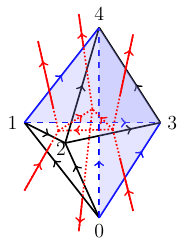}}}$
\caption{For triangulation $\mathcal T$ and dual lattice $\mathcal P$.}
\end{subfigure}
~
\begin{subfigure}[h!]{.45\textwidth}
\centering
$\vcenter{\hbox{\includegraphics[]{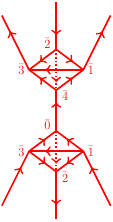}}}
\ \longrightarrow\ 
\vcenter{\hbox{\includegraphics[]{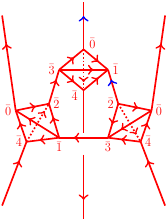}}}$
\caption{For resolved dual lattice $\tilde{\mathcal P}$.}
\end{subfigure}
\caption{(2-3)-H move.}
\label{fig:2-3A}
\end{figure*}

For 3D triangulation with a branching structure, there are 10 different kinds of (2-3) Pachner move. Eight have branching structures induced by global ordering (see Figs.~\ref{fig:23A}-\ref{fig:2-3A}).

The (2-3) move in \fig{fig:23A} is the standard move. Other (2-3) moves can be also derive from a sequence of the standard $F$ move and the (2-0) moves. In fact, we can also use (1-4) moves since they have already been obtained from standard $F$ moves and (2-0) moves in the last subsection. Here is an example of deriving the (2-3) move of adding link 14 in \fig{fig:23C}:
\begin{align}
\vcenter{\hbox{\includegraphics[scale=1]{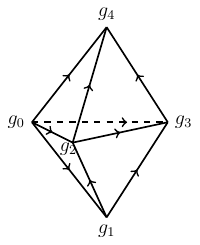}}}
\xrightarrow{\text{(1-4)}}
\vcenter{\hbox{\includegraphics[scale=1]{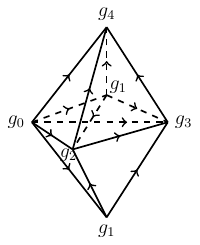}}}
\xrightarrow{\text{(2-0)}}
\vcenter{\hbox{\includegraphics[scale=1]{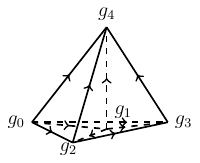}}}
=
\vcenter{\hbox{\includegraphics[scale=1]{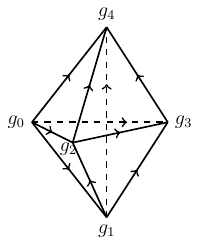}}}
\end{align}
The first step is adding a vertex labelled by $g_1$ at the center of the tetrahedron $0234$. This is the (1-4) move in \fig{fig:14B}, which is obtained from the standard $F$ move and (2-0) move in \fig{fig:other14} and has expression \eq{eq:other14}. The expression for the (2-3) FSLU transformation shown above is
\begin{align}\nonumber
F(0123)|_{\mathrm{Fig.~}(\ref{fig:23C})} &=
\nu_4^{-1}(01234) |G_b|^{1/2}
c^{ n_3(0124)}_{(0124)} \bar c^{\dagger n_3(0234)}_{(0234)} \bar c^{n_3(0123)}_{(0123)} \bar c^{n_3(0134)}_{(0134)} \bar c^{n_3(1234)}_{(1234)}
\cdot
\frac{1}{|G_b|^{1/2}}  c_{(0123)}^{\dagger n_3(0123)} \bar c_{(0123)}^{\dagger n_3(0123)} X[\tilde n_2]\\
&=
\nu_4^{-1}(01234) (-1)^{n_3(0123)}
c^{ n_3(0124)}_{(0124)} \bar c^{\dagger n_3(0234)}_{(0234)} c^{\dagger n_3(0123)}_{(0123)} \bar c^{n_3(0134)}_{(0134)} \bar c^{n_3(1234)}_{(1234)} X[\tilde n_2].
\end{align}

Different from the 2D case, when decorating Kitaev's Majorana chains to the dual lattice, the (2-3) moves may change the Majorana fermion parity. One can check the Kasteleyn orientation property for all loops in the transition graph of the left and right Majorana dimer states. A striking result is that, for all 10 kinds of (2-3) move, the move changes the Majorana fermion parity if and only if both link $012$ and link $234$ in lattice $\cP$ are decorated by Kitaev's Majorana chains. This leads to the total fermion parity conservation equation $\dd n_3 = \tilde n_2\smile\tilde n_2$. $2^6=64$ kinds of Kitaev's Majorana chain decoration configurations are summarized in \tab{tab:2-3}.

\begin{figure*}[h!]
\centering
\begin{subfigure}[h!]{.45\textwidth}
\centering
$\vcenter{\hbox{\includegraphics[]{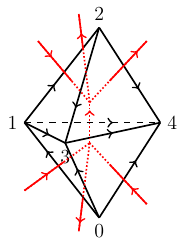}}}
\ \longrightarrow\ 
\vcenter{\hbox{\includegraphics[]{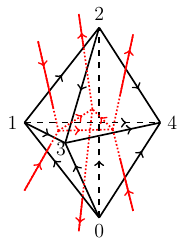}}}$
\caption{For triangulation $\mathcal T$ and dual lattice $\mathcal P$.}
\end{subfigure}
~
\begin{subfigure}[h!]{.45\textwidth}
\centering
$\vcenter{\hbox{\includegraphics[]{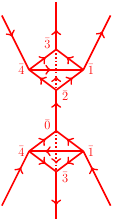}}}
\ \longrightarrow\ 
\vcenter{\hbox{\includegraphics[]{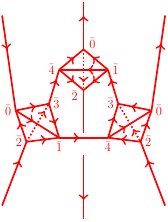}}}$
\caption{For resolved dual lattice $\tilde{\mathcal P}$.}
\end{subfigure}
\caption{(2-3)-D move.}
\end{figure*}

\begin{figure*}[h!]
\centering
\begin{subfigure}[h!]{.45\textwidth}
\centering
$\vcenter{\hbox{\includegraphics[]{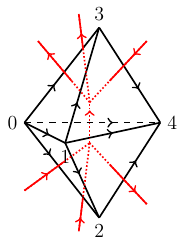}}}
\ \longrightarrow\ 
\vcenter{\hbox{\includegraphics[]{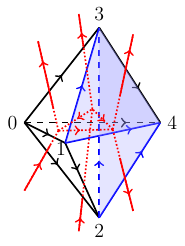}}}$
\caption{For triangulation $\mathcal T$ and dual lattice $\mathcal P$.}
\end{subfigure}
~
\begin{subfigure}[h!]{.45\textwidth}
\centering
$\vcenter{\hbox{\includegraphics[]{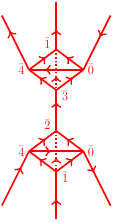}}}
\ \longrightarrow\ 
\vcenter{\hbox{\includegraphics[]{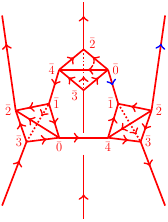}}}$
\caption{For resolved dual lattice $\tilde{\mathcal P}$.}
\end{subfigure}
\caption{(2-3)-F move.}
\end{figure*}

\begin{figure*}[h!]
\centering
\begin{subfigure}[h!]{.45\textwidth}
\centering
$\vcenter{\hbox{\includegraphics[]{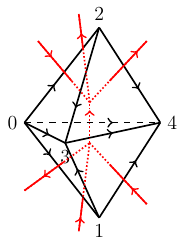}}}
\ \longrightarrow\ 
\vcenter{\hbox{\includegraphics[]{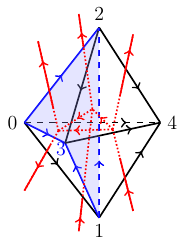}}}$
\caption{For triangulation $\mathcal T$ and dual lattice $\mathcal P$.}
\end{subfigure}
~
\begin{subfigure}[h!]{.45\textwidth}
\centering
$\vcenter{\hbox{\includegraphics[]{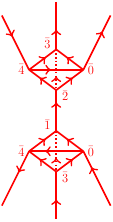}}}
\ \longrightarrow\ 
\vcenter{\hbox{\includegraphics[]{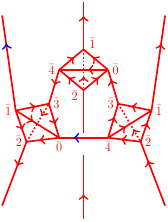}}}$
\caption{For resolved dual lattice $\tilde{\mathcal P}$.}
\end{subfigure}
\caption{(2-3)-G move.}
\end{figure*}

%
%

\begin{table}[h!]
\centering
\begin{tabular}{ |c|c|c|c| } 
\hline
out-leg string $\#$ & configuration $\#$ & configuration-1 & configuration-2 \\
\hline\hline
0 & $1\times2$ & 
$\vcenter{\hbox{\includegraphics[]{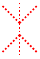}}}
\rightarrow
\vcenter{\hbox{\includegraphics[]{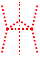}}}$
 & 
$\vcenter{\hbox{\includegraphics[]{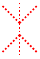}}}
\rightarrow
\vcenter{\hbox{\includegraphics[]{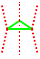}}}$
\\
\hline
 & $3\times2$ & 
$\vcenter{\hbox{\includegraphics[]{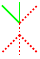}}}
\rightarrow
\vcenter{\hbox{\includegraphics[]{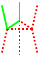}}}$
 & 
$\vcenter{\hbox{\includegraphics[]{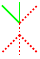}}}
\rightarrow
\vcenter{\hbox{\includegraphics[]{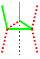}}}$
 \\ 
\multirow{4}{*}{2} & $3\times2$ & 
$\vcenter{\hbox{\includegraphics[]{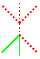}}}
\rightarrow
\vcenter{\hbox{\includegraphics[]{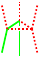}}}$
 & 
$\vcenter{\hbox{\includegraphics[]{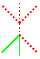}}}
\rightarrow
\vcenter{\hbox{\includegraphics[]{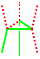}}}$
\\
& $3\times2$ & 
$\vcenter{\hbox{\includegraphics[]{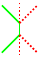}}}
\rightarrow
\vcenter{\hbox{\includegraphics[]{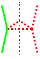}}}$
 & 
$\vcenter{\hbox{\includegraphics[]{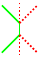}}}
\rightarrow
\vcenter{\hbox{\includegraphics[]{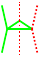}}}$
\\
& $6\times2$ & 
$\vcenter{\hbox{\includegraphics[]{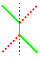}}}
\rightarrow
\vcenter{\hbox{\includegraphics[]{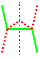}}}$
 & 
$\vcenter{\hbox{\includegraphics[]{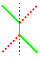}}}
\rightarrow
\vcenter{\hbox{\includegraphics[]{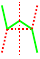}}}$
\\
\hline
 & $3\times2$ &
$\vcenter{\hbox{\includegraphics[]{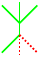}}}
\rightarrow
\vcenter{\hbox{\includegraphics[]{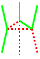}}}$
 & 
$\vcenter{\hbox{\includegraphics[]{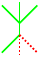}}}
\rightarrow
\vcenter{\hbox{\includegraphics[]{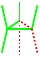}}}$
\\
\multirow{4}{*}{4} & $3\times2$ & 
$\vcenter{\hbox{\includegraphics[]{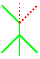}}}
\rightarrow
\vcenter{\hbox{\includegraphics[]{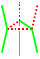}}}$
 & 
$\vcenter{\hbox{\includegraphics[]{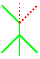}}}
\rightarrow
\vcenter{\hbox{\includegraphics[]{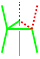}}}$
\\
& $3\times2$ & 
$\vcenter{\hbox{\includegraphics[]{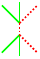}}}
\rightarrow
\vcenter{\hbox{\includegraphics[]{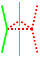}}}$
 & 
$\vcenter{\hbox{\includegraphics[]{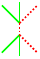}}}
\rightarrow
\vcenter{\hbox{\includegraphics[]{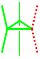}}}$
 \\
& $6\times2$ & 
$\vcenter{\hbox{\includegraphics[]{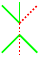}}}
\rightarrow
\vcenter{\hbox{\includegraphics[]{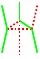}}}$
 & 
$\vcenter{\hbox{\includegraphics[]{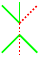}}}
\rightarrow
\vcenter{\hbox{\includegraphics[]{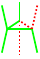}}}$
 \\
\hline
6 & $1\times2$ & 
$\vcenter{\hbox{\includegraphics[]{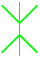}}}
\rightarrow
\vcenter{\hbox{\includegraphics[]{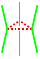}}}$
 & 
$\vcenter{\hbox{\includegraphics[]{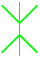}}}
\rightarrow
\vcenter{\hbox{\includegraphics[]{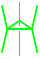}}}$
\\
\hline
\end{tabular}
\caption{$2^6=64$ kinds of configurations of Kitaev's Majorana chain decorations (green line) on the dual lattice $\mathcal P$ for each (2-3) move. The second column counts the number of configurations of the same symmetry type. When four green lines meet at one point, we should resolve the decoration according to the conventions in main text.}
\label{tab:2-3}
\end{table}

\section{Abelian group structure of the general group super-cohomology solutions}
\label{Appen:group}

Assume $(\nu_4,n_3,\tilde n_2)$ and $(\nu_4',n_3',\tilde n_2')$ are two solutions of general group super-cohomology theory, i.e., they satisfy the equations
\begin{align}\label{eq:n4}
\dd \nu_4 &= \mathcal O[n_3,\tilde n_2],\\\label{eq:n3}
\dd n_3 &= \tilde n_2^2,\\\label{eq:n2}
\dd \tilde n_2 &= 0,
\end{align}
and
\begin{align}\label{eq:n4p}
\dd \nu_4' &= \mathcal O[n_3',\tilde n_2'],\\\label{eq:n3p}
\dd n_3' &= \tilde n_2'^2,\\\label{eq:n2p}
\dd \tilde n_2' &= 0.
\end{align}
We want to construct another solution triple $(\nu_4^\mathrm{tot},n_3^\mathrm{tot},\tilde n_2^\mathrm{tot})$ from ``adding'' $(\nu_4,n_3,\tilde n_2)$ and $(\nu_4',n_3',\tilde n_2')$. The new solution satisfy
\begin{align}\label{eq:n4tot}
\dd \nu_4^\mathrm{tot} &= \mathcal O[n_3^\mathrm{tot},\tilde n_2^\mathrm{tot}],\\\label{eq:n3tot}
\dd n_3^\mathrm{tot} &= (\tilde n_2^\mathrm{tot})^2,\\\label{eq:n2tot}
\dd \tilde n_2^\mathrm{tot} &= 0.
\end{align}
If there exists such construction, then the solutions of the general group super-cohomology theory is closed under this ``adding" and form an Abelian group (if the ``adding'' is commutative). The ``adding'' operation corresponds to stacking of two FSPT states physically.

The expressions of $n_3^\mathrm{tot}$ and $\tilde n_2^\mathrm{tot}$ are very simple. One can show that if we choose
\begin{align}\label{eq:add_n3}
n_3^\mathrm{tot} &= n_3 + n_3' + \tilde n_2 \smile_1 \tilde n_2',\\\label{eq:add_n2}
\tilde n_2^\mathrm{tot} &= \tilde n_2 + \tilde n_2',
\end{align}
then Eqs.~(\ref{eq:n3tot}) and (\ref{eq:n2tot}) can be derived from Eqs.~(\ref{eq:n3}), (\ref{eq:n2}), (\ref{eq:n3p}) and (\ref{eq:n2p}).

Assume that $\nu_4^\mathrm{tot}$ has the expression 
\begin{align}
\nu_4^\mathrm{tot} = \nu_4\cdot \nu_4' \cdot e^{i\pi \theta[n_3,n_3',\tilde n_2,\tilde n_2']},
\end{align}
where $\theta[n_3,n_3',\tilde n_2,\tilde n_2']$ is some unknown functional of $n_3,n_3',\tilde n_2,\tilde n_2'$. As $n_3,n_3',\tilde n_2$, and $\tilde n_2'$ are cochains that are invariant under $G_b$ action, $\theta[n_3,n_3',\tilde n_2,\tilde n_2']$, and $\nu_4^\mathrm{tot}$ are also invariant.

The obstruction in \eq{eq:n4} is denoted by
\begin{align}
\mathcal O[n_3,\tilde n_2] = (-1)^{n_3 \smile_1 n_3+n_3 \smile_2 \dd n_3+f[\tilde n_2]},
\end{align}
where $\mathcal O_\gamma=(-1)^{f[\tilde n_2]}$ is the obstruction coming from the Majorana fermions.
From Eqs.~(\ref{eq:n4}), (\ref{eq:n4p}) and (\ref{eq:n4tot}), the equation for $\theta$ is
\begin{align}\nonumber
(-1)^{\dd \theta[n_3,n_3',\tilde n_2,\tilde n_2']} &= \frac{\dd \nu_4^\mathrm{tot}}{\dd \nu_4\cdot \dd \nu_4'}\\
&= (-1)^{(n_3^\mathrm{tot} \smile_1 n_3^\mathrm{tot}+n_3^\mathrm{tot} \smile_2 \dd n_3^\mathrm{tot}) - (n_3 \smile_1 n_3+n_3 \smile_2 \dd n_3) - (n_3' \smile_1 n_3'+n_3' \smile_2 \dd n_3')+f[\tilde n_2^\mathrm{tot}]-f[\tilde n_2]-f[\tilde n_2']}.
\end{align}
After substituting Eqs.~(\ref{eq:add_n3}) and (\ref{eq:add_n2}) to the right hand side of the equation above, we obtain the (mod 2) linear equation
\begin{align}\label{eq:dx}\nonumber
\dd y[\tilde n_2,\tilde n_2'] &\equiv
(\tilde n_2 \smile_1 \tilde n_2')\smile_1 (\tilde n_2 \smile_1 \tilde n_2')
+ (\tilde n_2 \smile_1 \tilde n_2') \smile_2 (\tilde n_2\smile \tilde n_2'+\tilde n_2'\smile \tilde n_2)\\
&\quad +(\tilde n_2^2+\tilde n_2'^2)\smile_3 (\tilde n_2\smile \tilde n_2'+\tilde n_2'\smile \tilde n_2)
+\tilde n_2^2\smile_3 \tilde n_2'^2
+f[\tilde n_2+\tilde n_2']-f[\tilde n_2]-f[\tilde n_2'],\quad \text{(mod 2)}
\end{align}
where $y[\tilde n_2,\tilde n_2']$ is is related to $\theta[n_3,n_3',\tilde n_2,\tilde n_2']$ by
\begin{align}\label{eq:x_theta}
\theta[n_3,n_3',\tilde n_2,\tilde n_2'] &= y[\tilde n_2,\tilde n_2'] + n_3\smile_2n_3' + (\tilde n_2\smile_1 \tilde n_2')\smile_2(n_3+n_3') + (n_3+n_3')\smile_3 (\tilde n_2\smile\tilde n_2'+\tilde n_2'\smile\tilde n_2) + \tilde n_2^2\smile_3 n_3'.
\end{align}

In summary, if $y[\tilde n_2,\tilde n_2']$ is a solution of \eq{eq:dx}, then the Abelian group structure of the group supercohomology solutions is
\begin{align}\label{eq:group}
(\nu_4,n_3,\tilde n_2) + (\nu_4',n_3',\tilde n_2') = (\nu_4 \nu_4' e^{i\pi \theta[n_3,n_3',\tilde n_2,\tilde n_2']}, n_3 + n_3' + \tilde n_2 \smile_1 \tilde n_2', \tilde n_2 + \tilde n_2'),
\end{align}
where $\theta[n_3,n_3',\tilde n_2,\tilde n_2']$ is defined by \eq{eq:x_theta}.

\subsection{Explicit construction of $\theta[n_3,n_3',\tilde n_2,\tilde n_2']$}
To construct $\theta[n_3,n_3',\tilde n_2,\tilde n_2']$ in the \eq{eq:group}, we need only to find a solution $y[\tilde n_2,\tilde n_2']$ of \eq{eq:dx}. Note that \eq{eq:dx} is a mod 2 equation when evaluating for arbitrary $(g_0,g_1,g_2,g_3,g_4,g_5)\in G_b^6$. $y[\tilde n_2,\tilde n_2'](g_0,g_1,g_2,g_3,g_4)$ is automatically symmetric under $G_b$ because it depends on $g_i$ through $\tilde n_2$ and $\tilde n_2'$.

As defined previously, $y[\tilde n_2,\tilde n_2']$ is a functional of $\tilde n_2$ and $\tilde n_2'$. When evaluating at $(g_0,g_1,g_2,g_3,g_4)$, it is a function $y[\{\tilde n_2(g_i,g_j,g_k)\},\{\tilde n_2'(g_{i},g_{j},g_{k})\}]$ ($0\le i<j<k \le 4$). It seems that $y$ has $2\times C_5^3=20$ arguments. However, as $\tilde n_2$ is a cocycle with equation $\dd \tilde n_2=0$ (mod 2), $\{\tilde n_2(g_i,g_j,g_k)\}$ ($0\le i<j<k \le 4$) are not fully independent. After solving the equations $\dd n_2 (g_i,g_j,g_k,g_l)=0$ ($0\le i<j<k<l\le 4$) explicitly, we find that the number of independent cocycles $\tilde n_2(g_i,g_j,g_k)$ ($0\le i<j<k \le 4$) is six. In fact, $2^6$ is exactly the number of Kitaev chain decoration configurations of a 4-simplex or a (2-3) Pachner move (see \tab{tab:2-3}). We can choose the six independent cocycles to be $ \tilde n_2(012), \tilde n_2(013), \tilde n_2(014), \tilde n_2(023), \tilde n_2(024), and \tilde n_2(034)$ [again, $(012)$ means $(g_0,g_1,g_2)$, etc.], respectively. The other four cocycles, including $\{ \tilde n_2(123), \tilde n_2(124), \tilde n_2(134), and \tilde n_2(234) \}$, respectively, can be derived from the previous six. Therefore, $y[\{\tilde n_2(g_i,g_j,g_k)\},\{\tilde n_2'(g_{i},g_{j},g_{k})\}]$ ($0\le i<j<k \le 4$) has only 12 independent arguments and can thus be written as
\begin{align}
y[\tilde n_2(012), \tilde n_2(013), \tilde n_2(014), \tilde n_2(023), \tilde n_2(024), \tilde n_2(034),\tilde n_2'(012), \tilde n_2'(013), \tilde n_2'(014), \tilde n_2'(023), \tilde n_2'(024), \tilde n_2'(034)].
\end{align}
Each $\tilde n_2$ or $\tilde n_2'$ can take two values, and so the number of unknown variables of $y$ is $q=2^{12}=4096$.

The number of equations in Eq. (\ref{eq:dx}) is the number of all possible cocycles $\tilde n_2$ and $\tilde n_2'$, i.e., the number of different cochain patterns $\tilde n_2$ and $\tilde n_2'$ satisfying $\dd \tilde n_2(g_i,g_j,g_k,g_l)=\dd\tilde n_2'(g_i,g_j,g_k,g_l)=0$ ($0\le i<j<k<l\le 5$). One can solve the cocyle equation and find that the number of different $\tilde n_2(g_i,g_j,g_k)$ ($0\le i<j<k \le 5$) patterns is $2^{10}$. This is the number of Kitaev chain decoration configurations of a 5-simplex or the hexagon equation. Thus, the number of equations is $p=2^{20}=1048576$.

The right-hand side of \eq{eq:dx} involves $f(\tilde n_2)$. As $\mathcal O_\gamma=(-1)^{f[\tilde n_2]}$ can be $\pm 1$ or $\pm i$, $f(\tilde n_2)$ takes value in $\{0, \frac{1}{2}, 1, \frac{3}{2}\}$. To make the right-hand side of \eq{eq:dx} integers, we can multiply \eq{eq:dx} by 2, and scale $y$ by
\begin{align}
y=\frac{x}{2}.
\end{align}
The final equation for $x$ in the matrix form is
\begin{align}\label{eq:Ax_b}
Ax\equiv b, \quad \text{mod 4}.
\end{align}
This is an overdetermined system of linear equations with $p=2^{20}$ equations and $q=2^{12}$ unknowns.

Using the procedure list in the next subsection, we find that the above linear equation indeed has solutions. We can use the solution $x$ to construct $\theta$. And the group structure of the general super-cohomology solutions is given by \eq{eq:group}.

\begin{figure*}[th!]
\centering
\includegraphics[width=.4\textwidth]{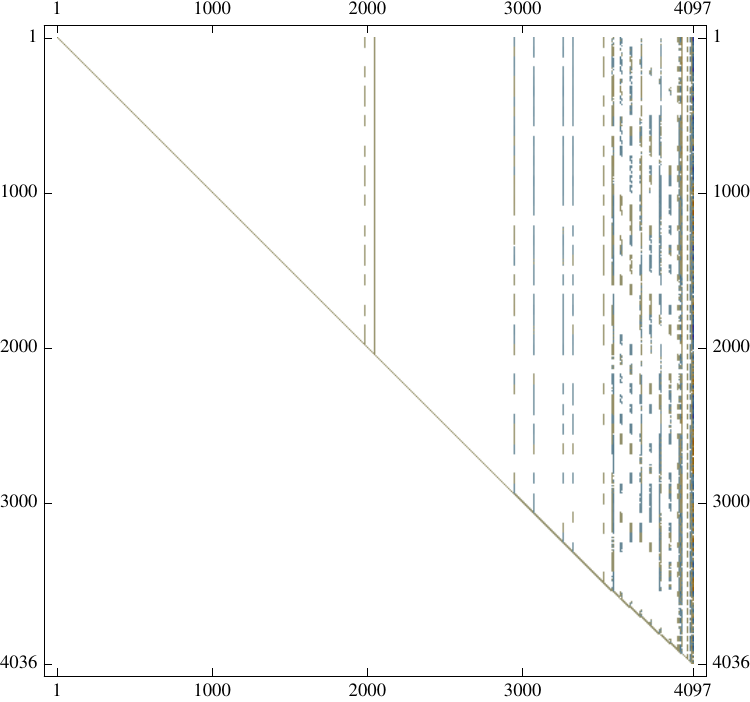}
\caption{Graphic display of the unique Hermite normal form of the augmented matrix $(A,b)$. We omit all the zero rows below the nonzero ones (the number of zero rows is $p-r-1=1044540$). White represents zero matrix elements. The last element at $(r+1,q+1)= (4036, 4097)$ is 4, indicating the existence of solutions for \eq{eq:Ax_b}. The first nonzero element in each row of the Hermite normal form of $A$ is either 1 or 2. Thus, the Abelian group structure of the general group super-cohomology solutions takes value in $\mathbb Z_8$ rather than in $\mathbb Z_4$.}
\label{fig:Hermite}
\end{figure*}

\subsection{Solving system of linear equations in $\mathbb R/4\mathbb Z$}

In this subsection, we discuss how to solve equations of the form
\begin{align}\label{eq:Axb}
Ax\equiv b, \quad \text{mod 4}.
\end{align}
Here, $A\in M_{p,q}(\mathbb Z)$ is a $p\times q$ ($p=2^{20}, q=2^{12}$) matrix with elements in $\mathbb Z$ (the matrix form of coboundary operator ``$\dd$" has elements in $\mathbb Z$). On the right hand side, $b\in \mathbb Z^p$ is a $\mathbb Z$-valued vector with length $p$. However, $x \in \mathbb R^q$ is a $\mathbb R$-valued vector with length $q$. All the equations hold modulo 4. Note that this equation is not a system of linear equations in ring $\mathbb Z/4\mathbb Z$ because elements of $x$ can take value in $\mathbb R/4\mathbb Z$, not merely in $\mathbb Z/4\mathbb Z$. In fact, we find that the above equation in our case has no solution of $x$ in $\mathbb Z/4\mathbb Z$, but it does have a solution in $\mathbb R/4\mathbb Z$.

To solve the mod 4 equation, we introduce an integer variable $n$ such that \eq{eq:Axb} becomes an equation in $\mathbb R$ without modulo 4.
\begin{align}\label{eq:Axb_Z}
Ax=b+4n.
\end{align}
Now the equation holds in $\mathbb R$ and the unknowns are $x\in \mathbb R^q$ and $n\in \mathbb Z^p$.

The basic idea to solving \eq{eq:Axb_Z} is Gaussian elimination. As $x$ takes value in $\mathbb R$, we can use the usual Gaussian elimination (column reduction) in field $\mathbb R$. However, as $n$ takes value in $\mathbb Z$, the row reduction of Gaussian elimination can only be performed in ring $\mathbb Z$ (the invertible scalars are only $\pm 1$). Performing Gaussian elimination in $\mathbb Z$ to solve linear equations is actually a way of finding the Hermite normal form of $A$, which is the standard technique in solving systems of linear Diophantine equations. The detailed procedures for solving \eq{eq:Axb_Z} are as follows.

We first find the Hermite normal form of $A$:
\begin{align}
UA=H,
\end{align}
where $U\in M_{p,p}(\mathbb Z)$ is a unimodular matrix with elements in $\mathbb Z$. As an upper triangular integer matrix, $H$ is the Hermite normal form of $A$. The equation becomes
\begin{align}\label{eq:Hxb}
Hx=b'+4n'
\end{align}
where
\begin{align}
b'&=Ub,\\
n'&=Un.
\end{align}
As $U$ is a unimodular matrix that is invertible in $\mathbb Z$, we can construct a unique solution $(x,n)$ of the original equation \eq{eq:Axb_Z} from a solution $(x,n')$ of the new equation \eq{eq:Hxb}, and vice versa.

Now we perform column Gaussian elimination for $H$ in $\mathbb R$, such that
\begin{align}
HV=
\begin{pmatrix}
I_r &0\\
0 & 0
\end{pmatrix},
\end{align}
where $V$ is an invertible matrix in $\mathbb R$, and $I_r$ is the identity matrix with size $r\times r$. The equation becomes
\begin{align}\label{eq:Iyb}
\begin{pmatrix}
I_r &0\\
0 & 0
\end{pmatrix}
x'=b'+4n'
\end{align}
where
\begin{align}\label{eq:xy}
x'=V^{-1}x
\end{align}
The solution $(x',n')$ of the new equation \eq{eq:Iyb} can be also used to construct a unique solution $(x,n')$ of \eq{eq:Hxb} because $V$ is invertible in $\mathbb R$ (recall that $x$ has elements in $\mathbb R$).

\eq{eq:Iyb} has solution $(x',n')$ ($x'\in \mathbb R^q,n'\in \mathbb Z^p$) if and only if the last $p-r$ elements of $b'$ are all 0 (mod 4), i.e.,
\begin{align}
b'_{p-r}\in (4\mathbb Z)^{p-r},
\end{align}
where we have used the notation
$b'=
\begin{pmatrix}
b'_r\\
b'_{q-r}
\end{pmatrix}
$.
If it is true, one can construct all of the solutions of $x'$ in \eq{eq:Iyb} and use them to obtain all $x$ by using \eq{eq:xy}.

In summary, to know whether \eq{eq:Axb} has solution $(x,n)$ ($x\in \mathbb R^p,n\in \mathbb Z^p$), we need to find the Hermite normal $H$ of $A$ ($UA=H$). First, denote the number of nonzero rows of $H$ by $r$. The next step is to check whether the last $p-r$ elements of $Ub$ are all 0 (mod 4). In practice, calculating the unimodular matrix $U$ is very time consuming. We only need to find the Hermite normal form of the augmented matrix $(A,b)$. The original equation (\ref{eq:Axb}) has the solution if and only if the element in the $(r+1)$-th row and $(q+1)$-th column of the Hermite normal form is 0 (mod 4).

We use SageMath \cite{sagemath} to calculate the Hermite normal form of the augmented matrix $(A,b)$ with size $p\times (q+1)=2^{20}\times (2^{12}+1)$. The rank of $A$ is $r=4035$, and the matrix element at the place $(r+1,q+1)= (4036, 4097)$ of the Hermite normal form is 4 (see \fig{fig:Hermite}). Therefore, the highly overdetermined \eq{eq:Ax_b} indeed has solutions. We can use the solutions to construct the group structure of the general group super-cohomology solutions by \eq{eq:group}.

We further check to see that the first nonzero element in each row of the Hermite normal form of $A$ is either 1 or 2. Thus, the solution $x$ can be chosen to have elements in $\frac{1}{2}\mathbb Z$ (the equation has no solution if we restrict $x$ in $\mathbb Z^q$). Also, the phase factor $e^{i\pi\theta}$ in $\nu_4^\mathrm{tot}$ takes values in $\mathbb Z_8 \cong \{ e^{in\pi/4} | n\in \mathbb Z\}$. This implies that the cohomology operation defined by the obstruction is in fact in $\mathbb Z_8$ rather than $\mathbb Z_4$. This is a very interesting hierarchical structure: the obstruction in the special group super-cohomology has a $\mathbb Z_2$ phase factor that is the fermion sign, the Majorana fermion decoration gives a $\mathbb Z_4$ phase factor obstruction from the Kitaev chain ($p$-wave superconductor), and the Abelian group structure of general group super-cohomology theory takes value in $\mathbb Z_8$.

\bibliography{bibfile.bib}

\end{document}